\documentclass[leqno]{amsart} 

\usepackage{amsmath}
\usepackage{amsfonts}
\usepackage{hyperref}
\usepackage{enumerate}

\theoremstyle{plain}
\newtheorem{thm}{Theorem}[section]{\bf}{\it}
\newtheorem{prop}[thm]{Proposition}{\bf}{\it}
\newtheorem{lemma}[thm]{Lemma}{\bf}{\it}
\newtheorem{lem}[thm]{Lemma}{\bf}{\it}
\newtheorem{cor}[thm]{Corollary}{\bf}{\it}
\newtheorem*{lem3.1'}{Lemma 3.1'}{\bf}{\it}
\newtheorem*{lem3.4'}{Lemma 3.4'}{\bf}{\it}
\newtheorem*{lem3.5'}{Lemma 3.5'}{\bf}{\it}

\newtheorem{defn}[thm]{Definition}{\bf}{\rm}
\newtheorem{rem}[thm]{Remark}{\it}{\rm}
\newtheorem{remark}[thm]{Remark}{\it}{\rm}

\newtheorem*{acknowledgement}{Acknowledgement}

\newenvironment{pf}{\par\medskip\noindent\textit{Proof}:\,}{\hspace*{\fill}\qed\medskip\par\noindent}
\newenvironment{pf*}[1]{\par\medskip\noindent\textit{#1}\,:}{\hspace*{\fill}\qed\medskip\par\noindent}  

\numberwithin{equation}{section}

\newcommand{\sphere}{{\mathbb S}}

\newcommand{\Span}{\operatorname{Span}}
\newcommand{\dist}{{\operatorname{dist}}}

\newcommand{\Ran}{{\operatorname{Ran}}}
\newcommand{\Tr}{{\operatorname{Tr}}}
\newcommand{\N}{{\mathbb N}}
\newcommand{\R}{{\mathbb R}}
\newcommand{\C}{{\mathbb C}}

\begin{document}

\title[Sharp Regularity for Wave Functions]{Sharp Regularity
       Results for Many-Electron Wave Functions\footnote{\copyright\ 2003 by the
       authors. This article may be reproduced in its entirety for
       non-commercial purposes.}}

\author[S. Fournais and M. \& T. Hoffmann-Ostenhof and 
       T. \O. S\o rensen]{S\o ren Fournais, Maria Hoffmann-Ostenhof, 
       Thomas Hoffmann-Ostenhof and
       Thomas \O stergaard S\o rensen}
\address[S. Fournais]
        {CNRS and Laboratoire de
         Math\'{e}matiques, 
         UMR CNRS 8628,
         Universit\'{e} Paris-Sud - B\^{a}t 425,
         F-91405 Orsay Cedex,
         France}
\address[M. Hoffmann-Ostenhof]
        {Institut f\"ur Mathematik,
         Universit\"at Wien,         
         Nordbergstra\ss e 15,
         A-1090 Vienna,
         Austria} 
\address [T. Hoffmann-Ostenhof]        
         {The Erwin 
         Schr\"odinger International 
         Institute for Mathematical Physics,
         Boltzmanngasse 9,
         A-1090 Vienna,
         Austria}
\address[T. Hoffmann-Ostenhof]
        {Institut f\"ur Theoretische Chemie, 
         Universit\"at Wien,
         W\"ahringer\-stra\ss e 17,
         A-1090 Vienna,                                              
         Austria}
\address[T. \O stergaard S\o rensen (on leave from)]
         {Department of 
         Mathematical Sciences,
         Aalborg University,
         Fredrik Bajers Vej 7G,
         DK-9220 Aalborg East,
         Denmark}
\address[T. \O stergaard S\o rensen (present address)]
         {Mathematisches Institut,
         Universit\"at M\"unchen,
         Theresienstra\ss e 39,
         D-80333 Munich,
         Germany}
\address[]{}

\email[S. Fournais]{soeren.fournais@math.u-psud.fr}
\email[M. Hoffmann-Ostenhof]{maria.hoffmann-ostenhof@univie.ac.at}
\email[T. Hoffmann-Ostenhof]{thoffman@esi.ac.at}
\email[T. \O. S\o rensen]{sorensen@mathematik.uni-muenchen.de}


\begin{abstract} 
We show that electronic wave functions $\psi$ of atoms and molecules 
have a representation $\psi=\mathcal F \phi$, where $\mathcal F$ is an explicit 
universal
factor, locally Lipschitz, and independent of the eigenvalue and the 
solution $\psi$ itself, and $\phi$ has locally bounded second derivatives.
This representation turns
out to be optimal as can already be demonstrated with the help of hydrogenic
wave functions. 
The proofs of these results are, in an essential way, based on a new 
elliptic regularity result which is  of independent interest.
Some identities that can be interpreted as cusp conditions 
 for second order derivatives of $\psi$ are 
derived.
\end{abstract}
\maketitle
\section{Introduction}
\subsection{Motivation and results}
\label{chap:intro}
 The non-relativistic quantum mechanical Hamiltonian of an \(N\)-elec\-tron 
 molecule with L  fixed nuclei is given by
\begin{equation*}
  H_{N,L}(\mathbf X,\mathbf Z)=-\Delta+V(\mathbf X,\mathbf Z)+U(\mathbf X,\mathbf Z),
\end{equation*} 
where  $V$, the Coulombic potential, is given by 
\begin{equation}\label{V}
  V\equiv V(\mathbf X,\mathbf
  Z)=-\sum_{j=1}^N\sum_{k=1}^L\frac{Z_k}{|X_k-x_j|}
  +\sum_{1\le i<j\le N} \frac{1}{|x_i-x_j|},
\end{equation}
and the internuclear repulsion $U$ by
\begin{equation*}
U(\mathbf X,\mathbf Z)=\sum_{1\le k<\ell\le L}\frac{Z_kZ_\ell}{|X_k-X_\ell|}.
\end{equation*}
 The latter is merely an additive term that will be neglected in the 
sequel and we will henceforth consider
\begin{equation}\label{H}
H\,\equiv\, H_{N,L}(\mathbf X,\mathbf Z)-U(\mathbf X,\mathbf Z).
\end{equation}
Above, $\mathbf x =(x_1,x_2,\dots, x_N)\in \mathbb R^{3N}$
denotes the positions of the $N$ electrons, with
$x_j=(x_{j,1},x_{j,2},x_{j,3})\in \mathbb R^3$ 
the position of the \(j^{\text{\it th}}\) electron. 
The positions of the $L$ nuclei with the postive charges  
$\mathbf Z=(Z_1,Z_2\dots, Z_L)\in \mathbf R_+^L$ are denoted 
by $\mathbf X = (X_1,X_2,\dots, X_L)\in\mathbf R^{3L}$ where 
$X_{k}=(X_{k,1},X_{k,2},X_{k,3})\in\mathbb R^3$ is the (fixed)
position of the \(k^{\text{\it th}}\) nucleus with charge $Z_{k}$, and
it is assumed that $X_{\ell}\neq X_k$ for $\ell\neq k$.  The Laplacian
corresponding to the $j^{\text{\it th}}$  electron is
$\Delta_j=\sum_{i=1}^3\frac{\partial^2}{\partial x_{j,i}{}^{2}}$ and 
so the Laplacian on $\mathbb R^{3N}$ is given by
$\Delta=\sum_{j=1}^N\Delta_j$. We also introduce the $3N$-dimensional
gradient by $\nabla=(\nabla_1,\ldots,\nabla_N)$.

The operator $H$ is selfadjoint on $L^2(\R^{3N})$ with operator domain 
$\mathcal D(H)=W^{2,2}(\R^{3N})$
\cite{Kato:1966}, and  it depends parametrically on
 \(\mathbf X\) and \(\mathbf Z\). 
In the case of an \(N\)-electron atom with (one) nucleus of charge $Z$ fixed
at the origin \(0\in\mathbb R^{3}\), \eqref{H} becomes
\begin{align}\label{Hatom}
  H&\equiv H_{N}(Z)=-\Delta +V\\
  &=\sum_{j=1}^N\Big(-\Delta_j-\frac{Z}{|x_j|}\Big)+
  \sum_{1\le i<j\le N}\frac{1}{|x_i-x_j|}.
  \nonumber
\end{align}

Generations of chemists and physicists have devoted a good
part of their research to the analysis of various problems
related to $H_{N,L}(\mathbf X,\mathbf Z)$. Most of the present
day understanding of atoms and molecules is based on the analysis 
of problems directly related to this operator, see any textbook in
atomic and molecular quantum mechanics. 

One of the central problems is the eigenvalue problem
\begin{equation}\label{S}
H\psi=E\psi, \: E\in \mathbb R,\:\: \psi\in L^2(\mathbb R^{3N}).
\end{equation}
Since the electrons are Fermions the \(N\)-electron wave function $\psi$
has to satisfy the Pauli Principle. This can be achieved in a spinless 
formulation  by requiring that $\psi$ transforms according to certain
irreducible representations of the symmetric group $\mathfrak S^N$. 
Our present work will not require any symmetry assumptions on $\psi$. More
precisely, we will consider local properties of distributional
solutions (locally \(L^{1}\)) in a domain \(\Omega\subseteq\R^{3N}\)
to $H\psi=E\psi$ where $E$ can be any real number. 

Within mathematics and mathematical physics 
 Schr\"odinger operators as \eqref{H} are studied  mostly from an operator 
theoretical point of view,
 see  the textbooks \cite{CFKS:1987}, \cite{Kato:1966},
 \cite{RS:1978}, and
\cite{Thirring:1981} as well as the recent
survey \cite{Simon:2000}. 

The PDE-aspects of \eqref{S} have been studied in relatively 
few works. We first note the following: Let $\Sigma(\mathbf X)$ denote the set of points
in $\mathbb R^{3N}$ where 
the potential  $V$ defined in \eqref{V} is singular.     
The function $V$ is real analytic in $\mathbb R^{3N}\setminus \Sigma (\mathbf X)$ and hence 
by classical results (see \cite[Section 7.5, pp. 177-180]{H0}),  so is $\psi$. 

Therefore a basic question is how to characterize the effect of the singularities 
of $V$ on the local behaviour of a solution $\psi$ of \eqref{S}.

In 1957 Kato \cite{Kato:1957} showed that 
a solution $\psi$  satisfying \eqref{S} is continuous in all of
\(\mathbb R^{3N}\) with locally
bounded first derivatives, i.e., \(\psi\) is locally Lipschitz. 
He also analyzed how $\psi$ behaves near the so-called
two-particle coalescence points, i.e., those points in $\Sigma(\mathbf X)$  
where exactly one term  in the sums representing $V$  (see \eqref{V})
is unbounded.
 
Generalizations with new insights for those points in \(\R^{3N}\) where more than one
term in \eqref{V} is singular   were obtained in \cite{HHS:1994} and 
more recently in \cite{HHO:2001}.
We mention that the present authors in \cite{FHHO:2002a},
\cite{FHHO:2002b}, and \cite{FHHO:2003} 
studied  the smoothness of the electron density, a question related 
to the present investigation; we shall not discuss this further here. 

Suppose we have a
solution $\psi$ to \(H\psi=E\psi, E\in\R\), with $H$
as in \eqref{H} or \eqref{Hatom}.
We want to find a  representation for $\psi$  
\begin{equation*}
\psi(x_1,\ldots, x_n)=\mathcal F(x_1,\ldots, x_n)\,\phi(x_1,\ldots, x_n)
\end{equation*}
 such that $\phi$ is  as smooth as possible and
$\mathcal F$ is a universal (i.e., not depending on \(\psi\) or \(E\))
positive factor 
reflecting the  behaviour 
of the potential $V$ near $\Sigma(\mathbf X)$.
This means that
for any two solutions $\psi_1,\psi_2$ of a fixed Schr\"odinger
operator \eqref{H} (or \eqref{Hatom}) the function
$\mathcal F$ will be the same, i.e.,
\begin{equation*}
\psi_1=\mathcal F\phi_1,\:\: \psi_2=\mathcal F \phi_2.
\end{equation*}   
Since it is already known from one-electron atoms 
that $\psi$ is just 
locally Lipschitz, 
$\mathcal F$ cannot be smoother than that.
We shall see that by choosing $\mathcal F$ in a special
way one can say a lot more. Let us first recall some of the ideas developed in 
\cite{HHO:2001}.

Suppose $\psi$ is a solution to $(-\Delta+V)\psi=E\psi$ in
\(\Omega\subseteq \R^{3N}\). 
Set $\psi=e^F\phi$, then $\phi$ satisfies 
\begin{equation}\label{expF}   
\Delta \phi+2\nabla F\cdot\nabla \phi+\big(\Delta F+|\nabla
F|^2+(E-V)\big)\phi=0. 
\end{equation}
Now assume  
$H=-\Delta+V$ is given by \eqref{H}. The specific nature of the 
Coulomb potential makes it possible to find an explicit $F$ such that
$\Delta F=V$, 
namely
\begin{equation*}
F({\mathbf x})\equiv F_2({\mathbf
  x}):=-\frac{1}{2}\sum_{j=1}^N\sum_{\ell=1}^{L}Z_{\ell}|X_\ell-x_j|\:  
+\:\frac{1}{4}
\sum_{1\le i<j\le N}|x_i-x_j|.
\end{equation*}
We have given $F$ an index \(2\) to indicate that $F_2$ is a sum of
 functions each only depending on the coordinates of two particles. 
If we insert $F_2$ into \eqref{expF}
we obtain 
\begin{equation*}
\Delta\phi_2+\nabla F_2\cdot\nabla \phi_2 +\big(|\nabla F_2|^2+E\big)\phi_2=0,
\end{equation*}
where we have also given $\phi$ an index \(2\) to show that it is associated 
with $F_2$. 
The regularity properties of \(\phi_{2}\) are now determined 
by the regularity of 
$\nabla F_2$, respectively, $|\nabla F_2|^2$. Since $\nabla F_2$ is
locally bounded, 
standard elliptic regularity theory (see Section~\ref{chap:GT})
gives us that 
\begin{equation}\label{C1a}
\phi_2\in C^{1,\alpha}(\Omega)\quad\text{ for }\quad\alpha\in(0,1).
\end{equation}
(For the definition of the H\"older-spaces \(C^{k,\alpha}\), see
Definition~\ref{def:Holder}). 
Since $\nabla F_2$ is just bounded and not continuous,
one cannot in general expect anything better than \eqref{C1a}.
Note that since $\psi=e^{F_2}\phi_2$ we have 
\begin{equation}\label{cusp1}
\nabla \psi-(\nabla F_2)\psi\in C^{\alpha}(\Omega)\quad\text{
  for }\quad\alpha\in(0,1). 
\end{equation}
This is a general formulation of Kato's cusp condition \cite{Kato:1957} 
which plays an
important role in the numerical treatment of \eqref{S}. (Here, and in
the sequel, by a `cusp condition' we 
understand a condition a solution \(\psi\) has to satisfy at a point in the 
singular set \(\Sigma(\bf X\))).

We are now ready to state our main result about the regularity
of $\psi$.
\begin{thm}\label{thm:main:Jastrow}
Suppose $\psi$ is a solution to 
$H\psi=E\psi$ in \(\Omega\subseteq\R^{3N}\) where $H$ is given by \eqref{H}. 
Define \(y_{i,\ell}=x_{i}-X_{\ell}\), \(i\in\{1,\ldots,N\},
\ell\in\{1,\ldots,L\}\).
Let 
\begin{equation}\label{F23}
\mathcal F=e^{F_2+F_3}  
\end{equation}
with
\begin{align}
 \label{F2}
  &F_2({\mathbf
  x})=-\frac{1}{2}\sum_{\ell=1}^{L}\sum_{i=1}^NZ_{\ell}|y_{i,\ell}| 
  +\frac{1}{4}
  \sum_{1\le i<j\le N}|x_i-x_j|,\\
 \label{F3}
  &F_3({\mathbf x})=
C_0\sum_{\ell=1}^L\sum_{1\le i<j\le N}Z_\ell\,
  (y_{i,\ell}\cdot y_{j,\ell})\,\ln\big(|y_{i,\ell}|^2+
  |y_{j,\ell}|^2)\big),
\end{align}
where $C_0=\frac{2-\pi}{12\pi}$.

Then 
\begin{equation}\label{Factor}
  \psi=\mathcal F\phi_3
\end{equation}
with 
\begin{equation}\label{C11}
  \phi_3\in C^{1,1}(\Omega).
\end{equation}

Furthermore this representation is optimal in the following sense:
There is no other function $\widetilde{\mathcal F}$ depending only on 
 on $\mathbf X, \mathbf Z$ and on $N$, but   not on
$ \psi$ or $ E$ itself, such that $\psi =\widetilde{\mathcal F}\phi$ with 
$\phi$ having more regularity than $C^{1,1}(\Omega)$. 
\end{thm} 
\begin{rem}
  \(\, \)
  \begin{enumerate}[\rm (i)]
\item
Of course one can consider more
 general Hamiltonians, for instance
   molecular Hamiltonians
   where the nuclei are 
 allowed to move. Kato~\cite{Kato:1957} considered this case. Our results, suitably 
 modified, extend to this 
 situation. We concentrate on the model
 with fixed nuclei since this is the `standard model' in molecular  physics.
\item
For the proof of Theorem \ref{thm:main:Jastrow} a special regularity result (see
Theorem~\ref{thm:abstract}) 
for solutions of the Poisson equation $\Delta u=g$ will be vital.
Roughly speaking, if $g\in L^\infty$ has a certain multiplicative structure,
we can show that $u\in C^{1,1}$, and not only \(u\in C^{1,\alpha},
\alpha\in(0,1)\) as in general (see
Proposition~\ref{prop:GTbis}). This result is of independent interest.
  \item 
Note that each term in the sum $F_2$ is either a term involving 
the coordinates of one electron and one nucleus, or the coordinates of
two electrons, whereas
the terms in $F_3$ involve the coordinates of two electrons and one
nucleus. In the representation \eqref{F3} of  $F_3$ no terms involving  
the  coordinates of three electrons occur; see
Section~\ref{chap:C-1-1} for details. 
\\ 
\noindent
  The fact that no terms involving the coordinates of four and more particles
  show up in $F_3$ stems from the 
  fact that in the summands contributing to $|\nabla F_2|^2$  only
  terms involving  at most three particle coordinates occur (again, see
  Section~\ref{chap:C-1-1} for details).     
  \item
An immediate consequence of Theorem \ref{thm:main:Jastrow} is the following
shar\-pe\-ning of \eqref{cusp1}:
\begin{equation*}
\nabla \psi-\psi(\nabla F_2+\nabla F_3)\in C^{0,1}(\Omega).
\end{equation*}
 \item 
  Attempts to approximate  many-particle wave functions by  a product 
  as in \eqref{Factor} are common in computational chemistry and 
  physics. There, such an $\mathcal F$ is usually  called a `Jastrow
  factor'. 
\end{enumerate}
\end{rem}
It is also interesting to consider the regularity of $\psi$ near 
the zero-set $\mathcal N(\psi)=\{\mathbf x\in\mathbb
R^{3N}\:|\:\psi=0\}$ of $\psi$. 
A simple argument shows that Theorem \ref{thm:main:Jastrow}
actually implies that 
$\nabla \psi\::\mathcal N(\psi)\mapsto \mathbb R^{3N}$ is locally Lipschitz, whereas
$\nabla \psi$ is just bounded in $\Sigma(\mathbf X)\setminus \mathcal
N(\psi)$. By `locally Lipschitz'  we here mean the following: For all
closed balls \(K\subset\R^{3N}\), there is a constant \(C=C(K)\) such
that \(|\nabla\psi(\mathbf x)-\nabla\psi(\mathbf y)|\leq C(K)|\mathbf
x-\mathbf y|\) for all \(\mathbf x,\mathbf y\in \mathcal N(\psi)\cap K\).
Indeed, writing
$\nabla\psi=\psi\nabla(F_2+F_3)+\exp(F_2+F_3)\nabla\phi_3$, we get,
for \({\mathbf 
  x}\in\mathcal N(\psi)\), that \(\nabla\psi(\mathbf
x)=\exp(F_{2}(\mathbf x)+F_{3}(\mathbf x))\nabla\phi_{3}(\mathbf x)\)
since \(\nabla(F_2+F_3)\) is bounded. The assertion follows, since
both \(\exp(F_2+F_3)\) and \(\nabla\phi_{3}\) are Lipschitz in \(\R^{3N}\).

In \cite{HHN:1995} it was shown for a wide class of potentials
that at their zero-sets real valued distributional solutions (which
for these potentials are then actually continuous functions) to
$(-\Delta+V)u=0$ are, 
roughly speaking, by one degree smoother than
away from their zero sets. So the observation above 
extends these results to the Coulombic case. 
The potentials considered in \cite{HHN:1995}
were of Kato type, $K^{n,\delta}$,
where $n$ is the dimension (in our case, \(n=3N\)) and $\delta\in(0,2)$; see \cite{Simon:1982}
for  definitions and many far-reaching results concerning these
potentials.
In \cite{Simon:1982} (see also \cite{Simon:1982bis}) it was shown that solutions are locally
\(C^{\delta}\) for \(\delta<1\) and \(C^{1,\delta-1}\) for
\(\delta\in(1,2)\). However, since the Coulomb potential \(V\) in
\eqref{V} is in $K^{3N,\delta}$ for all $\delta<1$, but not in
\(K^{3N,1}\) these results are not sharp and actually weaker than
Kato's result.
  
  It is not surprising that logarithms occur in~\eqref{F3}. Such 
terms have been
  considered in classical work by Fock~\cite{Fock} for the atomic case; see 
  Morgan \cite{Morgan:1986} for an analysis of these `Fock-expansions' for 
  two-electron atoms. That paper also contains many
  references to earlier work on such expansions.
\begin{pf*}{Proof of the optimality of the representation \eqref{Factor}}
It suffices to find a simple example. 
Consider the  one electron atom whose Hamiltonian is given on $\mathbb R^3$
by 
\begin{align*}
  H=-\Delta-\frac{Z}{|x|}\quad,\quad x=(x_{1},x_{2},x_{3})\in\mathbb R^{3}.
\end{align*}
With \(\psi_1(x)=e^{-\frac{Z}{2}|x|}\) and
\(\psi_2=x_1e^{-\frac{Z}{4}|x|}\) we have
\begin{align*}
  H\psi_{1}={}-\tfrac{Z^{2}}{4}\,\psi_{1}
  \qquad,\qquad
  H\psi_{2}={}-\tfrac{Z^{2}}{16}\,\psi_{2}.
\end{align*}
Write
$\psi_1=\mathcal F \phi^{(1)}$ and $\psi_2=\mathcal F \phi^{(2)}$.
Now $\psi_1>0$ and if we had an $\mathcal F$ which would allow more regularity of 
the \(\phi^{(i)}\)'s,\, then 
\begin{equation}\label{quotient}
  \frac{\phi^{(2)}}{\phi^{(1)}}=
  \frac{\psi_2}{\psi_1}=x_1e^{\frac{Z}{4}|x|}
\end{equation}
would be better behaved than just $C^{1,1}$. But near the origin 
the right hand side of \eqref{quotient} behaves like 
$x_1(1+\tfrac{Z}{4}|x|)$ and this is just $C^{1,1}$, i.e., the second
derivatives 
are bounded 
but not continuous.
\end{pf*}
The results in Theorem \ref{thm:main:Jastrow} are not well suited
for obtaining {\it a~priori} estimates. In particular  neither $F_2$ nor
$F_3$  stay bounded as  $|\mathbf x|$ tends to infinity so that 
if, say, $\psi\in L^2(\mathbb R^{3N})$ then $\phi_3$ is not necessarily in  
$L^2(\mathbb R^{3N})$. These shortcomings will be dealt with below in a similar
way as in \cite{HHO:2001}. 
\begin{defn}\label{Cutoff}
Let \(\chi\in C_{0}^{\infty}(\R)\), \(0\leq \chi\leq1\), with
\begin{align}
  \label{eq:def_cutoff}
  \chi(x)=
  \begin{cases}
    1& \text{ for } |x|\leq1 \\
    0& \text{ for } |x|\geq2.
  \end{cases}
\end{align}
We define
\begin{equation}\label{F23cut}
  F_{\text{\rm cut}}=F_{2,\text{\rm cut}}+F_{3,\text{\rm cut}},
\end{equation}
where 
\begin{align}
 \label{eq:def_F2cut}
  &F_{2,\text{\rm cut}}({\mathbf
  x})=-\frac{1}{2}\sum_{\ell=1}^{L}\sum_{i=1}^NZ_{\ell}\,
  \chi(|y_{i,\ell}|)\,|y_{i,\ell}|     
  \\&\qquad\qquad\qquad\qquad\qquad\qquad
  +\frac{1}{4}
  \sum_{1\le i<j\le N}\chi(|x_i-x_j|)\,|x_i-x_j|,\nonumber
  \\
 \label{eq:def_F3cut}
  &F_{3,\text{\rm cut}}({\mathbf x})=
  \\&\quad C_0\sum_{\ell=1}^L\sum_{1\le i<j\le N}Z_\ell\,
  \chi(|y_{i,\ell}|)\chi(|y_{j,\ell}|)
  ( y_{i,\ell}\cdot y_{j,\ell})\,\ln\big(|y_{i,\ell}|^2+
  |y_{j,\ell}|^2)\big),
  \nonumber
\end{align}
and where $C_0$ is the constant from \eqref{F3}.
We also introduce $\phi_{3,\text{\rm cut}}$ by
\begin{equation}\label{phcut}
  \psi =e^{F_{\rm cut}}\phi_{3,\text{\rm cut}}.
\end{equation}
\end{defn}
\begin{thm}\label{thm:main:apriori}
Suppose $\psi$ is a  solution to $H\psi=E\psi$ in \(\R^{3N}\). 
Then for all \(0<R<R'\) there exists a constant $C(R,R')$, not
depending on $\psi$ nor \(\mathbf x_0\in\R^{3N}\), such that for  
any second order derivative,
\begin{equation*}
\partial^2=\frac{\partial^2}
{\partial x_{i,k}\partial x_{j,\ell}},\:\:i,j=1,2,\dots, N, \:\:\: 
k,\ell =1,2,3,
\end{equation*}
the following estimate holds:
\begin{equation}\label{apriori}
\|\partial^2\psi -\psi\,\partial^2\!
F_{\text{\rm cut}}\|_{L^\infty(B_{3N}(\mathbf x_0,R))}\le
C(R,R')\|\psi\|_{L^\infty(B_{3N}(\mathbf x_0,R'))}.
\end{equation}
\end{thm}
\begin{rem}
  \label{rem:aprioris}
  Theorem~\ref{thm:main:apriori} strengthens results obtained in
  \cite{HHO:2001}. More precisely, to prove
  Theorem~\ref{thm:main:apriori} we will show that
  \begin{equation}\label{eq:a_priori}
    \|\phi_{3,\text{\rm cut}}\|_{C^{1,1}(B_{3N}(\mathbf x_0,R))}\le 
     C(R, R')\|\phi_{3,\text{\rm cut}}\|_{L^\infty(B_{3N}(\mathbf
     x_0,R'))}. 
 \end{equation}
 The estimate \eqref{apriori} is then a trivial consequence of
 \eqref{eq:a_priori}. (On the other hand, \eqref{apriori} and
 \eqref{eq:a_priori_1der} imply \eqref{eq:a_priori}).

 The estimate \eqref{eq:a_priori} is a strengthening of Proposition~\ref{c-1-alpha}
 below to \(\alpha=1\). We state and prove the proposition here, since
 we need it in the proof of \eqref{eq:a_priori}. It essentially
 follows from ideas in \cite{HHO:2001}.
 \begin{prop}\label{c-1-alpha}
   Suppose $\psi$ is a  solution to $H\psi=E\psi$ in \(\R^{3N}\). 
   Then for all \(0<R<R'\) and all \(\alpha\in(0,1)\) there exists a
   constant $C(\alpha,R,R')$, not 
   depending on $\psi$ nor \(\mathbf x_0\in\R^{3N}\), such that, with
   \(\phi_{3,\text{\rm cut}}\) defined as above,
   \begin{equation}\label{eq:a_priori_1der}
    \|\phi_{3,\text{\rm cut}}\|_{C^{1,\alpha}(B_{3N}(\mathbf x_0,R))}\le 
   C\|\phi_{3,\text{\rm cut}}\|_{L^\infty(B_{3N}(\mathbf x_0,R'))}.
 \end{equation}
 \end{prop}
 \begin{pf*}{Proof of Proposition~\ref{c-1-alpha}}
   Note first that with 
   \(\psi=e^{F_{2,\text{\rm cut}}}\phi_{2,\text{\rm cut}}\),
   \eqref{expF} and \(\Delta F_{2}=V\) gives
   \begin{align}
     \label{eq:phi_2_cut}
     \Delta \phi_{2,\text{\rm cut}}&+2\nabla F_{2,\text{\rm cut}}\cdot\nabla
     \phi_{2,\text{\rm cut}}\\
     &+\big(\Delta(F_{2,\text{\rm cut}}-F_{2})
     +|\nabla F_{2,\text{\rm cut}}|^2+E\big)\phi_{2,\text{\rm cut}}=0. \nonumber 
   \end{align}
   It follows from the form of \(F_{2,\text{\rm cut}}\) and \(F_{2}\)
   (see \eqref{eq:def_F2cut}, \eqref{eq:def_cutoff}, and \eqref{F2})
   that the coefficients in \eqref{eq:phi_2_cut} above are uniformly
   bounded in \(\R^{3N}\). Therefore, \eqref{eq:a_priori_1der}, with
   \(\phi_{2,\text{\rm cut}}\) instead of \(\phi_{3,\text{\rm cut}}\),
   follows from Proposition~\ref{prop:GTbis}. To get \eqref{eq:a_priori_1der} with
   \(\phi_{3,\text{\rm cut}}\), note that 
   \(\phi_{3,\text{\rm cut}}=e^{-F_{3,\text{\rm
   cut}}}\phi_{2,\text{\rm cut}}\), and that \(F_{3,\text{\rm
   cut}}\in C^{1,\alpha}(\R^{3N})\) and has compact support (see
   \eqref{eq:def_F3cut} and \eqref{eq:def_cutoff}). 
 \end{pf*}
\end{rem}
We point out some consequences of Theorem~\ref{thm:main:apriori} which 
can be viewed as cusp conditions for second order derivatives of \(\psi\).
Indeed,  we can relate the singularities of the second order derivatives 
of $F_{\text{\rm cut}}$ with those of the second order derivatives of 
$\psi$ in a precise way, thereby obtaining certain identities. 
Here we only explicitly state some representative cases. 
\begin{cor}\label{2Cusp}
Let $\psi$ be a 
solution to $H\psi=E\psi$ in \(R^{3N}\) with $H$ given by
\eqref{H}. 
\begin{enumerate}[\rm (i)]
\item
Let $1\le i<j\le N$, and 
fix any point $\mathbf z_0=(z_{1},\ldots,z_{N})\in \mathbb R^{3N}$
  with \(z_{i}=z_{j}\equiv z\).

Then
\begin{equation}\label{lim12}
  \lim_{R\to0}\Big\|\big(|x_i-x_j|\:\nabla_i\cdot\nabla_j\:\psi
  \big)+\frac{1}{2}\psi(\mathbf
  z_0)\Big\|_{L^{\infty}(B_{3N}(\mathbf z_{0},R))}=0.
\end{equation}
\item
Let \(1\leq i\leq N\), \(1\leq \ell\leq L\), and 
fix any point $\mathbf z_0=(z_{1},\ldots,z_{N})\in \mathbb R^{3N}$
with \(z_{i}=X_{\ell}, z_{j}\neq X_{\ell}, j\neq i\). 

Then
\begin{equation}\label{lim1}
  \lim_{R\to0}\Big\|\big(|x_i-X_\ell|\:\Delta_i\psi\big)
  +Z_\ell\;\psi(\mathbf z_{0})\Big\|_{L^{\infty}(B_{3N}(\mathbf z_{0},R))}=0.
\end{equation}
\end{enumerate}
\end{cor} 
\begin{pf}
In order to show \eqref{lim12} we first show that 
\begin{equation}\label{limF12}
  \lim_{\mathbf x\rightarrow \mathbf z_{0}}|x_i-x_j|\:\nabla_i\cdot\nabla_j\,  
F_{\text{\rm cut}}(\mathbf x)=-\frac{1}{2}.
\end{equation} 
It suffices to consider the limits for $|x_i-x_j|\:\nabla_i\cdot\nabla_jF_2$ and 
$|x_i-x_j|\:\nabla_i\cdot\nabla_jF_3$.
An easy calculation shows that 
\begin{equation*}
\lim_{\mathbf x\rightarrow \mathbf
  z_{0}}\:|x_i-x_j|\nabla_i\cdot\nabla_j\,F_2(\mathbf x)= 
-\frac{1}{2}. 
\end{equation*}
If $z\neq X_\ell$ for all $\ell$ then $\nabla_i\cdot\nabla_j\;F_3$ is
smooth near \(\mathbf z_{0}\). We therefore  
only need to consider the case $z=X_\ell$. We have 
\begin{align*}
  \nabla_i&\cdot\nabla_j F_3=\\
  &C_0Z_\ell\nabla_i\cdot\nabla_j\big\{\big((
  x_i-X_\ell)\cdot (x_j-X_\ell)\big)
  \ln\big(|x_i-X_\ell|^2+|x_j-X_\ell|^2\big)\big\}\\
  &=3C_0Z_\ell\, \ln\big(|x_i-X_\ell|^2+|x_j-X_\ell|^2\big)+\eta,
\end{align*} 
where $\eta$ is bounded in a neighbourhood of $\mathbf z_0$. Noting that 
\begin{align*}
  |x_i-x_j|\le \sqrt2\:\big(|x_i-X_\ell|^2+|x_j-X_\ell|^2\big)^{1/2}, 
\end{align*}
we see that
\begin{equation*}
\lim_{\mathbf x\rightarrow \mathbf
  z_{0}}\:|x_i-x_j|\:\nabla_i\cdot\nabla_j 
F_3(\mathbf x)=0.
\end{equation*}
Using the triangle inequality we obtain 
\begin{align*}
  &\Big\|\:|x_i-x_j|\:(\nabla_i\cdot\nabla_j\psi)+\frac{1}{2}
  \psi(\mathbf z_0)\Big\|_{L^{\infty}(B_{3N}(\mathbf z_{0},R))}
   \\
   &\le \Big\||x_i-x_j|\:\big((\nabla_i\cdot\nabla_j\psi)-
   (\nabla_i\cdot\nabla_j F_{\text{\rm
  cut}})\psi\big)\Big\|_{L^{\infty}(B_{3N}(\mathbf z_{0},R))}\\  
   &\qquad\qquad+\Big\||x_i-x_j|\: (\nabla_i\cdot\nabla_j\,F_{\text{\rm
   cut}})\psi
   +\frac{1}{2}\psi(\mathbf z_{0})\Big\|_{L^{\infty}(B_{3N}(\mathbf z_{0},R))}.
\end{align*}
This, \eqref{apriori}, and \eqref{limF12} imply \eqref{lim12}.

The proof of
\eqref{lim1} is similar. Just note
that  
\begin{equation*}
|x_i-X_\ell|\:\Delta_iF_2=
-Z_\ell+|x_i-X_\ell|\Big(\:\sum_{j\neq i}\frac{1}{2|x_j-x_i|}-
\sum_{k\neq\ell}^L\frac{Z_k}{|x_i-X_k|}\Big).
\end{equation*}
\end{pf}
\subsection{Organisation of the paper}
\label{chap:organisation}
For simplicity we shall only give the proofs of Theorems~\ref{thm:main:Jastrow} and
\ref{thm:main:apriori}  for the atomic case (i.e., \(\ell=1, X_{1}=0\)
and \(Z_{1}=Z\), see \eqref{Hatom}). Indeed, no additional complications
arise for molecules. Also, we only give the proof of
Theorem~\ref{thm:main:Jastrow} in the case \(\Omega=\R^{3N}\).

In subsection~\ref{chap:notation} we define some notation to be used in
the entire paper. Section~\ref{chap:GT} contains standard elliptic regularity
results in
subsection~\ref{subsection4.1}. Subsection~\ref{subsection4.2}
contains in particular the elliptic regularity result
Theorem~\ref{thm:abstract}, which is proved in
subsection~\ref{subsection4.3}. Theorem~\ref{thm:abstract} is the
essential new mathematical input necessary for the proofs of
Theorems~\ref{thm:main:Jastrow} and \ref{thm:main:apriori}. These
proofs are given in Section~\ref{chap:C-1-1}---the proof of
Theorem~\ref{thm:main:Jastrow} in subsection~\ref{chap:proof:Jastrow} and
that of Theorem~\ref{thm:main:apriori} in
subsection~\ref{chap:proof:apriori}. The Appendices~\ref{chap:kappa},
\ref{chap:nu}, and \ref{chap:P_2_two_elec} contain the construction of
solutions to certain Poisson equations. These solutions is another
important ingredient for the proofs of the main theorems.
\subsection{Notation}
\label{chap:notation}
Throughout the paper, constants occuring in inequalities will be
denoted by the symbol \(C\), although their actual value might change
from line to line. 

For \(x\in\R^{n}\) (\(n\geq2\)) we write \(x=r\omega\), with
\(r=|x|\), \(\omega=x/|x|\in\sphere^{n-1}\), the unit sphere in
  \(\R^{n}\). Denote by \(B_{n}(x,r)\) the open ball of radius
  \(r>0\) around \(x\).

We denote by \(Y_{l,m}(\omega)\) the normalised (in
\(L^{2}(\sphere^{n-1})\)) real valued spherical harmonics of
degree \(l, l\in\N_{0}\), with \(m=1,\ldots,h(l)-1\), where 
\begin{align}
  \label{eq:h(l)}
  h(l)=\frac{(2l+n-2)(l+n-3)!}{(n-2)!\,l\,!}.
\end{align}
Then \(\{Y_{l,m}\}_{l,m}\) constitutes an orthonormal basis in
\(L^{2}(\sphere^{n-1})\).

The \(Y_{l,m}\)'s are the eigenfunctions for \(\mathcal{L}^{2}\), the
Laplace-Beltrami operator on \(\sphere^{n-1}\):
\begin{align*}
  \mathcal{L}^{2}Y_{l,m}=l(l+n-2)Y_{l,m},
\end{align*}
where \({}-\frac{\mathcal{L}^{2}}{r^{2}}\) is the angular part of the
  Laplacian in \(\R^{n}\), so
  \begin{align*}
    {}-\Delta={}-\frac{\partial^{2}}{\partial
      r^{2}}-\frac{n-1}{r}\frac{\partial}{\partial
      r}+\frac{\mathcal{L}^{2}}{r^{2}}.
  \end{align*}

We define \(\mathcal{P}_{l,m}^{(n)}\) to be the orthogonal projection
in \(L^{2}(\sphere^{n-1})\) 
on \(Y_{l,m}\):
\begin{align*}
  \big(\mathcal{P}_{l,m}^{(n)}f\big)(\omega)
  =Y_{l,m}(\omega)\int_{\sphere^{n-1}}Y_{l,m}(\omega)f(\omega)\,d\omega
  \quad,\quad
  f\in L^{2}(\sphere^{n-1}),
\end{align*}
and 
\begin{align}
\label{def:proj}
  \mathcal{P}_{l}^{(n)}=\sum_{m=0}^{h(l)-1}\mathcal{P}_{l,m}^{(n)}.
\end{align}
We denote 
\(\mathfrak{h}_{l}^{(n)}=\Ran(\mathcal{P}_{l}^{(n)})\).

By abuse of notation, for a function \(\ f:\R^{n}\to\C\) we write
\(f(r\omega)=f(x)\), and, whenever 
\(f(r_{0}\,\cdot):\sphere^{n-1}\to\C\) is in
\(L^{2}(\sphere^{n-1})\) for some \(r_{0}\in(0,\infty)\), we write
\begin{align*}
  \big(\mathcal{P}_{l,m}^{(n)}f\big)(r_{0}\omega)
  =Y_{l,m}(\omega)\int_{\sphere^{n-1}}Y_{l,m}(\omega)f(r_{0}\omega)\,d\omega 
  \equiv f_{l,m}(r_{0})Y_{l,m}(\omega).
\end{align*}
\section{Elliptic regularity}
\label{chap:GT}
In this section we collect results on regularity of solutions to
second order elliptic equations needed for the proof of
Theorems~\ref{thm:main:Jastrow}. 
and \ref{thm:main:apriori}. 
The results fall in two parts, known
ones (in subsection~\ref{subsection4.1}) and new ones, developed for
our purpose, and of interest in themselves. The latter are in
subsection~\ref{subsection4.2}. The result of main interest is
Theorem~\ref{thm:abstract}, which is proved in subsection~\ref{subsection4.3}.
\subsection{Known results}
\label{subsection4.1}
$\, $

We start by recalling the definition of H\"older continuity:
\begin{defn}
  \label{def:Holder}
  Let \(\Omega\) be a domain in \(\R^{n}\), \(k\in\N\), and
  \(\alpha\in(0,1]\). We say that a function \(u\) belongs to
  \(C^{k,\alpha}(\Omega)\) whenever
  \(u\in C^{k}(\Omega)\), and for all \(\beta\in\N^{n}\) with
  \(|\beta|=k\), and all open balls $B_{n}(x_{0},r)$ with
  $\overline{B_{n}(x_{0},r)}\subset\Omega $, we have
  \begin{align*}
    \sup_{x,y\in B_{n}(x_{0},r),\,x\neq y}
    \!\!\!\!\!\!\!\!\!
    \frac{|D^{\beta}u(x)-D^{\beta}u(y)|}{|x-y|^{\alpha}}
    \leq C(x_{0},r).
  \end{align*}
  For any domain \(\Omega'\), with \(\Omega'\subset\subset\Omega\),
  we define the following norms:
  \begin{align}
    \nonumber
    \|u\|_{C^{k,\alpha}(\Omega')}&=\sum_{|\beta| \leq k}\|D^{\beta}
    u\|_{L^{\infty}(\Omega')}
    +[u]_{k,\alpha,\Omega'},
    \\\nonumber
    [u]_{k,\alpha,\Omega'}
    &=\sum_{|\beta| = k}\sup_{x,y\in\Omega',\,x\neq y}
    \frac{|D^{\beta}u(x)-D^{\beta} u(y)|}{|x-y|^{\alpha}}. 
  \end{align}
  For \(k=0\) we use the notation \(C^{\alpha}(\Omega)\equiv
  C^{0,\alpha}(\Omega)\) and \([u]_{\alpha,\Omega'}\equiv[u]_{0,\alpha,\Omega'}\).

Furthermore, for a function \(u\in C^{\alpha}(\R^{n}\setminus\{0\})\)
we define
\begin{align}
  \|u\|_{C^{\alpha}(\sphere^{n-1})}&=\sup_{\sphere^{n-1}}|u| + 
  [u]_{\alpha,\sphere^{n-1}}, \\
  [u]_{\alpha,\sphere^{n-1}}&=
    \sup_{x,y\in\sphere^{n-1},\,x\neq y}\frac{|u(x)-u(y)|}{|x-y|^{\alpha}}.
   \nonumber
\end{align}
\end{defn}
We will need the following result on elliptic
regularity in order to conclude
that the solutions of elliptic second order equations with
bounded coefficients are $C^{1,\alpha}$. The proposition is a
reformulation of Corollary~8.36 in Gilbarg and Trudinger~\cite{GandT},
adapted for our purposes:
\begin{prop}
  \label{prop:GTbis}
  Let \(\Omega_{0}\) be a bounded domain in \({\mathbb
  R}^{n}\) and suppose \(u\in
  W^{1,2}(\Omega_{0})\) is a weak solution of \(\Delta u
  +\sum_{j=1}^{n}b_{j}D_{j}u+Wu=g\) in \(\Omega_{0}\), where
  \(b_{j},W,g\in L^{\infty}(\Omega_{0})\). Then \(u\in
  C^{1,\alpha}(\Omega_{0})\) for all \(\alpha\in(0,1)\) and
  for any domains \(\Omega',\Omega\),
  \(\overline{\Omega'}\subset\Omega\),
  \(\overline{\Omega}\subset\Omega_{0}\) we have 
  \begin{align*}
    \|u\|_{C^{1,\alpha}(\Omega')}\leq
    C\big(\sup_{\Omega}|u|+\sup_{\Omega}|g|\big)
  \end{align*}
  for \(C=C(\alpha,n,M,\dist(\Omega',\partial\Omega))\), with
  \begin{align}
    \max \{1,\max_{j=1,\ldots,n}\|b_{j}\|_{L^{\infty}(\Omega)},
 \|W\|_{L^{\infty}(\Omega)},\|g\|_{L^{\infty}(\Omega)}\} \leq M.
    \nonumber
  \end{align}
\end{prop}

We further need results concerning the regularity of solutions of the
Poisson equation. These regularity properties are based on the
regularity properties of the Newton potential of the
inhomogeneity. For our further considerations we recall here the
properties of this function.

Let \(g\in L^{\infty}(\Omega)\) for \(\Omega\) a bounded domain in
\(\R^{n}\), \(n\geq2\). The Newton potential of \(g\) is the function
\(w\) defined on \(\R^{n}\) 
by
\begin{align}
\label{Newton}
  w(x)&=\int_\Omega\Gamma(x-y)g(y)dy
  \intertext{with}
  \Gamma(x)&=\left\{
  \begin{array}{ll}
     \frac{1}{2\pi}\ln(|x|),& n=2, \\
     \frac{1}{(2-n)\;
     |\sphere^{n-1}|}\;|x|^{2-n}, & n\geq 3. \\
  \end{array}\right.
  \nonumber
\end{align}
From \cite[Theorem 10.2 and 10.3]{Li-loss} we have
\begin{prop}
  \label{prop:newton}
  Let \(\Omega\subset\R^{n}, n\geq2\), be a bounded domain, then:
\begin{enumerate}[\rm (i)]
\item If \(g\in L^{\infty}(\Omega)\), then \(w\in
  C^{1,\alpha}(\Omega)\) for all \(\alpha\in(0,1)\), and
  \(\Delta w=g\) in \(\Omega\) holds in the distributional sense.
\item If \(g\in C^{k,\alpha}(\Omega)\) for some \(k\in\N\) and some
  \(\alpha\in(0,1)\), then \(w\in C^{k+2,\alpha}(\Omega)\).
\end{enumerate}
\end{prop}
Since every solution to the Poisson equation can be written as a sum
of the Newton potential of the inhomogeneity and a harmonic function,
the above implies in particular the following well-known result:
\begin{prop}
  \label{prop:GT}
  Let \(g\in C^{k,\alpha}(\Omega_{0})\) for some \(k\in\N\) and some
  \(\alpha\in(0,1)\), and assume \(u\) is a weak solution to \(\Delta
  u=g\) in \(\Omega_{0}\).

  Then \(u\in C^{k+2,\alpha}(\Omega_{0})\). 
  Furthermore, for any domains \(\Omega',\Omega\),
  \(\overline{\Omega'}\subset\Omega\),
  \(\overline{\Omega}\subset\Omega_{0}\), we have  
  \begin{align}
    \|u\|_{C^{k+2,\alpha}(\Omega')}\leq
    C\big(\sup_{\Omega}|u|+
    \|g\|_{C^{k,\alpha}(\Omega)}\big),
  \end{align}
with  \(C=C(n, k, \alpha,\dist(\Omega',\partial\Omega))\).
\end{prop}

The next lemma, which is taken from Gilbarg and Trudinger~\cite[Lemma
4.2]{GandT}, is essential for the proof of the main regularity result in
subsection~\ref{subsection4.2}.

\begin{lem}
\label{GT42}
Let \(\Omega\) be a bounded domain in \(\R^{n}, \: n\ge 2\)  and let 
\(g\in C^\alpha(\Omega)\cap L^{\infty}(\Omega)\) for some \(\alpha\in
(0,1]\).

Then the Newton potential \(w\) of \(g\) (given in \eqref{Newton})
satifies, 
for \(x\in\Omega\)
and \(i,j=1,2,\dots,n\),
\begin{align}
\label{Dij}
  D_{ij}w(x)&=\int_{\Omega_0}D_{ij}\Gamma(x-y)\big(g(y)-g(x)\big)\,dy
  \nonumber\\
  &\quad\quad\quad
  -g(x)\int_{\partial\Omega_0}D_i\Gamma(x-y)\nu_j(y)\,d\sigma(y).
\end{align}
Here, \(\Omega_0\) is any bounded domain containing \(\Omega\) for which  
the divergence theorem holds, and \(g\) is extended to vanish outside 
\(\Omega\). In the last integral, \(d\sigma\) denotes the surface
measure of \(\partial\Omega_{0}\), and \(\nu_{j}\) the \(j\)-th
coordinate of its (outwards directed) normal vector.
\end{lem}

\subsection{New results}
\label{subsection4.2}
$\, $

We here collect a number of more explicit regularity results needed in
the proof of Theorems~\ref{thm:main:Jastrow} and \ref{thm:main:apriori}.

The following result shows that one can push the $C^{1,\alpha},
\alpha\in(0,1)$,  in  
Proposition~\ref{prop:GTbis} to $C^{1,1}$ in certain cases.
\begin{thm}\label{thm:abstract}
Let \(g\in L^\infty(\mathbb R^k)\), \(k\geq2\), be a homogeneous
function of degree \(0\)  
which has the properties $g\in C^\alpha(\R^{k}\setminus\{0\})$ and 
$g|_{\sphere^{k-1}}$ is orthogonal 
to \(\mathfrak{h}_{2}^{(k)}\)
(the subspace of \(L^2(\sphere^{k-1})\) spanned by the spherical 
harmonics of degree \(2\)). Let \(f\in C^\alpha(\R^{d})\) for some 
\(d\ge0\) and let \(u\in
C^{1,\alpha}(\R^{k+d})\) be a weak solution of the equation
\begin{equation}
  \label{eq:Poisson}
  \Delta u(x',x'')=g(x')f(x'')
\end{equation}
where \(x'\in\R^{k}, \:x''\in\mathbb R^d\),
\(\Delta=\Delta_{x'}+\Delta_{x''}\).

Then \(u\in W_{\rm loc}^{2,\infty}(\R^{n})\), \(n=k+d\), and the
following {\it a~priori} 
estimate holds:\\
For all balls \(B_{n}(z,R)\) and \(B_{n}(z,R_1)\) in \(\R^{n}\) where $0<R<R_{1}$, 
\(z\in\R^{n}\),
\begin{align}
  \label{eq:apriori}
  \sup_{B_{n}(z,R)}|D_{ij}u|\leq C\,\Big(\sup_{B_{n}(z,R_1)}|u|
  &+\big(\sup_{\sphere^{k-1}}|g|\,\big)\:\|f\|_{C^\alpha(\pi_{d}
    B_{n}(z,R_1))}   
  \nonumber\\
  &
  +\big(\!\!\!\!\sup_{\pi_{d}B_{n}(z,R_1)}\!\!\!\!\!\!|f|\,\big)
  \;\|g\|_{C^\alpha(\sphere^{k-1})}\Big)
\end{align}
with \(C=C(n,\alpha,R,R_{1})\). 
Here  \(\pi_{d}(x',x'')=x''\) for \(x'\in\R^{k}$, $x''\in\R^{d}\) for
\(d>0\); for \(d=0, \pi_{d}(x')=0\). 
\end{thm}
\begin{rem}
\label{rem:abstract} 
\(\, \)  
\begin{enumerate}[\rm (i)]
\item \label{abstract(ii)}
The case $d=0$ means that $f$ is a constant and the terms in \eqref{eq:apriori}
with $f$ then equal this constant.
\item\label{abstract(iii)}
The reason for the condition \(k\geq2\) will become clear in the proof
of the theorem, when Lemma~\ref{GT42} is applied.
\item\label{abstract(iv)}
Note that if \(k=0, d\geq2\), one has stronger conclusions:
Equation \eqref{eq:Poisson} becomes
\(\Delta u(y)=f(y)\) with \(f\in C^{\alpha}(\R^{d})\), so by
Proposition~\ref{prop:GT}, \(u\in C^{2,\alpha}(\R^{d})\). The {\it a~priori}
estimate analogous to \eqref{eq:apriori} is then a consequence of
H\"older-estimates for \(u\) (see e.~g., \cite[Corollary
6.3]{GandT}).
\item\label{abstract(v)}
Using the standard fact (\cite[Theorem 4 in 5.8]{Evans}) that
\(W^{2,\infty}_{\text{\rm 
    loc}}(\R^{n})\) \(=C^{1,1}_{\text{\rm loc}}(\R^{n})\) (with equivalent
norms) we may 
    replace the term \(\sup_{B_{n}(z,R)}|D_{ij}u|\) by \([u]_{1,1,B_{n}(z,R)}\)
    on the left hand side in \eqref{eq:apriori}.
\item\label{abstract(vi)}
For the special solution to \eqref{eq:Poisson} given by the Newton
potential of $g f$, the estimate \eqref{eq:apriori} holds
without the term $\sup_{B_n(z,R_1)} |u|$ on the right hand
side (see \eqref{eq:apriori1}).
\end{enumerate}
\end{rem}

Since the proof of Theorem~\ref{thm:abstract} is a bit lenghty we
present it separately in subsection~\ref{subsection4.3}.

The following proposition, on solutions to
Poisson's equation, when the inhomogeneity \(f\) in \(\Delta u=f\) is
a homogeneous function, is needed often in the paper.
\begin{prop}
  \label{prop:homogen}
  Assume that the function \(g\) satisfies\par\noindent
  \(g(r\omega)=r^{k}G(\omega)\) with
  \(G\in L^{\infty}(\sphere^{n-1})\) and
  \(\mathcal{P}_{k+2}^{(n)}G=0\).

  Then there exists a solution \(u\) to 
  \begin{align}
    \label{eq:homogen}
     \Delta u=g \quad \text{ on } \quad B_{n}(0,R)\subset\R^{n},
  \end{align}
  satisfying \(u(r\omega)=r^{k+2}U(\omega)\) with \(U\in
  C^{1,\alpha}(\sphere^{n-1})\) for all \(\alpha\in(0,1)\).
\end{prop}
 \begin{pf}
    Let
   \begin{align*}
     g_{l,m}(r)=\int_{\sphere^{n-1}}g(r\omega)Y_{l,m}(\omega)\,d\omega
    =r^{k}\int_{\sphere^{n-1}}G(\omega)Y_{l,m}(\omega)\,d\omega
    =r^{k}g_{l,m}.
   \end{align*}
   Then (see \eqref{eq:h(l)} for \(h(l)\))
   \begin{align*}
     g(r\omega)=r^{k}\sum_{l=0,l\neq k+2}^{\infty}\sum_{m=0}^{h(l)-1}
     g_{l,m}Y_{l,m}(\omega),
   \end{align*}
   since \(g_{k+2,m}=0\) for all \(m\).
 
   Now define 
   \begin{align}
     \label{eq:def_U}
     U(\omega)=\sum_{l=0,l\neq k+2}^{\infty}\sum_{m=0}^{h(l)-1}
     \frac{g_{l,m}}{b_{l}(n,k)}\,Y_{l,m}(\omega)
  \end{align}
  with \(b_{l}(n,k)\equiv(k+2)((k+2)+n-2)-l(l+n-2)\).
   Note that 
 \(b_{l}(n,k))\neq 0\) for \(l\neq k+2\).
   Since \(\sum_{l,m}g_{l,m}Y_{l,m}\in L^{2}(\sphere^{n-1})\) (since
   \(G\in L^{\infty}(\sphere^{n-1})\)) the sum \eqref{eq:def_U} therefore
   converges in \(L^{2}(\sphere^{n-1})\).

    Make the `Ansatz' \(u(r\omega)=r^{k+2}U(\omega)\), and denote for
    \(N\in\N\)
    \begin{align*}
      g_{N}(r\omega)&=
         \sum_{l=0,l\neq k+2}^{N}\sum_{m=0}^{h(l)-1}
         g_{l,m}r^{k}\,Y_{l,m}(\omega), \\
      u_{N}(r\omega)&=r^{k+2}
         \sum_{l=0,l\neq k+2}^{N}\sum_{m=0}^{h(l)-1}
         \frac{g_{l,m}}{b_{l}(n,k)}\,Y_{l,m}(\omega).
    \end{align*}
    Now let \(\phi\in C_{0}^{\infty}\big(B_{n}(0,R)\big)\), then, using that 
    \(\mathcal{L}^{2}Y_{l,m}=l(l+n-2)Y_{l,m}\),
    \begin{align}
      \label{eq:indskud}
      \int_{B_{n}(0,R)}\!\!\!\!\!\!\phi(\Delta u-g)\,
      dx=\int_{B_{n}(0,R)}\!\!\!\!\!\!(\Delta\phi)(u-u_{N})\,dx 
      +\int_{B_{n}(0,R)}\!\!\!\!\!\!\phi(g_{N}-g)\,dx.
    \end{align}
    Since \(u-u_{N}\to0, g-g_{N}\to 0\) (in \(L^{2}\) -
    sense) for \(N\to0\), the RHS of \ref{eq:indskud} tends to zero for
    \(N\to0\). Hence \(u=r^{k+2}U(\omega)\) solves \ref{eq:homogen} in
    the distributional sense. With \(w\) the Newton potential
    corresponding to \(g\) (see \ref{Newton}), we have \(w\in
    C^{1,\alpha}(B_{n}(0,R))\) due to Proposition~\ref{prop:newton}, and
    \(u-w\) is harmonic, so \(u\in C^{1, \alpha}(B_{n}(0,R))\). This
    implies that 
  \(U\in C^{1,\alpha}(\sphere^{n-1})\). 
  \end{pf}
We prove the following useful lemma:
\begin{lem}
  \label{lem:XdotG}
  Let \(G:U\to\R^{n}\) 
for \(U\subset\R^{n+m}\) a neighbourhood of
  a point \((0,y_{0})\in\R^{n}\times\R^{m}\). Assume \(G(0,y)=0\) for
  all \(y\) such that \((0,y)\in U\). Let
  \begin{align*}
    f(x,y)=\left\{\begin{array}{cc}
             \frac{x}{|x|}\cdot G(x,y)& x\neq 0, \\
             0& x=0. \\
  \end{array}\right. 
  \end{align*}
  Then, for \(\alpha\in(0,1]\), 
  \begin{align}
    \label{eq:lem_G=0}
     G\in C^{0,\alpha}(U;\R^{n})\Rightarrow f\in C^{0,\alpha}(U).
  \end{align}
Furthermore, $\| f \|_{C^{\alpha}(U)} \leq 2\| G \|_{C^{\alpha}(U)}$.
\end{lem}
\begin{pf}
Let \(\alpha \in (0,1]\). We need to estimate 
\(\frac{f(x_1,y_1)-f(x_2,y_2)}{|(x_1,y_1)-(x_2,y_2)|^{\alpha}}\). 

Suppose first that $x_2=0$. Then $f(x_2,y_2)=0$ and we get
  \begin{align*}
    \frac{|f(x_1,y_1)-f(0,y_2)|}{|(x_1,y_1)-(0,y_2)|^{\alpha}}
    &\leq\frac{\left|\frac{x_1}{|x_1|}\cdot G(x_1,y_1)\right|}{|x_1|^{\alpha}}
    \leq \left|\frac{x_1}{|x_1|}\right|\cdot\frac{|G(x_1,y_1)|}{|x_1|^{\alpha}}\\
    &\leq \| G \|_{C^{\alpha}(U)},
  \end{align*}
  since \(G\in C^{\alpha}(U;\R^{n})\) and \(G(0,y_1)=0\).

Next, suppose $0<|x_2| \leq |x_1|$. By the triangle inequality:
\begin{align*}
  \big|f(x_1,y_1)-f(x_2,y_2)\big|
  &\leq
  \Big|\frac{x_1}{|x_1|}\cdot\big(G(x_1,y_1)-G(x_2,y_2)\big)\Big|
  \\&\quad
  +
  \Big|( \frac{x_1}{|x_1|} - \frac{x_2}{|x_2|})\cdot G(x_2,y_2)\Big| .
\end{align*}
Using that $G$ is $C^{\alpha}$ and that
$G(0,y_2) = 0$, we get
\begin{align*}
  \big|f&(x_1,y_1)-f(x_2,y_2)\big|
  \\&\leq \| G \|_{C^{\alpha}(U)}\left(\big|(x_1,y_1)-(x_2,y_2)\big|^{\alpha}+
  \Big|\big(\frac{x_1}{|x_1|}-\frac{x_2}{|x_2|}\big)\Big|\ |x_2 |^{\alpha}\right).
\end{align*}
To control the last term---divided by
$\big|(x_1,y_1)-(x_2,y_2)\big|^{\alpha}$---we first derive a  
lower bound on $\big|(x_1,y_1)-(x_2,y_2)\big|^{\alpha}$:
\begin{align*}
  &\big|(x_1,y_1)-(x_2,y_2)\big|^2
  \geq|x_1-x_2|^2
  \\&=
  \big(|x_1|-|x_2|\big)^2 +|x_1|\,|x_2|\Big(\frac{x_1}{|x_1|}- 
  \frac{x_2}{|x_2|}\Big)^2
  \geq
  |x_1|\,|x_2|\Big(\frac{x_1}{|x_1|}-\frac{x_2}{|x_2|}\Big)^2.
\end{align*}
Therefore, using the assumption $0<|x_2| \leq|x_1|$,
\begin{align*}
  \big|(x_1,y_1) - (x_2,y_2)\big|^{\alpha} \geq |x_2|^{\alpha}\,
  \Big|\frac{x_1}{|x_1|} - \frac{x_2}{|x_2|}\Big| ^{\alpha}.
\end{align*}
This finishes the proof of the lemma.
\end{pf}
The following obvious lemma is used repeatedly throughout the paper:
\begin{lem}
  \label{lem:r2G}
  Assume \(f(r\omega)=r^2 G(\omega)\) with \(G\in
  C^{1,1}(\R^{n}\setminus\{0\})\cap L^{\infty}(\R^{n})\). 
  Then \(f\in C^{1,1}(\R^{n})\).
\end{lem}
\begin{pf}
  The first derivatives of \(f\) trivially exist and are continuous. Therefore
  it suffices to show that all derivatives of \(f\) of second order belong
  to \(L_{\rm loc}^{\infty}(\R^{n})\); the result then follows from
  Remark~\ref{rem:abstract} \eqref{abstract(v)}.
   \begin{align*}
    \frac{\partial^2 f}{\partial x_j\partial x_k}=2\delta_{j,k}
    G(\omega)+2\Big(x_j \frac{\partial G}{\partial
    x_k}+x_k\frac{\partial G}{\partial x_j}\Big) + r^2\frac{\partial^2
    G}{\partial x_j\partial x_k}\in L_{\rm loc}^{\infty}(\R^{n}),
  \end{align*}
  since \(G\in C^{1,1}(\R^{n}\setminus\{0\})\). This proves the lemma.
\end{pf}
Note that better regularity cannot be expected without assuming
continuity of \(G\) at \(x=0\). On the other hand, if \(G\) only
depends on \(\omega\in\sphere^{n-1}\), and \(G\) is continuous at
\(x=0\), then \(G\) is a constant.

\subsection{Proof of Theorem~\ref{thm:abstract}}
\label{subsection4.3}
\(\, \)

We  first investigate, for  \(x_0\in B_{n}(z,R_1)\), the behaviour of
the Newton potential $w$ as  
given in \eqref{Newton}, namely
\begin{align}
\label{y'}
  w(x_0)=\underset{B_{n}(z,R_1)}\int\Gamma(x_0-y)g(y')f(y'')\,dy
\end{align}
with \(y=(y',y'')\in\R^{k+d}=\R^{n}\).

Since $u$ and $w$ are $C^{1,\alpha}$ - solutions 
of \eqref{eq:Poisson} in $B_{n}(z,R_1)$ (see
Proposition~\ref{prop:GTbis}), 
$h=u-w$ is harmonic. Any harmonic function $h$ in a bounded domain
$\Omega$ satisfies the 
following {\it a~priori} estimate (see \cite[Theorem 2.10]{GandT}):
\begin{equation}\label{hap}
  \sup_K |D_{ij}h|\le \frac{C(n)}{\delta^2}\sup_\Omega |h|
  \quad,\quad i,j\in\{1,\ldots,n\},
\end{equation} 
with \(K\) compact, \(K\subset\Omega\subset\R^{n}\), and 
\(\delta=\dist(K,\,\partial\Omega)\).
So, by \eqref{y'} and \eqref{hap}, for \(x_{0}\in B_{n}(z,R)\) (recall
that \(h=u-w\))
\begin{align}
  \label{eq:enough_w}
  |D_{ij}u(x_{0})|&\leq\frac{C(n)}{(R_{1}-R)^{2}}
  \Big(\sup_{B_{n}(z,R_{1})}|u| 
  \nonumber\\
  &\qquad\qquad\qquad\quad+ C(n,R_{1})\big(\sup_{\sphere^{k-1}}|g|\big)
  \big(\sup_{\pi_{d}B_{n}(z,R_{1})}|f|\big)\Big)
  \nonumber\\
  &\quad+ |D_{ij}w(x_{0})|.
\end{align}

Therefore to prove that $u\in W^{2,\infty}_{\rm loc}(\mathbb R^n)$ and that $u$
satisfies \eqref{eq:apriori} it obviously remains to show that $w$ satisfies the 
{\it a~priori} estimate \eqref{eq:apriori}. This will be done via Lemma \ref{GT42}
and will finish the proof of Theorem~\ref{thm:abstract}.

We proceed as follows: Define $N=\{(x',x'')\in \mathbb R^n\:|\:x'=0\}$ and 
note that $|N|=0$ ($|N|$  denotes \(n\)-dimensional Lebesgue measure
of $N$) and that for every  
ball $B_{n}\subset\R^{n}$, $B_{n}\setminus N$ is still a domain.
For this the assumption \(k\geq2\) is vital (see also
Remark~\ref{rem:abstract} \eqref{abstract(iii)}).
Note also that (still with \(x_0\in B_{n}(z,R_1)\)) $w$ can be written as 
\begin{align}\label{(.)}
  w(x_0)=\int_{B_{n}(z,R_1)\setminus N}
  \!\!\!\!\!
  \Gamma(x_0-y)g(y')f(y'')\,dy.
\end{align}
Taking into account the H\"older continuity assumptions on
$g$ and $f$ it is easily seen that 
for every domain $\Omega\subset\R^{n}$, $g f\in C^{\alpha}(\Omega\setminus N)$.
Hence \eqref{(.)} and Proposition~\ref{prop:GT}
implies that 
$w\in C^{2,\alpha}\big(B_{n}(z,R_1)\setminus N\big)$.

Now we are ready to apply Lemma \ref{GT42}: Pick $\Omega=B_{n}(z,R_1)\setminus N$ and 
$\Omega_0=B_{n}(z,R_2)$ with $R_1<R_2$, then we obtain from
\eqref{Dij}, for \(x_0\in B_{n}(z,R_1)\setminus N\), that
\begin{align}\label{Dij1}
  D_{ij}w(x_0)&=\int_{B_{n}(z,R_2)}D_{ij}\Gamma(x_0-y)\big((gf)(y)-(gf)(x_0)\big)\,dy
  \nonumber\\
  &\quad\quad-(gf)(x_0)\int_{\partial B_{n}(z,R_2)}
  D_i\Gamma(x_0-y)\nu_j(y)\,d\sigma(y)
  \nonumber\\
  &\equiv I(x_{0})+J(x_{0}).
\end{align}
Here as before $gf$ is extended by zero outside $B_{n}(z,R_1)\setminus N$.
Noting again that $|N|=0$,  
we can use this integral
representation to derive the {\it a~priori} estimates on $D_{ij}w$. 
We want to show that for $0<R<R_1$  
\begin{align}
\label{eq:apriori1}
  &\sup_{B_{n}(z,R)}|D_{ij}w| \\
  &\quad\quad\leq C_2
   \bigg[\big(\sup_{\sphere^{k-1}} |g|\big)\,
          \|f\|_{C^{\alpha}(\pi_{d} B_{n}(z,R_1))}
          +\big(\!\!\!\!\!
          \sup_{\pi_{d}B_{n}(z,R_1)}
          \!\!\!\!|f|\big)\,
          \|g\|_{C^\alpha(\sphere^{k-1})}
   \bigg]\nonumber
\end{align}
where \(C_2=C_2(n,\alpha,R_1-R, R/R_1)\).
Inequality \eqref{eq:apriori1} together with
inequality \eqref{eq:enough_w} 
will yield the desired {\it a~priori} estimate \eqref{eq:apriori} and implies 
in particular that $u\in W^{2,\infty}_{\rm loc}(\R^{k+d})$. So to finish the 
proof of Theorem~\ref{thm:abstract} it remains to prove inequality
\eqref{eq:apriori1}. For this we have to estimate the integrals \(I(x_{0})\) and
\(J(x_{0})\) in \eqref{Dij1}. We state the estimates in the following
lemma (Lemma~\ref{lem:I&J}), which we then apply to prove inequality
\eqref{eq:apriori1}. The proof of Lemma~\ref{lem:I&J} is given
afterwards.
For convenience we shall henceforth use the following notation: $B=B_{n}(z,R),\,
B_j=B_{n}(z,R_j),\: j=1,2$.
\begin{lem}
  \label{lem:I&J}
  With \(I(x_{0})\) and \(J(x_{0})\) as in \eqref{Dij1} we have the
  estimates
  \begin{align}
    \label{est:I}
      |I(x_0)|&\le C(n)\; \Big(\frac{R_2}{R_{1}-R}\Big)^{n}
      \big(\sup_{\sphere^{k-1}}|g|\big)\,
      \big(\sup_{\pi_{d}B_{1}}|f|\big)
      \nonumber\\
      &\quad+C(n,\alpha)\; (R_{1}-R)^{\alpha}\big(\sup_{\sphere^{k-1}}|g|\big)
      \big[f\big]_{\alpha,\pi_{d} B_1}
      \nonumber\\
      &\quad+C(n,\alpha)\;\big(\sup_{\pi_{d}B_{1}}|f|\big)\;\|g\|_{C^{\alpha}(\sphere^{k-1})}, 
      \\
    \label{est:J}
      |J(x_{0})|&\leq
      C(n)\sup_{B_{n}(z,R)}|gf|\;\Big(\frac{R_2}{R_2-R}\Big)^{n-1}.
  \end{align}  
for \(x_0\in B_{n}(z,R)\setminus N\).
\end{lem}

Combining the inequalities \eqref{est:I} and \eqref{est:J} with \eqref{Dij1}
leads to the {\it a~priori} estimate 
\begin{align}\label{26}
  &\sup_{x_0\in B_{n}(z,R)}|D_{ij} w(x_0)|
  \\
  &\leq C(n)\Big[\Big(\frac{R_2}{R_1-R}\Big)^n+\Big(\frac{R_2}{R_2-R}\Big)^{n-1}
  \Big]\big(\sup_{\sphere^{k-1}}|g|\big)\:\big(\sup_{\pi_{d}B_{1}}|f|\big)
  \nonumber\\
  &+C(n,\alpha)\Big[ (R_1-R)^{\alpha}\big(\sup_{\sphere^{k-1}}|g|\big)\;\big[f\big]_{\alpha,
  \pi_{d}
  B_1}+\big(\sup_{\pi_{d}B_1}|f|\big)\;\|g\|_{C^{\alpha}(\sphere^{k-1})}\Big].
  \nonumber
\end{align}
Finally we pick $R_2=2R_1$ and obtain, with \(C=C(n,\alpha, R,R_1)\), 
\begin{align}\label{27}
  \sup_{B_{n}(z,R)}|D_{ij}w|
  \le C\;
  \Big[\big(\sup_{\sphere^{k-1}}|g|\big)&\;\|f\|_{C^{\alpha}(\pi_{d}B_1)}
  \\
  &+
  \big(\sup_{\pi_{d}B_{1}}|f|\big)\:\|g\|_{C^{\alpha}(\sphere^{k-1})}\Big].\nonumber
\end{align}
This finishes the proof of \eqref{eq:apriori1} and according to our 
previous considerations the proof of Theorem~\ref{thm:abstract}. It
remains to prove the estimates in Lemma~\ref{lem:I&J}.

\begin{pf*}{Proof of Lemma~\ref{lem:I&J}}
We start by proving the estimate \eqref{est:J} on \(J(x_{0})\). 
For $y\in\partial B_2$ and
$x_0\in B_{n}(z,R)\setminus N$ we have $|x_0-y|\ge R_2-R$. This, and
\begin{equation*}
  |D_i\Gamma(x_0-y)|\le \frac{C(n)}{|x_0-y|^{n+1}},
\end{equation*}
yields
\begin{align*}
  |J(x_{0})|&\leq
  |(g f)(x_0)|\;|\int_{\partial B_2}
  |D_i\Gamma(x_0-y)\;\nu_j(y)\;d\sigma(y)|
  \\&
  \le C(n)\sup_{B_{n}(z,R)}|gf |\;\Big(\frac{R_2}{R_2-R}\Big)^{n-1},
\end{align*}
verifying \eqref{est:J}.

It remains to prove the estimate \eqref{est:I} on \(I(x_{0})\). This is
more involved.
With \(R'=R_1-R\) and
\(\Omega=B_{n}(z,R_2)\setminus N\), write
\begin{align}
  \label{Jx0}
  I(x_0)=I_{1}(x_0,R')+I_{2}(x_0,R')
\end{align}
with
\begin{align*}
  I_{1}(x_0,R')&=\int_{\Omega\setminus
  B_{n}(x_0,R')}
    \!\!\!\!\!\!\!\!\!\!\!\!
  D_{ij}\Gamma(x_0-y)\big((gf)(y)-(gf)(x_0)\big) \,dy,\\ 
  I_{2}(x_0,R')&=\int_{B_{n}(x_0,R')}
  \!\!\!\!\!\!\!\!
  D_{ij}\Gamma(x_0-y)\big((gf)(y)-(gf)(x_0)\big)
  \,dy. 
\end{align*}

Clearly we have (since \(gf\equiv0\) on \(B_{2}\setminus B_{1}\)
and \(g\) is homogeneous)
\begin{align}\label{J1}
  |I_{1}(x_0,R')|&\leq 2\,\big(\sup_{B_1} |gf|\big)
  \int_{\Omega\setminus B_{n}(x_0,R')}
  \!\!\!\!\!\!
  |D_{ij}\Gamma(x_0-y)|\,dy
  \nonumber\\
  &\leq C(n)\big(\sup_{\sphere^{k-1}}|g|\big)\; 
  \big(\sup_{\pi_{d}B_{1}}|f|\big)\big(\frac{R_2}{R'}\big)^{n}.
\end{align}  
The estimate for $I_{2}(x_0,R')$ will be more involved and we need
several steps.

First notice that  
\begin{align}
\label{eq:der_Gamma}
  (D_{ij} \Gamma)(x)=\frac{P_2(x)}{|x|^{n+2}}\ ,
\end{align}
 where $P_2$ is a homogeneous harmonic polynomial 
of degree 2 (which clearly depends on the indices $i,j$; 
we suppress these for simplicity).  Use polar coordinates
 $x=r\omega, \;r=|x|,\; \omega=x/|x|$, 
and obtain (using $\int_{\sphere^{n-1}} P_2(\omega)\,d\omega=0$,
and \eqref{eq:der_Gamma}) that
\begin{equation}
  \label{J2}
  I_{2}(x_0,R')= C(n) \int_0^{R'}\int_{\sphere^{n-1}}r^{-1} 
  P_2(\omega)\big(gf\big)(x_0+r\omega)\,d\omega\,dr.
\end{equation}
Denote \(x\in\R^{n}\) by \(x=(x',x'')=r\omega=r(\omega',\omega'')\)
where \(\omega'\in\R^{k}\), \(\omega''\in\R^{d}\) (so that \(|\omega'|^2
+|\omega''|^2=1\); when \(d=0, \omega''\equiv0\)).
With this, write
\begin{align}\label{J2'}
  I_{2}(x_0,R')= C(n)\int_0^{R'}r^{-1}
  \big(I^{(1)}_{2}(x_0,r)+I_{2}^{(2)}(x_0,r)\big)\,dr
\end{align}
with (\(I_{2}^{(1)}=0\) when \(d=0\))
\begin{align*}
  I_{2}^{(1)}(x_0,r)&=\int_{\sphere^{n-1}}P_2(\omega)\,
    g(x_0'+r\omega')\big(f(x_0''+r\omega'') -f(x_0'')\big)\,d\omega,\\
  I_{2}^{(2)}(x_0,r)&=f(x_0'')\int_{\sphere^{n-1}}P_2(\omega)\,g(x_0'+r\omega')\,d\omega. 
\end{align*}
We need to estimate $|I_{2}^{(1)}|$ and $|I_{2}^{(2)}|$ such that we gain a suitable  
$r$-behaviour for small, respectively, large $r$ which will 
enable us to estimate  $|I_{2}(x_0, R')|$.
 
Firstly, due to Lemma \ref{GT42},
$(gf)(y)$ is defined to be zero for $y\in B_2\setminus B_1$ in
\(I(x_{0})\) and formula \eqref{Dij1} holds for $x_0\in B_1\setminus N$. 
Using this formula just for $x_0\in B_{n}(z,R)\setminus N$ we 
have $x_0+r\omega\in \overline{B_1}$ for all $r$ with $0\le r\le R'=R_1-R$ 
and therefore (up to the zero set $N\cup \partial B_1$) we can make use  of the 
H\"older continuity properties of $g$ and $f$ for 
the points $x_0$ and $x_0+r\omega$ in the integrals above.

Using the H\"older continuity of $f$ and the homogeneity of \(g\) we obtain
\begin{align}
  |I_{2}^{(1)}&(x_0,r)| 
  \nonumber\\
  &=r^\alpha \bigg |\int_{\sphere^{n-1}}|\omega''|^\alpha P_2(\omega)\;
  g(x_0'+r\omega')\;\frac{f(x_0''+r\omega'')-f(x_0'')}
  {r^\alpha|\omega''|^\alpha}\,d\omega\bigg|
  \nonumber\\
  &\leq C(n)\,r^\alpha \big(\sup_{\sphere^{k-1}} |g|\big)\: [f]_{\alpha,\: \pi_{d} B_1}
  \nonumber
\end{align}
(for \([f]_{\alpha,\:\pi_{d} B_1}\), see Definition~\ref{def:Holder}).
Hence 
\begin{align}
  \label{r-1I1}
  \int_0^{R'}r^{-1}|I_{2}^{(1)}(x_0,r)|\,dr
  \leq 
  \frac{C(n)}{\alpha}\,(R')^{\alpha}\big(\sup_{\sphere^{k-1}}|g|\big)
  \:[f]_{\alpha,\:\pi_{d} B_1}.
\end{align}
To estimate the second term in \eqref{J2'}, 
we write in the following $x_0'=|x_0'|\,\eta$ with $\eta\in\sphere^{k-1}$ and 
define $s$ by $r=|x_0'|s$. Then
\begin{align}
  \label{I=}
  \Big|\int_0^{R'}r^{-1}I_{2}^{(2)}&(x_0,r)\,dr\Big|
  \\
  &=\big|f(x_0'')\big|\:\Big|\int_0^{R'}r^{-1}
  \int_{\sphere^{n-1}}P_2(\omega)\,g(|x_0'|\eta
  +r\omega')\,d\omega\,dr\Big|
  \nonumber\\
  &=\big|f(x_0'')\big|\:\Big |\int_0^{\frac{R'}{|x_0'|}}
  s^{-1}\int_{\sphere^{n-1}}P_2(\omega)\,g(\eta+s\omega')\,d\omega\,ds\Big|
  \nonumber
\end{align}
where we used that $g$ is homogeneous of degree zero and $|x_0'|\neq 0$.
Because of the $s^{-1}$-term in  the \(s\)-integral we have to control 
the \(\omega\)-integral for $s\rightarrow 0$ and for
$|x_0'|\rightarrow 0$. 

Define, for \(0\le s_1<s_2\le\infty\), 
\begin{align}
  \label{K}
  K(s_1,s_2)=\int_{s_1}^{s_2}s^{-1}\int_{\sphere^{n-1}}P_2(\omega)
  g(\eta+s\omega')\,d\omega\,ds.
\end{align}
The behaviour of \(K\) for different regimes of \(s_{1}\) and
\(s_{2}\) is expressed in Lemma~\ref{lem:K} below. Applying it, we get
that (for all \(|x_0'|\in (0,\infty)\))
  \(\big|K\big(0,R'/|x_0'|\big)\big| \le C(n,\alpha)\;\|g\|_{C^{\alpha}(\sphere^{k-1})}\)
(for \(\|g\|_{C^{\alpha}(\sphere^{k-1})}\), see Definition~\ref{def:Holder}).

Since due to \eqref{I=} 
\begin{equation*}
  \Big|\int_0^{R'}r^{-1}I_{2}^{(2)}(x_0,r)\,dr\Big|=|f(x_0'')|\;\big|K\big(0,R'/|x_0'|\big)\big| 
\end{equation*}
we obtain 
\begin{equation}\label{I22}
  \Big|\int_0^{R'}r^{-1}I_{2}^{(2)}(x_0,r)\,dr\Big|\le C(n,\alpha)\:
  \big(\sup_{\pi_{d}B_{1}}|f|\big)\:\|g\|_{C^{\alpha}(\sphere^{k-1})}.
\end{equation}
Further via \eqref{J2'}, \eqref{r-1I1} and \eqref{I22} lead to 
\begin{align}\label{J23}
  |I_{2}(x_0,R')|&\le C(n,\alpha)\;(R')^{\alpha}\;\big(\sup_{\sphere^{k-1}}|g|\big)\;
  \big[f\big]_{\alpha,\; \pi_{d} B_{1}}
  \nonumber\\&\quad
  +C(n,\alpha)\;
  \big(\sup_{\pi_{d}B_{1}}|f|\big)\:\|g\|_{C^\alpha(\sphere^{k-1})}.
\end{align}
The estimate \eqref{est:I} now follows from \eqref{Jx0}, \eqref{J1},
and \eqref{J23}.

Proving Lemma~\ref{lem:K} below will finish the proof of Lemma~\ref{lem:I&J}.
\end{pf*}
\begin{lem}
  \label{lem:K}
  \(\, \) 
  With \(K\) as in \eqref{K} we have:
  \begin{enumerate}[\rm (i)]
  \item\label{K(i)} 
    \(s_{1}=0, s_{2}\leq1/2\):
    \begin{align}
      \label{eq:K(i)}
      \big|K(0,s_{2})\big|\leq C(n,\alpha)[g]_{\alpha,\sphere^{k-1}}.
    \end{align}
  \item \label{K(ii)} 
    \(1/2\leq s_{1}<s_{2}\leq4\):
    \begin{align}
      \label{eq:K(ii)}
      \big|K(s_{1},s_{2})\big|\leq C(n)\,\sup_{\sphere^{k-1}}|g|.
    \end{align}
  \item \label{K(iii)} 
    \(4\leq s_{1}\) and \(s_{2}\to\infty\):
    \begin{align}
      \label{eq:K(iii)}
      \big|K(s_{1},s_{2})\big|\leq
       C(n,\alpha)\big[g\big]_{\alpha,\sphere^{k-1}}
       +C(n)\,\sup_{\sphere^{k-1}}|g|.
    \end{align}
  \end{enumerate}
\end{lem}

\begin{pf*}{Proof of Lemma~\ref{lem:K}}
\(\, \)

\eqref{K(i)}:
Since $\int_{\sphere^{n-1}}P_2(\omega)\,d\omega=0$ and $g$ is homogeneous of
degree \(0\) we have  
\begin{align}
  K(0,s_2)=\int_0^{s_2}s^{-1}\int_{\sphere^{n-1}}P_2(\omega)
  \big(g\Big(\frac{\eta+s\omega'}
  {|\eta+s\omega'|}\Big)-g(\eta)\big)\,d\omega\,ds.
  \nonumber
\end{align}
Note that $|\eta+s\omega'|\ge 1-s\ge 1/2.$ Since 
$g\in C^{\alpha}(\sphere^{k-1})$ we obtain 
\begin{align}
  \label{K0s2<}
  \big|K(0,s_2)\big|\le C(n)\:\big[g\big]_{\alpha,\sphere^{k-1}}
  \int_0^{s_2}s^{-1}\Big|\frac{\eta+s\omega'}{|\eta+s\omega'|}
  -\eta\Big|^\alpha\:ds.
  \nonumber
\end{align}
This, and
\begin{equation*}
\Big|\frac{\eta+s\omega'}{|\eta+s\omega'|}-\eta\Big|
\le\frac{s+\big|1-|\eta+s\omega'|\;\big|}{|\eta+s\omega'|}
\le \frac{2s}{|\eta+s\omega'|}\le\frac{2s}{1-s}\le 4s
\end{equation*}
implies \eqref{eq:K(i)}.

\eqref{K(ii)}: This follows directly from the definition of \(K\) (see \eqref{K}).

\eqref{K(iii)}:
This is the most involved case. We write  the unit sphere $\sphere^{n-1}$ 
as the union of 
\begin{align}
  \label{Sbeta}
  \Sigma(s)=
   \big\{\omega\in S^{n-1}\:\big|\: 
   s|\omega'|\le \sqrt{s}\,\big\}
   =\big\{\omega\in S^{n-1}\:\big|\:
    |\omega'|\le \frac{1}{\sqrt{s}}\,\big\} 
\end{align}
and its complement $\Sigma(s)^{c}$ (when \(d=0, \Sigma(s)=\emptyset\)
for \(s>1\))
and write
\(K(s_1,s_2)=A_1+A_2+A_3\) where (when \(d=0, A_{2}=0\) for \(s_{1}>1\))
\begin{align}
  A_1&=\int_{s_1}^{s_2}s^{-1}\int_{\Sigma(s)^{c}}P_2(\omega)\:
  (g(\eta+s\omega')-g(s\omega'))\,d\omega\,ds,\nonumber\\
  A_2&=\int_{s_1}^{s_2}s^{-1}\int_{\Sigma(s)}P_2(\omega)\;(g(\eta+s\omega')
  -g(s\omega'))\,d\omega\,ds,\nonumber\\
  A_3&=\int_{s_1}^{s_2}s^{-1}\int_{\sphere^{n-1}}P_2(\omega)\;g(s\omega')\, 
  d\omega\,ds.\nonumber
\end{align}
The estimate \eqref{eq:K(iii)} is a direct consequence of
the following lemma. Proving it will finish the proof of Lemma~\ref{lem:K}.
\end{pf*}
\begin{lem}
  \label{lem:As}
  We have
\begin{align}
  \label{A1<}
  |A_1|&\le C(n,\alpha) \big[g\big]_{\alpha,\; \sphere^{k-1}},\\
  \label{A2<}
  |A_2|&\le C(n)\sup_{\sphere^{k-1}}|g|,\\
  A_3&=0.
\end{align}
\end{lem}

\begin{pf}
$\mathbf{A_1:}$ Note first that since $s|\omega'|\ge 2$ and $|\eta+s\omega'|\ge 1$ in
$\Sigma(s)^{c}$ we obtain, using the homogeneity of degree zero of $g$
and the H\"older continuity of $g$ on $\sphere^{k-1}$, that
\begin{equation*}
|A_1|\le C(n) \big[g\big]_{\alpha,\sphere^{k-1}} 
\int_{s_1}^{s_2}s^{-1}
\int_{\Sigma^c(s)}\Big|\frac{\eta+s\omega'}{|\eta+s\omega'|}-
\frac{s\omega'}{|s\omega'|}\Big|^\alpha\;d\omega \,ds.
\end{equation*}
Then by using the triangle inequality and that $s|\omega'|\ge \sqrt s\ge 2$,
we get
\begin{align*}
  \Big|\frac{\eta+s\omega'}{|\eta+s\omega'|}-\frac{s\omega'}{|s\omega'|}\Big|
  &\le \frac{1+\big|\,|s\omega'|-|\eta+s\omega'|\big|}{|\eta+s\omega'|}
  \\&
  \le
  \frac{2}{|\eta+s\omega'|}\le\frac{2}{\sqrt s-1}\le 
  \frac{4}{\sqrt s}
\end{align*}
which leads to 
\begin{equation*}
  |A_1|\le C(n,\alpha)\big[g\big]_{\alpha,\;\sphere^{k-1}}s_1^{-\frac{\alpha}{2}}
  \le C(n,\alpha) \big[g\big]_{\alpha,\; \sphere^{k-1}},
\end{equation*}
verifying \eqref{A1<}.
\\
\noindent
$\mathbf{A_2:}$ For \(d=0\), \(A_{2}=0\). For \(d>0\), the estimate \eqref{A2<}
 is a consequence of the following lemma, which is not hard to prove
 using polar coordinates in \(\R^{n}\) (we omit the proof):
\begin{lem}\label{Soeren}
Let $|\Sigma(s)|$ denote the $n-1$-dimensional surface measure of $\Sigma(s)$.
Then
\begin{equation}\label{Sbs}
\big|\Sigma(s)\big|\le C(n)s^{-1/2}.
\end{equation}
\end{lem}
From \eqref{Sbs} we immediately get \eqref{A2<}:
\begin{equation*}
  |A_2|\le\Big(\int_4^\infty s^{-1}\big|\Sigma(s)\big|\,ds\Big)\:C(n)\sup_{\sphere^{k-1}}|g|
  \le C(n)\sup_{\sphere^{k-1}}|g|.
\end{equation*}
\noindent
$\mathbf{A_3:}$ We have 
\begin{equation}\label{A3=0}
A_3=0
\end{equation}
as a consequence of the lemma below (when \(d=0\), \eqref{A3=0} is trivially
true, due to the assumptions on \(g\)), since, by assumption, 
$g|_{\sphere^{k-1}}$ is orthogonal 
to \(\mathfrak{h}_{2}^{(k)}\)
(the subspace of \(L^2(\sphere^{k-1})\) spanned by the sphe\-ri\-cal 
harmonics of degree \(2\)).

\begin{lem}
Let $0<k<n$ and suppose $\phi \in L^2({\mathbb S}^{k-1})$
is ortho\-gonal to
\(\mathfrak{h}_{2}^{(k)}\).

Let \(\tilde\phi\) denote the following `natural' extension of
\(\phi\): 
\begin{align*}
  \tilde\phi\Big(\frac{(x,y)}{|(x,y)|}\Big)=
  \begin{cases}
  \phi\Big(\frac{x}{|x|}\Big)&
  \text{ for } |x|\neq 0,
  \\
  0&\text{ for } |x|=0.
\end{cases}
\end{align*}
Then \(\tilde{\phi}\in L^2({\sphere}^{n-1})\) and
\(\tilde{\phi}\) is orthogonal to 
\(\mathfrak{h}_{2}^{(n)}\).
\end{lem}

\begin{pf}
Since \(\phi\) can be expanded  
in the natural basis of \(L^2({\mathbb
S}^{k-1})\) it suffices to consider a $\phi$ which is the
restriction to ${\mathbb S}^{k-1}$ of a harmonic, homogeneous
polynomial $P_{s}$ of degree $s\neq 2$. Then
$\tilde{P}_{s}(x,y)=P_{s}(x)$ for \((x,y)\in\R^{n}\) is 
a harmonic homogeneous
polynomial in $\R^{n}$ of degree $s
\neq 2$. Therefore $\tilde{\phi}$, being the restriction of \(\tilde
P_{s}\) to \(\sphere^{n-1}\), 
is orthogonal in \(L^2({\mathbb S}^{n-1})\) to \(\mathfrak{h}_{2}^{(n)}\).
\end{pf}

This finishes the proof of Lemma~\ref{lem:As}, and therefore finally
the proof of Theorem~\ref{thm:abstract}.
\end{pf}
\section{Proofs of Theorems~\ref{thm:main:Jastrow} and \ref{thm:main:apriori}}
\label{chap:C-1-1}
We recall that for notational simplicity we shall give the proofs
of Theorems~\ref{thm:main:Jastrow} and 
\ref{thm:main:apriori}  only for the atomic case.
\subsection{Proof of Theorem~\ref{thm:main:Jastrow}}
\label{chap:proof:Jastrow}
Let \(\psi\) satisfy \((H-V)\psi=0\) in \(\R^{3N}\), with \(V\) as in
\eqref{V}, and let \(F_{2}\) and \(F_{3}\) be given as in
\eqref{F2} and \eqref{F3}. Define \(\phi_{3}\) by the equation
\(\psi=e^{F_{2}+F_{3}}\phi_{3}\). 
Recall 
that \(\Delta F_{2}=V\). We now make use of Lemma~\ref{lem:F3} below
which, together with Theorem~\ref{thm:abstract}, is the main
ingredient in the proof of Theorem~\ref{thm:main:Jastrow}.
Due to this lemma, there exists a function
\(K_{3}:\R^{3N}\to\R\) such that \(\Delta K_{3}=-|\nabla F_{2}|^{2}\),
and \(G_{3}\equiv K_{3}-F_{3}\in C^{1,1}(\R^{3N})\). Define
\(\zeta_{3}\) by
\begin{align}
  \label{def:zeta3}
 \psi = e^{F_{2}+K_{3}} \zeta_{3}
\end{align}
that is, \(\zeta_3 = e^{-G_3} \phi_3\).
Since
\(G_{3}\in C^{1,1}(\R^{3N})\), 
it remains to prove  Lemma~\ref{lem:F3} below and
that \(\zeta_{3}\in C^{1,1}(\R^{3N})\), then
\(\phi_{3}\in C^{1,1}(\R^{3N})\)
will follow. 
\begin{lem}
  \label{lem:F3}
  There exists a function \(G_{3}:\R^{3N}\to\R\), \(G_{3}\in
  C^{1,1}(\R^{3N})\) such that the function
  \begin{align}
    \label{K3}
    K_{3}({\bf x})=K_{3}(x_1,\ldots,x_N)
    &=Z\frac{(2-\pi)}{12\pi}\sum_{1\leq j<k\leq
    N}(x_j\cdot x_k)\ln(x_j^2+x_k^2)\nonumber\\ 
    &\quad+G_{3}(\mathbf x)
  \end{align}
  solves the equation \(\Delta K_{3}=-|\nabla F_2|^2\), with \(F_2\) as
  in~\eqref{F2}.
\end{lem}
\begin{rem}
  Note that the function \((x \cdot y)\ln(x^2+y^2)\) belongs to
  \(C^{1,\alpha}(\R^6)\) for all \(\alpha\in(0,1)\), but not to \(C^{1,1}(\R^6)\).
\end{rem}
\begin{pf}
Note that
\begin{align}
  \label{eq:grad_F}
  \nabla
  F_{2}=-\frac{Z}{2}\Big(\frac{x_{1}}{|x_{1}|},\ldots,\frac{x_{N}}{|x_{N}|}\Big)
  +\frac14\Big(\sum_{j=2}^{N}\frac{x_{1}-x_{j}}{|x_{1}-x_{j}|},\ldots,
  \sum_{j=1}^{N-1}\frac{x_{N}-x_{j}}{|x_{N}-x_{j}|}\Big),
\end{align}
so that
\begin{align}
  \label{eq:grad_F2_squared}
  |\nabla F_{2}|^{2}
  &=\Big(\frac{NZ^{2}}{4}+\frac{N(N-1)}{16}\Big)
  -\frac{Z}{4}\sum_{1\leq j<k\leq N}\gamma_{2}(x_{j},x_{k})
  \nonumber\\
  &\quad+\frac{1}{8}\sum_{1\leq j< k<l\leq
    N}\gamma_{3}(x_{j},x_{k},x_{l}) 
  \nonumber\\
  &{}\equiv \Gamma_{1}+\Gamma_{2}({\bf x})+\Gamma_{3}({\bf x}),
\end{align}
with (\(x,y,z\in\R^{3}\))
\begin{align}
  \label{def:gamma_2&gamma_3}
  \gamma_{2}(x,y)&=
  \Big(\frac{x}{|x|}-\frac{y}{|y|}\Big)\cdot\frac{x-y}{|x-y|}, 
  \\
  \gamma_{3}(x,y,z)&=\frac{x-y}{|x-y|}\cdot\frac{x-z}{|x-z|}
  +\frac{y-x}{|y-x|}\cdot\frac{y-z}{|y-z|}
  +\frac{z-x}{|z-x|}\cdot\frac{z-y}{|z-y|}.\nonumber
\end{align}
Therefore it is natural to make the 'Ansatz'
\begin{align*}
  K_3=\hat\mu+\hat\kappa+\hat\nu,
\end{align*}
 and look for \(\hat\mu, 
\hat\kappa, \hat\nu\) solving
\begin{align*}
  \Delta\hat\mu=-\Gamma_1\quad,\quad
  \Delta\hat\kappa=-\Gamma_2\quad,\quad \Delta\hat\nu=-\Gamma_3. 
\end{align*}

First, it is easily seen that with \(\mu(x)=|x|^2, x\in\R^{3}\), the
function
\begin{align*}
  \hat\mu({\bf x})&
  =-\frac{1}{6}\Big(
  \sum_{j=1}^{N}\frac{Z^2}{4}\mu(x_{j})
  +\sum_{1\le j<k\le
    N}\frac{1}{16}\mu(x_{j}-x_{k})\Big), 
\end{align*} 
satisfies \(\Delta\hat\mu=-\Gamma_1, \hat\mu\in C^{\infty}(\R^{3N})\).

Further, it suffices to find functions \(\kappa\) and \(\nu\) such that
\begin{align}
  \label{eq:eq_for_kappa}
  \kappa(x,y)&= 
  \frac{2-\pi}{3\pi}\ (x\cdot y)\log(x^{2}+y^{2})+\kappa_{1}(x,y)\ ,
  \ \kappa_{1}\in C^{1,1}(\R^{6}),\nonumber\\
  \text{ with }\qquad
  &(\Delta_{x}+\Delta_{y})\kappa(x,y)
  =\gamma_{2}(x,y),
\end{align}
and \(\nu\in C^{1,1}(\R^{9})\) with
\begin{align}
  \label{eq:eq_for_nu}
  (\Delta_{x}&+\Delta_{y}+\Delta_{z})\nu(x,y,z)=
  \gamma_{3}(x,y,z),
\end{align}
since letting
\begin{align}
  \hat\kappa({\bf x})=
  \frac{Z}{4}\sum_{1\le j<k\le N}
  \kappa(x_{j},x_{k})\quad,\quad 
  \hat\nu({\bf x})=-\frac{1}{8}\sum_{1\leq j<k<l\leq
    N}\nu(x_{j},x_{k},x_{l}) 
  \nonumber
\end{align}
gives (\(\Delta=\sum_{j=1}^{N}\Delta_{j}\))
\begin{align}
  \Delta\hat\kappa({\bf x})&=\frac{Z}{4}\sum_{1\leq j<k\leq N}
  \big((\Delta_{j}+\Delta_{k})\kappa\big)(x_{j},x_{k})
  =-\Gamma_{2}({\bf x}),
  \nonumber
  \\
  \Delta\hat\nu({\bf x})&  
  =-\frac{1}{8}
  \sum_{1\leq j<k<l\leq
    N}\big((\Delta_{j}+\Delta_{k}+\Delta_{l})\nu\big)(x_{j},x_{k},x_{l})   
  =-\Gamma_{3}({\bf x}).
  \nonumber
\end{align}
The functions \(\kappa\) and \(\nu\) are constructed in
Appendices~\ref{chap:kappa} and \ref{chap:nu}.
Lemma~\ref{lem:F3} then follows from
Lemma~\ref{lem:construct_kappa} and Lemma~\ref{lem:construct_nu}.
\end{pf}
\begin{rem}
 \label{rem:sing_types}
Summarizing, one can say that only 
those points where the coordinates of
(at least) \(2\) electrons coincide with that of the nucleus
\((x_{i}=x_{j}=0)\) give
rise to the logarithmic terms in \(K_{3}\). These terms stem from the
function \(\kappa\) and are due to the type of singularity of the
\(\gamma_{2}\)-terms in \(|\nabla F_{2}|^{2}\). There is no such
contribution from the function \(\nu\), i.e., from the
\(\gamma_{3}\)-terms in \(|\nabla F_{2}|^{2}\). This is due to the
permutational symmetry of \(\nu\) with respect to the electron
coordinates as will be seen from the proof of
Lemma~\ref{lem:construct_nu}.
\end{rem}

To finish the proof of Theorem~\ref{thm:main:Jastrow}
it remains to prove that \(\zeta_{3}\in C^{1,1}(\R^{3N})\). 

Using \((H-E)\psi=0\) and \(H=-\Delta+V\),
we get the following equation
for \(\zeta_{3}\) (see \eqref{expF} and \eqref{def:zeta3}; set
\(F=F_{2}+K_{3}\) and \(\phi=\zeta_{3}\))
\begin{align}
  \label{eq:zeta3_1}
  \Delta\zeta_{3}+2\nabla\big(F_{2}&+K_{3}\big)\cdot\nabla\zeta_{3}
  \\
  &+\Big(\Delta\big(F_{2}+K_{3}\big)+|\nabla\big(F_{2}+K_{3}\big)|^{2}
  +(E-V)\Big)\zeta_{3}=0.
  \nonumber
\end{align}
Using \(\Delta F_{2}=V\) and \(\Delta K_{3}=-|\nabla F_{2}|^{2}\),
this reduces to the equation
\begin{align}
  \label{eq:zeta3_2}
   \Delta\zeta_{3}+2\nabla\big(F_{2}&+K_{3}\big)\cdot\nabla\zeta_{3}
  \\
  &+\big(|\nabla K_{3}|^{2}+2\nabla F_{2}\cdot\nabla
  K_{3}+E\big)\zeta_{3}=0.
  \nonumber
\end{align}
This eliminated one of the terms in the equation for \(\zeta_{3}\)
that was only in \(L^{\infty}(\R^{3N})\), and not continuous, namely
\(|\nabla F_{2}|^{2}\).

To deal with the two remaining ones (containing \(\nabla F_{2}\)),
re-ar\-range the 
equation~\eqref{eq:zeta3_2}: 
\begin{align}
   \label{eq:zeta3_3}
    \Delta\zeta_{3}+\nabla F_{2}\cdot\Big(2\nabla\zeta_{3} 
    &+2\zeta_{3}\nabla K_{3}\Big)
    \\&  
    +\Big(|\nabla K_{3}|^{2}+E\Big)\zeta_{3}
    +2\nabla K_{3}\cdot\nabla\zeta_{3}
    =0.\nonumber
\end{align}
Define \(\Psi=(\Psi_{1},\ldots,\Psi_{N}):\R^{3N}\to\R^{3N}\) by
\begin{align}
  \label{def:Psi}
  \Psi(x_{1},\ldots,x_{N})=2\nabla\zeta_{3} 
    +2\zeta_{3}\nabla K_{3}.
\end{align}
That is,
\(\Psi_{j}=(\Psi_{j,1},\Psi_{j,2},\Psi_{j,3}):\R^{3N}\to\R^{3}\) with
\begin{align}
  \label{def:PsiJI}
  \Psi_{j,i}=2\frac{\partial\zeta_{3}}{\partial x_{j,i}}
  +2\zeta_{3}\frac{\partial K_{3}}{\partial x_{j,i}}
  \ ,\ j\in\{1,\ldots,N\}\ ,\ i\in\{1,2,3\}.
\end{align}
Then 
\begin{align}
  \label{eq:whyPsi}
  \nabla F_{2}\cdot\big(2\nabla\zeta_{3}+2\zeta_{3}\nabla K_{3}\big)
  =\nabla F_{2}\cdot\Psi.
\end{align}
Since \(K_{3},\zeta_{3}\in C^{1,\alpha}(\R^{3N})\) for all
\(\alpha\in(0,1)\), we have \(\Psi_{j,i}\in C^{\alpha}(\R^{3N})\) for
all \(j\in\{1,\ldots,N\}, i\in\{1,2,3\}\) and \(\alpha\in(0,1)\). 

Next, let \(\hat\Psi_{j,i}:\R^{3(N-1)}\to\R\) be defined by
\begin{align}
  \label{def:PsiHats}
   \hat\Psi_{j,i}(x_{1},\ldots,x_{j-1},x_{j+1},\ldots,x_{N})
   =\Psi_{j,i}(x_{1},\ldots,x_{j-1},0,x_{j+1},\ldots,x_{N}),
\end{align}
that is, by setting \(x_{j}\) equal to zero in \(\Psi_{j,i}\).

Furthermore, define, for \(j<k\), \(j,k\in\{1,\ldots,N\}\), the
functions \(\Phi_{(j,k)}:\R^{3N}\to\R^{3}\) by
\begin{align}
  \label{def:PhiJK}
  &\Phi_{(j,k)}(x_{1},\ldots,x_{N})=
  \\\nonumber
  &\Psi_{j}(x_{1},\ldots,x_{j-1},\tfrac{1}{2}(x_{j}+x_{k}),
  x_{j+1},\ldots,x_{k-1},\tfrac{1}{2}(x_{j}+x_{k}),x_{k+1},
  \ldots,x_{N})
  \\\nonumber&\quad-\\\nonumber
  &\Psi_{k}(x_{1},\ldots,x_{j-1},\tfrac{1}{2}(x_{j}+x_{k}),
  x_{j+1},\ldots,x_{k-1},\tfrac{1}{2}(x_{j}+x_{k}),x_{k+1},
  \ldots,x_{N}).
\end{align}
The proof of Theorem~\ref{thm:main:Jastrow} will follow from the
following two lemmas:
\begin{lem}
  \label{lem:ExplPoisson}
  Let \(\hat\Psi_{j,i},\Phi_{(j,k),i}\) be defined according to
  \eqref{def:PsiJI}, \eqref{def:PsiHats} and \eqref{def:PhiJK}.
  Assume the functions \(u_{j,i}, v_{(j,k),i}\) solve the equations
  \begin{align}
    \label{eq:u_s}
    \Delta u_{j,i}&=\frac{Z}{2}\frac{x_{j,i}}{|x_{j}|}\ 
    \hat\Psi_{j,i},
    \\
    \label{eq:v_s}
    \Delta v_{(j,k),i}&={}-\frac{1}{4}
    \frac{x_{j,i}-x_{k,i}}{|x_{j}-x_{k}|}\ 
    \Phi_{(j,k),i}.
  \end{align}
  Then \(u_{j,i}, v_{(j,k),i}\in C^{1,1}(\R^{3N})\). 
\end{lem}
\begin{lem}
  \label{lem:regRest}
  Let \(\hat\Psi_{j},\Phi_{(j,k)}\) be defined according to
  \eqref{def:PsiJI}, \eqref{def:PsiHats} and \eqref{def:PhiJK}.
  Then the functions
  \begin{align}
    \label{eq:regRest1}
    &\frac{1}{4} \frac{x_{j}-x_{k}}{|x_{j}-x_{k}|}\cdot 
    \Big\{\big(\Psi_{j}-\Psi_{k}\big)-\Phi_{(j,k)}\Big\},
    \\
    \label{eq:regRest2}
    &\frac{Z}{2}\frac{x_{j}}{|x_{j}|}\cdot\big(\Psi_{j}-\hat\Psi_{j}\big)
  \end{align}
   all belong to \(C^{\alpha}(\R^{3N})\) for all \(\alpha\in(0,1)\).
\end{lem}

Let us first finish the proof of Theorem~\ref{thm:main:Jastrow}, using
the two lemmas.

Let the function \(U:\R^{3N}\to\R\) be defined by
\begin{align}
  \label{def:U}
  U=\sum_{i=1}^{3}\sum_{j=1}^{N}u_{j,i}
  +\sum_{i=1}^{3}\sum_{1\leq j<k\leq N}v_{(j,k),i}
\end{align}
with the functions \(u_{j,i}, v_{(j,k),i}\) solving the equations
\eqref{eq:u_s} and \eqref{eq:v_s}.
Then
\begin{align}
  \label{eq:U}
  \Delta U=
  \sum_{j=1}^{N}\frac{Z}{2}\frac{x_{j}}{|x_{j}|}\cdot
  \hat\Psi_{j}\ 
  -\sum_{1\leq j<k\leq N}\frac{1}{4}\frac{x_{j}-x_{k}}{|x_{j}-x_{k}|}
  \cdot\Phi_{(j,k)},
\end{align}
and, due to Lemma~\ref{lem:ExplPoisson}, \(U\in C^{1,1}(\R^{3N})\).

Let \(W=\zeta_{3}-U\), then due to \eqref{eq:zeta3_3}, 
\eqref{eq:U}, and the form of \(\nabla F_{2}\) (see \eqref{eq:grad_F})
\begin{align}
  \label{eq:W}
  \Delta W&=\sum_{j=1}^{N}-\frac{Z}{2}\frac{x_{j}}{|x_{j}|}\cdot
  \big(\Psi_{j}-\hat\Psi_{j}\big)
  -\Big(|\nabla K_{3}|^{2}+E\Big)\zeta_{3}
  -2\nabla K_{3}\cdot\nabla\zeta_{3}
  \nonumber\\
  &\quad-\sum_{1\leq j<k\leq N}\frac{1}{4}\frac{x_{j}-x_{k}}{|x_{j}-x_{k}|}
  \cdot\Big\{\big(\Psi_{j}-\Psi_{k}\big)-\Phi_{(j,k)}\Big\}.
\end{align}
Using the fact that \(K_{3}, \zeta_{3}\in C^{1,\alpha}(\R^{3N})\), and
Lemma~\ref{lem:regRest}, we conclude that the RHS in \eqref{eq:W}
belongs to \(C^{\alpha}(\R^{3N})\) for all \(\alpha\in(0,1)\) . Due to
Proposition~\ref{prop:GT}, \(W\in C^{2,\alpha}(\R^{3N})\) for all
\(\alpha\in(0,1)\), and so \(\zeta_{3}=W+U\in C^{1,1}(\R^{3N})\)
(since \(U\in C^{1,1}(\R^{3N})\) as mentioned above).

This finishes the proof that \(\zeta_{3}\in C^{1,1}(\R^{3N})\),
and therefore \(\phi_{3}=e^{G_{3}}\zeta_{3}\in C^{1,1}(\R^{3N})\),
since \(G_{3}\in C^{1,1}(\R^{3N})\).

To finish the proof of Theorem~\ref{thm:main:Jastrow}, it therefore
remains to prove Lemma~\ref{lem:ExplPoisson} and
Lemma~\ref{lem:regRest}.
\begin{pf*}{Proof of Lemma~\ref{lem:ExplPoisson}}
Firstly, for \(u_{j,i}\), this is a straightforward application of
Theorem~\ref{thm:abstract}, with \(k=3, d=3(N-1)\) and
\begin{align*}
  g&\equiv\frac{x_{j,i}}{|x_{j}|} \quad,\quad x'\equiv x_{j}\in\R^{3},\\
  f&\equiv \frac{Z}{2}\ \hat\Psi_{j,i}\quad,\quad
  x''\equiv(x_{1},\ldots,x_{j-1},x_{j+1},\ldots,x_{N})\in\R^{3(N-1)}.
\end{align*}
It has already been noted that \(\Psi_{j,i}\in C^{\alpha}(\R^{3N})\)
for all \(\alpha\in(0,1)\) and therefore (see \eqref{def:PsiHats})
also \(\hat\Psi_{j,i}\in C^{\alpha}(\R^{3(N-1)})\) for all
\(\alpha\in(0,1)\). Clearly, \(\frac{x_{j,i}}{|x_{j}|}\in
  C^{\infty}(\R^{3}\setminus\{0\}) \subset
  C^{\alpha}(\R^{3}\setminus\{0\})\), and
  \(\mathcal{P}_{2}^{(3)}\big(\frac{x_{j,i}}{|x_{j}|}\big)=0\), due to
  the anti-symmetry of the function \(\frac{x_{j,i}}{|x_{j}|}\). 
Therefore, all assumptions of Theorem~\ref{thm:abstract} are
fullfilled and it follows that \(u_{j,i}\in C^{1,1}(\R^{3N})\). 

Secondly, for \(v_{(j,k),i}\), we make an orthogonal
change of coordinates: \(a=\frac{1}{\sqrt{2}}(x_{j}-x_{k}),
b=\frac{1}{\sqrt{2}}(x_{j}+x_{k})\), the other coordinates remaining
unchanged. Due to the specific definition of \(\Phi_{(j,k),i}\), this
brings us to a setup exactly as the one above for \(u_{j,i}\). Since
the orthogonal change of coordinates does not change the regularity,
the conclusion follows as before.

This finishes the proof of  Lemma~\ref{lem:ExplPoisson}.
\end{pf*}
\begin{pf*}{Proof of Lemma~\ref{lem:regRest}} First, note that
  the function \(G_{j}=\Psi_{j}-\hat\Psi_{j}\) satisfies
\(G_{j}\in C^{\alpha}(\R^{3N})\) for all \(\alpha\in(0,1)\), and 
\begin{align*}
  G_{j}(x_{1},&\ldots,x_{j-1},x_{j}=0,x_{j+1},\ldots,x_{N})=0
  \\
  &\text{ for all }
  (x_{1},\ldots,x_{j-1},x_{j+1},\ldots,x_{N})\in\R^{3(N-1)}.
\end{align*}
Therefore, due to Lemma~\ref{lem:XdotG}, 
\begin{align*}
  \frac{Z}{2}\frac{x_{j}}{|x_{j}|}\cdot\big(\Psi_{j}-\hat\Psi_{j}\big)
  \in C^{\alpha}(\R^{3N})\quad\text{for all}\quad\alpha\in(0,1).
\end{align*}

Secondly, for the function
\begin{align*}
  \frac{1}{4}\frac{x_{j}-x_{k}}{|x_{j}-x_{k}|}\cdot
  \Big\{\big(\Psi_{j}-\Psi_{k}\big)-\Phi_{(j,k)}\Big\},
\end{align*}
the same orthogonal change
of coordinates as in the proof of Lem\-ma~\ref{lem:ExplPoisson} 
brings us in the same situation as the above, again due to the
specific definition of \(\Phi_{(j,k)}\). The conclusion follows as above.

This finishes the proof of Lemma~\ref{lem:regRest}.
\end{pf*}
This finishes the proof of Theorem~\ref{thm:main:Jastrow}.\hspace*{\fill}\qed
\subsection{Proof of Theorem~\ref{thm:main:apriori}}
\label{chap:proof:apriori}
By Remark~\ref{rem:aprioris} is suffices to prove that
\eqref{eq:a_priori} holds. 
We proceed similarly to
the proof of Theorem~\ref{thm:main:Jastrow}, but here we need to
estimate carefully all the involved quantities uniformly
(i.e., independently of ${\mathbf x_0}\in\R^{3N}$).
For notational simplicity, we will prove \eqref{eq:a_priori} only in
the case \(R'=2R\). 

For the proof we
need the following regularised version of Lem\-ma~\ref{lem:F3}.
\begin{lem3.1'}
There exists a function \(G_{3,\text{\rm cut}}:\R^{3N}\to\R\),
\(G_{3,\text{\rm cut}}\in C^{1,1}(\R^{3N})\),
such that the function 
\begin{align}
\label{def:K_3_cut}
  K_{3,{\rm cut}}(\mathbf x) =& Z\,\frac{2-\pi}{12 \pi} \sum_{1 \leq j < k
  \leq N} (x_j\cdot x_k)\chi(|x_j|) \chi(|x_k|) \ln(x_j^2
  + x_k^2) \nonumber\\
  &+
  G_{3,\text{\rm cut}}(\mathbf x)=F_{3,\text{\rm cut}}(\mathbf
  x)+G_{3,\text{\rm cut}}(\mathbf x) 
\end{align}
(for \(F_{3,\text{\rm cut}}\), see \eqref{eq:def_F3cut}) solves the equation
$$
\Delta K_{3,{\rm cut}} = - |\nabla F_{2,{\rm cut}} |^2 +
r_{{\rm cut}},
$$
with $F_{2,{\rm cut}}$ as defined in \eqref{eq:def_F2cut}
and $r_{{\rm cut}} \in C^{\alpha}(\mathbb R^{3N})$ for all
$\alpha \in (0,1)$. Furthermore, \(G_{3,\text{\rm cut}}\) can be chosen
such that for all \(\rho>0\) the following estimate holds:
\begin{align}
  \label{eq:G_3-r-est}
  \| G_{{3,\rm cut}}\|_{C^{1,1}(B_{3N}({\mathbf x_0},\rho))}+
  \| r_{{\rm cut}}\|_{C^{\alpha}(B_{3N}({\mathbf x_0},\rho))} \leq C,
\end{align}
for some constant $C=C(\rho) > 0$ independent of 
${\mathbf x_0} \in {\mathbb R}^{3N}$.
\end{lem3.1'}

\begin{pf}
The proof of Lemma~3.1' is analogous to that of
Lemma~\ref{lem:F3}.
Instead of $\mu, \kappa, \nu$ we will use functions
$\mu_{{\rm cut}}, \kappa_{{\rm cut}}$ and $\nu_{{\rm cut}}$
to be defined presently. With
$\chi$ being the function defined in
\eqref{eq:def_cutoff} we define 
\begin{align}
  \mu_{{\rm cut}}(x) &=\chi(|x|) \mu(x)=\chi(|x|)|x|^{2}, \\
  \label{def:kappa_cut}
  \kappa_{{\rm cut}}(x,y) &= 
  \chi(|x|)\chi(|y|)\kappa(x,y)\\
  &\quad-\frac{1}{4}\chi(3|y|)\big(1-\chi(|x|)\big)\Big(|y|^{2}\frac{x\cdot
  y}{|x||y|}\Big)\nonumber\\
  &\quad-\frac{1}{4}\chi(3|x|)\big(1-\chi(|y|)\big)\Big(|x|^{2}\frac{x\cdot
  y}{|x||y|}\Big)\nonumber\\
  & \equiv\chi(|x|)\chi(|y|)\frac{2-\pi}{3\pi}(x\cdot
  y)\ln(x^{2}+y^{2})+\kappa_{1,\text{\rm cut}}(x,y).
\nonumber
\end{align}
(Note that \(\kappa_{1,\text{\rm cut}}(x,y)\neq\chi(|x|)\chi(|y|)\kappa_{1}(x,y)\)).
Let \(\nu_{{\rm cut}}\) be as in Lem\-ma~\ref{lem:construct_nu_cut}, we
then have
\begin{align}
  \label{eq:nu_cut}
  &\Delta\nu_{{\rm cut}} = \gamma_3 +
  h_{\nu},\\
  &\| \nu_{{\rm cut}} \|_{C^{1,1}(B_{9}((x_0,y_0,z_0),\rho))}+
  \| h_{\nu} \|_{C^{\alpha}(B_{9}((x_0,y_0,z_0),\rho))} \leq C,
  \nonumber
\end{align}
with  $\gamma_3$ as in \eqref{def:gamma_2&gamma_3}
and with \(C\) independent of \((x_0,y_0,z_0)\in\R^{9}\) and \(\rho>0\).

For \(\mu_{\text{\rm cut}}\), note that
\begin{align}
  \label{eq:mu_cut}
  \Delta\mu_{\text{\rm cut}}&=\Delta|x|^{2}+\Delta(\mu_{\text{\rm
  cut}}-|x|^{2})\\
  &=6-\Delta\big((1-\chi(|x|))|x|^{2}\big)
  \equiv6-h_{\mu},\nonumber
\end{align}
where obviously, 
\begin{align}
  \label{est:h_mu}
  \| \mu_{{\rm cut}} \|_{C^{1,1}(B_{3}(x_0,\rho))}+
  \big\|h_{\mu}\big\|_{C^{\alpha}(B_{3}(x_{0},\rho))}\leq C,
\end{align}
with \(C\) independent of \(x_{0}\in\R^{3}\) and \(\rho>0\).

For \(\kappa_{\text{\rm cut}}\), using 
\(\Delta\kappa=\gamma_{2}\) (see \eqref{def:gamma_2&gamma_3} 
and \eqref{eq:eq_for_kappa}), that 
\(\Delta_{y}(|y|^{2}\frac{x\cdot y}{|x||y|})=4\frac{x\cdot y}{|x||y|}\),
and the 
support properties of \(\chi\), we have that
\begin{align}
  \label{eq:kappa_cut1}
  \Delta\kappa_{\text{\rm cut}}
  &=\gamma_{2}
  -\big\{1-\chi(|x|)\chi(|y|)\big\}\big(1-\chi(3|x|)-\chi(3|y|)\big)\gamma_{2}\nonumber\\  
  &-\Big\{\chi(3|y|)\big(1-\chi(|x|)\big) + \chi(3|x|)\big(1-\chi(|y|)\big)\Big\}
  \big(\gamma_{2}+\frac{x\cdot y}{|x||y|}\big)\nonumber\\
  &+R_1+R_2+R_3,\nonumber\\
  &\equiv \gamma_{2}-
  \mathcal{H}\gamma_{2}-\mathcal{G}\big(\gamma_{2}+\frac{x\cdot
  y}{|x||y|}\big) +R_1+R_2+R_3,
\end{align}
where
\begin{align*}
  R_1&=\chi(|y|)\kappa\Delta_{x}\chi(|x|)
  +\chi(|y|)2\nabla_{x}\chi(|x|)\cdot\nabla_{x}\kappa\nonumber\\
  &\,+\chi(|x|)\kappa\Delta_{y}\chi(|y|)
  +\chi(|x|)2\nabla_{y}\chi(|y|)\cdot\nabla_{y}\kappa,\\
  R_2&=-\frac{1}{4}
  \chi(3|y|)|y|^{2}\frac{y}{|y|}\cdot\Delta_{x}\big((1-\chi(|x|))\frac{x}{|x|}\big)\nonumber\\
  &-\frac{1}{4}\big(\Delta_{y}\chi(3|y|)\big)|y|^{2}\frac{y}{|y|}
  \cdot\big((1-\chi(|x|)\frac{x}{|x|}\big)\nonumber\\
  &-\frac{1}{2}\big(\nabla_{y}\chi(3|y|)\big)\cdot\nabla_{y}\Big(\frac{x\cdot
  y}{|x||y|}|y|^{2}\big(1-\chi(|x|)\big)\Big),
\end{align*}
and where \(R_{3}\) is \(R_{2}\) with \(x\) and \(y\) interchanged.

Using that \(\kappa\in C^{1,\alpha}(\R^{6})\) for all
\(\alpha\in(0,1)\), and the support properties of \(\chi\), it is
easily seen that
\begin{align}
  \label{est:R_j}
  \|R_{j}\|_{C^{\alpha}(B_{6}((x_{0},y_{0}),\rho))}\leq C,
\end{align}
with a constant \(C\) independent of \((x_{0},y_{0}))\in\R^{6}\) and \(\rho>0\).

Since for all \((x,y)\in\R^{6}\), 
\begin{align*}
  \big|\nabla\gamma_{2}\big|\leq
  6\sqrt{2}\Big(\frac{1}{|x|}+\frac{1}{|y|}\Big)
  \quad,\quad
 \Big|\nabla\big(\gamma_{2}+\frac{x\cdot
  y}{|x||y|}\big)\Big|\leq\frac{8\sqrt{2}}{|x-y|}
\end{align*}
we get, using the support properties of \(\mathcal{H}\) and
\(\mathcal{G}\), that
\begin{align*}
  \|\mathcal H\,\nabla\gamma_{2}\|_{L^{\infty}(\R^{6})}\leq C\quad,\quad
  \big\|\mathcal G\nabla\big(\gamma_{2}+\tfrac{x\cdot
  y}{|x||y|}\big)\big\|_{L^{\infty}(\R^{6})}\leq C.
\end{align*}
Again using the support properties of \(\mathcal{H}\) and
\(\mathcal{G}\), this implies that
\begin{align}
  \label{est:F&G}
  &\|\mathcal H\,\gamma_{2}\|_{C^{0,1}(B_{6}((x_{0},y_{0}),\rho))}\leq C,\\
  &\big\|\mathcal G\big(\gamma_{2}+\tfrac{x\cdot
  y}{|x||y|}\big)\big\|_{C^{0,1}(B_{6}((x_{0},y_{0}),\rho))}\leq C, \nonumber
\end{align}
with a constant \(C\) independent of \((x_{0},y_{0})\in\R^{6}\) and \(\rho>0\).

From \eqref{eq:kappa_cut1},  \eqref{est:R_j},  and  \eqref{est:F&G} we
get
\begin{align}
   \label{eq:kappa_cut}
  \Delta\kappa_{\text{\rm cut}}=\gamma_{2}+h_{\kappa}\quad,\quad
  \|h_{\kappa}\|_{C^{\alpha}(B_{6}((x_{0},y_{0}),\rho))}\leq C,
\end{align}
with a constant \(C\) independent of \((x_{0},y_{0})\in\R^{6}\) and \(\rho>0\).
Note that (see \eqref{def:kappa_cut} and \eqref{eq:formula_kappa})
\begin{align*}
  \kappa_{1,\text{\rm
  cut}}(x,y)=\chi(|x|)\chi(|y|)\Big((x^{2}+y^{2})
  G_{\kappa_{1}}\big(\frac{(x,y)}{|(x,y)|}\big)\Big)\ ,\
  G_{\kappa_{1}}\in C^{1,1}(\sphere^{5}).
\end{align*}
Therefore, due to the compact support of \(\chi\), 
\begin{align}
  \label{eq:est_kappa_1_cut}
  \|\kappa_{1,\text{\rm
  cut}}\|_{C^{1,1}(B_{6}((x_{0},y_{0}),\rho))}\leq C
\end{align}
with \(C\) independent of \((x_{0},y_{0})\in\R^{6}\) and \(\rho>0\).

Observe that 
\begin{align}
\label{eq:square_prod}
  |\nabla F_{2}|^2= |\nabla F_{2,{\rm cut}} |^2 
  +\nabla(F_{2}-F_{2,{\rm cut}})\cdot\nabla(F_{2}+
  F_{2,{\rm cut}})
\end{align}
 and that
\begin{align*}
  \nabla(F_{2}- F_{2,{\rm cut}})&\cdot\nabla(F_{2}+
  F_{2,{\rm cut}})\\
  &=
  \sum_{j=1}^{N}\nabla_{j}(F_{2}- F_{2,{\rm
    cut}})\cdot\nabla_{j}(F_{2}+
  F_{2,{\rm cut}})\\
  &=\sum_{j=1}^{N}\vec{b}_{j}\cdot\frac{x_{j}}{|x_{j}|}
  +\sum_{1\leq j<k\leq
  N}\vec{b}_{(j,k)}\cdot\frac{x_{j}-x_{k}}{|x_{j}-x_{k}|},
\end{align*}
where 
\begin{align*}
  \vec{b}_{j}&=-\frac{Z}{2}\Big\{1+\chi(|x_{j}|)+\chi'(|x_{j}|)|x_{j}|\Big\}
  \Big(\vec{a}_{j}+\sum_{l=1,l\neq j}^{N}\vec{a}_{(j,l)}\Big),\\
  \vec{b}_{(j,k)}&=\frac{1}{4}\Big\{1+\chi(|x_{j}-x_{k}|)
  +\chi'(|x_{j}-x_{k}|)|x_{j}-x_{k}|\Big\}\times\\
  &\quad\times
   \Big(\vec{a}_{j}-\vec a_{k}+\sum_{l=1,l\neq
  j}^{N}\vec{a}_{(j,l)}-\sum_{l=1,l\neq  k}^{N}\vec{a}_{(k,l)}\Big), \\ 
 \vec{a}_{j}&=-\frac{Z}{2}\Big\{\big(1-\chi(|x_{j}|)\Big)
 -\chi'(|x_{j}|)|x_{j}|\Big\}\frac{x_{j}}{|x_{j}|},\\
  \vec{a}_{(j,k)}&=\frac{1}{4}\Big\{\big(1-\chi(|x_{j}-x_{k}|)\big)
  -\chi'(|x_{j}-x_{k}|)|x_{j}-x_{k}|\Big\}\frac{x_{j}-x_{k}}{|x_{j}-x_{k}|}.
\end{align*}
Clearly (using the support properties of \(\chi\)), for all \(\beta\in\mathbb N^{3N}\), 
\begin{align}
  \label{eq:est_vec}
  \|\partial^{\beta}\vec{b}_{j}\|_{L^{\infty}(\R^{3N})}+
    \|\partial^{\beta}\vec{b}_{(j,k)}\|_{L^{\infty}(\R^{3N})}\leq C(\beta).
\end{align}
Define 
\begin{align*}
  G_{1,{\rm cut}}&=
  \frac{1}{4}\sum_{j=1}^{N}\vec{b}_{j}\cdot\Big(|x_{j}|^{2}\frac{x_{j}}{|x_{j}|}\Big)\, 
  \chi(|x_{j}|)\\
  &+\frac{1}{4}\sum_{1\leq j<k\leq
  N}\vec{b}_{(j,k)}\cdot\Big(\,\Big|\frac{x_{j}-x_{k}}{\sqrt{2}}\Big|^{2}
  \frac{x_{j}-x_{k}}{|x_{j}-x_{k}|}\Big)\, \chi(|x_{j}-x_{k}|). 
\end{align*}
Then, due to \eqref{eq:est_vec} and the support properties of
\(\chi\), 
\begin{align}
  \label{est:c-1-1_G_1}
   \| G_{{1,\rm cut}}\|_{C^{1,1}(B_{3N}({\mathbf x_0},\rho))}\leq C,
\end{align}
for some constant $C=C(\rho) > 0$ independent of 
${\mathbf x_0} \in {\mathbb R}^{3N}$.

Using
\(\Delta(|x_{j}|^{2}\frac{x_{j}}{|x_{j}|})=4\frac{x_{j}}{|x_{j}|}\)
and
\(\Delta(|\frac{x_{j}-x_{k}}{\sqrt{2}}|^{2}\frac{x_{j}-x_{k}}{|x_{j}-x_{k}|})
=4\frac{x_{j}-x_{k}}{|x_{j}-x_{k}|}\), 
we see that
\begin{align}
  \label{eq:G_1_cut}
  \Delta G_{1,{\rm cut}}=\nabla(F_{2}-F_{2,{\rm cut}})\cdot\nabla(F_{2}+
  F_{2,{\rm cut}})+R,
\end{align}
with
\begin{align*}
  R&= \frac{1}{4}\sum_{j=1}^{N}\Delta\big(\chi(|x_{j}|)
  \vec{b}_{j}\big)\cdot\Big(|x_{j}|^{2}\frac{x_{j}}{|x_{j}|}\Big)\\
  &+\frac{1}{2}\sum_{j=1}^{N}\sum_{i=1}^{3}\nabla_{j}\big(\chi(|x_{j}|)
  \vec{b}_{j,i}\big)\cdot\nabla_{j}\Big(|x_{j}|^{2}\frac{x_{j,i}}{|x_{j}|}\Big)\\
  &+\frac{1}{4}\sum_{1\leq j<k\leq
  N}\Delta\big(\chi(|x_{j}-x_{k}|)\vec{b}_{(j,k)}\big)
  \cdot\Big(\,\Big|\frac{x_{j}-x_{k}}{\sqrt{2}}\Big|^{2}
  \frac{x_{j}-x_{k}}{|x_{j}-x_{k}|}\Big)\\
 &+\frac{1}{2}\sum_{j<k
  }\sum_{i=1}^{3}\nabla_{j}\big(\chi(|x_{j}-x_{k}|)\vec{b}_{(j,k),i}\big)
  \cdot\nabla_{j}\Big(\,\Big|\frac{x_{j}-x_{k}}{\sqrt{2}}\Big|^{2}
  \frac{x_{j,i}-x_{k,i}}{|x_{j}-x_{k}|}\Big)\\
  &+\sum_{j=1}^{N}\vec{b}_{j}\cdot\frac{x_{j}}{|x_{j}|}\big(1-\chi(|x_{j}|)\big)\\
  &+\sum_{1\leq j<k\leq
  N}\vec{b}_{(j,k)}\cdot\frac{x_{j}-x_{k}}{|x_{j}-x_{k}|}
  \big(1-\chi(|x_{j}-x_{k}|)\big).
\end{align*}
From \eqref{eq:est_vec} and the support properties of \(\chi\), we see
that
\begin{align}
  \label{eq:R_cut_est}
  \| R\|_{C^{0,1}(B_{3N}({\mathbf x_0},\rho))} \leq C,
\end{align}
for some constant \(C\) independent of ${\mathbf x_0} \in {\mathbb
  R}^{3N}$ and \(\rho>0\). 

Define
\begin{align}
  \label{def:G_3_cut}
  G_{{3,\rm cut}}=G_{{1,\rm cut}}+G_{{2,\rm cut}}
\end{align}
with
\begin{align*}
  G_{{2,\rm cut}}&=\hat\mu_{\rm cut}+\hat \kappa_{1,\rm cut}+\hat\nu_{\rm cut},\\
  \hat\mu_{\rm cut}({\bf x})&
  =-\frac{1}{6}\Big(
  \sum_{j=1}^{N}\frac{Z^2}{4}\mu_{\rm cut}(x_{j})
  +\sum_{1\le j<k\le
    N}\frac{1}{16}\mu_{\rm cut}(x_{j}-x_{k})\Big), \\
 \hat \kappa_{1,\rm cut}({\bf x})&=
  \frac{Z}{4}\sum_{1\le j<k\le N}
  \kappa_{1,\rm cut}(x_{j},x_{k}),\nonumber\\
  \hat\nu_{\rm cut}({\bf x})&=-\frac{1}{8}\sum_{1\leq j<k<l\leq
    N}\nu_{\rm cut}(x_{j},x_{k},x_{l}).
  \nonumber
\end{align*} 
Then, with \(K_{3,\rm cut}\) defined as in \eqref{def:K_3_cut}, we
have, using \eqref{eq:G_1_cut}, \eqref{eq:nu_cut},
\eqref{eq:kappa_cut}, \eqref{eq:mu_cut}, \eqref{eq:square_prod},
\eqref{eq:grad_F2_squared} 
\begin{align}
  \Delta K_{3,\rm cut} &= |\nabla F_{2}|^2 - |\nabla F_{2,{\rm cut}}
  |^2 - \Gamma_{1} - \Gamma_{2} - \Gamma_{3} + r_{\text{\rm
  cut}}\nonumber\\
  &=- |\nabla F_{2,{\rm cut}}|^2+ r_{\text{\rm cut}},
\end{align}
where, due to \eqref{eq:nu_cut}, \eqref{eq:kappa_cut},
\eqref{eq:R_cut_est}, \eqref{est:h_mu},
\begin{align}
  \label{est:r_cut}
  \| r_{{\rm cut}}\|_{C^{\alpha}(B_{3N}({\mathbf x_0},\rho))} \leq C,
\end{align}
for some constant $C=C(\rho) > 0$ independent of 
${\mathbf x_0} \in {\mathbb R}^{3N}$.

Also, using \eqref{def:G_3_cut}, \eqref{eq:nu_cut}, \eqref{est:h_mu},
  \eqref{est:c-1-1_G_1} 
and \eqref{eq:est_kappa_1_cut},
\begin{align}
   \label{est:G_3_cut}
   \| G_{{3,\rm cut}}\|_{C^{1,1}(B_{3N}({\mathbf x_0},\rho))}\leq C,
\end{align}
for some constant \(C\) independent of ${\mathbf x_0} \in {\mathbb
  R}^{3N}$ and \(\rho>0\). 
Now, \eqref{eq:G_3-r-est} follows from \eqref{est:r_cut} and
  \eqref{est:G_3_cut}. This finishes the proof of Lemma~3.1'. 
\end{pf}
Let $K_{3,{\rm cut}}$ be the function constructed in Lemma~3.1'
above. Define (see \eqref{def:K_3_cut}, \eqref{F23cut}, and
\eqref{phcut}) 
\begin{align}
\zeta_{3,{\rm cut}} = e^{-F_{2,{\rm cut}} - K_{3,{\rm cut}}}
\psi=e^{-G_{3,{\rm cut}}}\phi_{3,{\rm cut}}.
\end{align}
Since for all \(\rho>0\) (using Lemma~3.1') 
\begin{align*}
  \| F_{3,{\rm cut}}- K_{3,{\rm cut}}
  \|_{C^{1,1}(B_{3N}({\mathbf x_0}, \rho))}=\|G_{3,\text{\rm cut}}
  \|_{C^{1,1}(B_{3N}({\mathbf x_0}, \rho))}
\end{align*}
 is bounded independently
of ${\mathbf x_0}$, to prove \eqref{eq:a_priori} is
equivalent to proving
\begin{align}
  \label{eq:a_priori_zeta}
  \| \zeta_{3,{\rm cut}} \|_{C^{1,1}(B_{3N}({\mathbf x_0}, R))}
  \leq
  C(R) \| \zeta_{3,{\rm cut}} \|_{L^{\infty}(B_{3N}({\mathbf x_0},
  2R))}.
\end{align}
Using that \(\zeta_{3,\text{\rm cut}}=e^{-G_{3,\text{\rm
        cut}}}\phi_{3,\text{\rm cut}}\), the estimate
        \eqref{eq:G_3-r-est} (twice), and
     the bound \eqref{eq:a_priori_1der}, we get, for all \(0<\rho<\rho'\),
  \begin{align}
    \label{eq:whatever2}
    \|\zeta_{3,\text{\rm cut}}\|_{C^{1,\alpha}(B_{3N}({\mathbf
        x_0},\rho))} 
    \leq
    C \|\zeta_{3,{\rm cut}}\|_{L^{\infty}(B_{3N}({\mathbf
    x_0}, \rho'))},
  \end{align}
with \(C=C(\rho,\rho')\).
Proving \eqref{eq:a_priori_zeta} is improving \eqref{eq:whatever2} to
\(\alpha=1\). 

The function $\zeta_{3,{\rm cut}}$ satisfies the equation
\begin{align*}
  &\Delta \zeta_{3,{\rm cut}} + 2 \big( \nabla F_{2,{\rm cut}} + \nabla
  K_{3,{\rm cut}}\big) \cdot \nabla \zeta_{3,{\rm cut}} \\
  &+
  \big( \Delta F_{2,{\rm cut}} + \Delta K_{3,{\rm cut}} +
  |\nabla F_{2,{\rm cut}} + \nabla K_{3,{\rm cut}}|^2 + (E-V)
  \big)\zeta_{3,{\rm cut}}=0.
\nonumber
\end{align*}
We can rewrite this as
\begin{align}
\label{eq:2nd_eq_for_zetaBIS}
  \Delta \zeta_{3,{\rm cut}} +
  2 \nabla F_{2,{\rm cut}} \cdot \big(\nabla \zeta_{3,{\rm cut}} &+
  \zeta_{3,{\rm cut}} \nabla K_{3,{\rm cut}}\big) \\
  &+ r_{1,{\rm cut}} \cdot
  \nabla \zeta_{3,{\rm cut}}
  + r_{2,{\rm cut}}\zeta_{3,{\rm cut}}=0,
\nonumber
\end{align}
with (since \(\Delta F_2=V\) and \(\Delta K_{3,\text{\rm
    cut}}=-|\nabla F_{2,\text{\rm cut}}|^{2}+r_{\text{\rm cut}}\))
\begin{align*}
  r_{1,\text{\rm cut}}&=2\nabla K_{3,\text{\rm cut}},\\
   r_{2,\text{\rm cut}}&=\Delta F_{2,\text{\rm cut}}+r_{\text{\rm cut}}
  +|\nabla K_{3,\text{\rm cut}}|^{2}+(E-V)\\
  &=\Delta(F_{2,\text{\rm cut}}-F_{2})+r_{\text{\rm cut}}+|\nabla
  K_{3,\text{\rm cut}}|^{2}+E. 
\end{align*}
By the construction of \(F_{2}\) and \(F_{2,\text{\rm cut}}\) (see 
\eqref{F2}, \eqref{eq:def_cutoff}, and \eqref{eq:def_F2cut}) it is
clear that for all \(\rho>0\)
\begin{align*}
  \|\Delta(F_{2,\text{\rm cut}}-F_{2})\|_{C^{\alpha}(B_{3N}({\mathbf
  x_0}, \rho))}
  \leq C,
\end{align*}
with \(C=C(\rho)\) independent of \(\mathbf x_0\in\R^{3N}\). 
Due to Lemma~3.1' (see also \eqref{eq:def_cutoff}), \(\nabla
K_{3,\text{\rm cut}}\) is \(C^{\alpha}\), and we have for all
\(\rho>0\) 
\begin{align}
  \label{eq:est_nabla_K_3}
  \|\nabla K_{3,\text{\rm cut}}\|_{C^{\alpha}(B_{3N}({\mathbf x_0},
  \rho))}\leq C,
\end{align}
with \(C=C(\rho)\) independent of \({\mathbf x_0}\in\R^{3N}\). This,
together with \eqref{eq:G_3-r-est}, means that
\begin{align}
  \label{eq:est_r_j}
  \| r_{j,{\rm cut}} \|_{C^{\alpha}(B_{3N}({\mathbf x_0}, \rho))} \leq
  C,\qquad j=1,2,
\end{align}
where $C=C(\rho)$ is independent of ${\mathbf x_0}\in\R^{3N}$.

In order to finish the proof, we introduce a localisation.
Let \(f:\R\to\R, 0\leq f\leq1\), be decreasing and such that
\(f(t)=1\) for \(t<0\) and \(f(t)=0\) for \(t>1\), and define, for
\(\rho>0, \lambda>1\), 
\begin{align}
  \theta(x)\equiv\theta_{\rho,\lambda}(x)
  =f\big(\tfrac{1}{\lambda-1}(\tfrac{|x-x_{0}|}{\rho}-1)\big).  
\end{align}
(So \(\theta(x)=1\) on \(B_{3N}(x_{0},\rho)\) and \(\theta(x)=0\)
outside \(B_{3N}(x_{0},\lambda\rho)\)).

Clearly the derivatives of \(\theta\) are bounded
independently of ${\mathbf x_0}$. 
Below, all constants \(C=C(\rho)\) also depend on \(\lambda>1\); we
omit this dependence in the notation.
On the set
$B_{3N}({\mathbf x_0}, \rho))$, $\theta \zeta_{3,{\rm cut}}$
satisfies the following equation:
\begin{align}
\label{eq:3rd_eq_for_zeta}
  \Delta (\theta \zeta_{3,{\rm cut}}) +
  2 \nabla F_{2,{\rm cut}} &\cdot \big(\nabla (\theta \zeta_{3,{\rm
  cut}}) +
  (\theta \zeta_{3,{\rm cut}}) \nabla K_{3,{\rm cut}}\big)
  \\
  &+ r_{1,{\rm cut}} \cdot
  \nabla (\theta \zeta_{3,{\rm cut}})
  + r_{2,{\rm cut}}( \theta \zeta_{3,{\rm cut}})=0.
  \nonumber 
\end{align}
Using \eqref{eq:3rd_eq_for_zeta}  we will deduce that
\begin{align}
\label{eq:a_priori_chizeta}
\|\theta_{R,\sqrt{2}}\, \zeta_{3,{\rm cut}}\|_{C^{1,1}(B_{3N}({\mathbf x_0},
R))}
\leq
C(R) \|\zeta_{3,{\rm cut}}\|_{L^{\infty}(B_{3N}({\mathbf
x_0}, 2R))},
\end{align}
from which \eqref{eq:a_priori_zeta} clearly follows (since
\(\theta\equiv 1\) on \(B_{3N}(\mathbf x_{0},R)\)).
To prove Theorem~\ref{thm:main:apriori}, it therefore remains to prove
\eqref{eq:a_priori_chizeta}.

\begin{pf*}{Proof of \eqref{eq:a_priori_chizeta}}
Let $\Psi_{j,i,\text{\rm cut}}$ be defined as $\Psi_{j,i,}$ 
was in \eqref{def:PsiJI} but with
$\zeta_3, K_3$ replaced by $\theta\zeta_{3,{\rm cut}}$,
$K_{3,{\rm cut}}$, that is (\(j\in\{1,\ldots,N\}, i\in\{1,2,3\}\)),
\begin{align}
  \label{eq:3.11bis}
  \Psi_{j,i,\text{\rm cut}} = 2 \frac{\partial (\theta\zeta_{3,{\rm
  cut}})}{\partial x_{j,i}} + 2 (\theta\zeta_{3,{\rm cut}})
  \frac{\partial K_{3,{\rm cut}}}{\partial x_{j,i}}.
\end{align}
(Here, \(\theta\equiv\theta_{R,\sqrt{2}}\)).
We define \(\hat{\Psi}_{j,i,\text{\rm cut}}\), \(\Phi_{(j,k),i,\text{\rm
    cut}}\)
analogously to \(\hat{\Psi}_{j,i}\), \(\Phi_{(j,k),i}\)
defined in \eqref{def:PsiHats} and
\eqref{def:PhiJK}.
Using \eqref{eq:est_nabla_K_3} and \eqref{eq:whatever2} we get that
for all \(0<\rho<\rho'\), 
  \begin{align}
    \label{eq:alpha-bound_PsiCut}
    \big\|\Psi_{j,i,\text{\rm cut}}\big\|_{C^{\alpha}(B_{3N}({\mathbf x_0},
    \rho))}&\leq C(\rho)\|\theta\zeta_{3,\text{\rm
    cut}}\|_{C^{1,\alpha}(B_{3N}({\mathbf x_0}, \rho))} \\ 
   &\leq C(\rho,\rho',R)\|\zeta_{3,\text{\rm
    cut}}\|_{L^{\infty}(B_{3N}({\mathbf x_0}, \rho'))}. 
    \nonumber
  \end{align}
We then have the following result, similar to
Lemma~\ref{lem:ExplPoisson}: 
\begin{lem3.4'}
Let $u_{j,i,\text{\rm cut}}, v_{(j,k),i,\text{\rm cut}}$ be the
solutions to the equations
\eqref{eq:u_s}, \eqref{eq:v_s} (with \(\hat{\Psi}_{j,i}\), \(\Phi_{(j,k),i}\)
replaced by \(\hat{\Psi}_{j,i,\text{\rm cut}}\),
\(\Phi_{(j,k),i,\text{\rm cut}}\))
given by the Newton potential on \(B_{3N}({\mathbf x_0}, \sqrt{2}R)\).

Then, for all \(\rho<\sqrt{2}R<\rho'\), there
exists a constant $C=C(\rho,\rho',R)$ (independent of ${\mathbf x_0}\in\R^{3N}$)
such that
\begin{align}\label{eq:est_uijCut}
  \| u_{j,i} \|_{C^{1,1}(B_{3N}({\mathbf x_0}, \rho))} & \leq
  C \|\zeta_{3,{\rm cut}}\|_{L^{\infty}(B_{3N}({\mathbf
  x_0}, \rho'))}, \\
  \label{eq:est_vjkCut}
  \| v_{(j,k),i} \|_{C^{1,1}(B_{3N}({\mathbf x_0}, \rho))} & \leq
  C \|\zeta_{3,{\rm cut}}\|_{L^{\infty}(B_{3N}({\mathbf
  x_0}, \rho'))}.
\end{align}
\end{lem3.4'}

\begin{pf}
Using Theorem~\ref{thm:abstract}  and Remark~\ref{rem:abstract}
\eqref{abstract(v)} and \eqref{abstract(vi)}, we get the {\it
a~priori} estimate
\begin{align}
  \label{eq:whatever3}
  \| u_{j,i,\text{\rm cut}} \|_{C^{1,1}(B_{3N}({\mathbf x_0}, \rho))}
  &\leq
  C \Big(  \sup \left| \frac{x_{j,i}}{|x_j|} \right|
  \| \hat{\Psi}_{j,i,\text{\rm cut}} \|_{C^{\alpha}(\pi_{3N-3}
    B_{3N}({\mathbf 
  x_0}, \sqrt{2}R))}\nonumber \\
   &+\Big(
  \sup_{\pi_{3N-3} B_{3N}({\mathbf
  x_0}, \sqrt{2}R))} |\hat{\Psi}_{j,i,\text{\rm cut}} |\Big)
  \Big\| \frac{x_{j,i}}{|x_j|} \Big\|_{C^{\alpha}({\mathbb S}^2)}
  \Big).\nonumber\\ 
\end{align}
Using \eqref{eq:3.11bis} and \eqref{eq:est_nabla_K_3} we have
\begin{align*}
  \| \hat{\Psi}_{j,i,\text{\rm cut}} \|_{C^{\alpha}(\pi_{3N-3}
    B_{3N}({\mathbf 
  x_0}, \sqrt{2}R))}&\leq
  \| \Psi_{j,i,\text{\rm cut}} \|_{C^{\alpha}((\pi_{3N-3}
    B_{3N}({\mathbf x_0}, \sqrt{2}R))\times\R^{3}})\\
  &\!\!\!\!\!\!\leq C \|\theta\zeta_{3,{\rm cut}}
  \|_{C^{1,\alpha}((\pi_{3N-3}B_{3N}({\mathbf x_0},
    \sqrt{2}R))\times\R^{3})}. 
\end{align*}
This, the compact support of \(\theta\), and \eqref{eq:whatever3}
implies the estimate
\begin{align}
   \label{eq:whatever1}
   \| u_{j,i,\text{\rm cut}} \|_{C^{1,1}(B_{3N}({\mathbf x_0}, \rho))}
   \leq  C \,\|\zeta_{3,{\rm cut}}
   \|_{C^{1,\alpha}(B_{3N}({\mathbf x_0}, \sqrt{2}R))}.
\end{align}
Combining \eqref{eq:whatever1} and \eqref{eq:whatever2}, we arrive at
\eqref{eq:est_uijCut}.
This finishes the proof of the estimate \eqref{eq:est_uijCut} for
$u_{j,i,\text{\rm cut}}$. 

The analogous estimate  \eqref{eq:est_vjkCut}
for $v_{(j,k),i,\text{\rm cut}}$ is proved 
in the same manner using the same coordinate transformation
as in the proof of Lemma~\ref{lem:ExplPoisson} (see also the proof of
Lemma~3.5' below). We omit the details.
\end{pf}
\begin{lem3.5'}
Let $\Psi_{j,i,\text{\rm cut}}$ be defined by \eqref{eq:3.11bis} and let
$\hat{\Psi}_{j,i,\text{\rm cut}}$ and $\Phi_{(j,k),i,\text{\rm cut}}$ be
defined by \eqref{def:PsiHats} 
and \eqref{def:PhiJK}
(with $\Psi_{j,i}$ replaced by $\Psi_{j,i,\text{\rm cut}}$).
Then the
functions defined by \eqref{eq:regRest1}  and
\eqref{eq:regRest2} (again, with an extra index `{\rm cut}') belong to
$C^{\alpha}({\mathbb R}^{3N})$ for all $\alpha \in (0,1)$.
Furthermore, for any \(\rho<\sqrt{2}R<\rho'\), 
their $C^{\alpha}$-norms on the ball
$B_{3N}({\mathbf x_0}, \rho)$ are bounded  by 
\begin{align}
  \label{bound:Lemma3.5bis}
  C \|\zeta_{3,\text{\rm
    cut}}\|_{L^{\infty}(B_{3N}({\mathbf x_0}, \rho'))}  
\end{align}
with \(C=C(\rho,\rho',R)\) independent of \({\mathbf x_0}\in\R^{3N}\).
\end{lem3.5'}
\begin{pf}
  That the functions belong to $C^{\alpha}({\mathbb R}^{3N})$ for all
  $\alpha \in (0,1)$ follows like in the proof of
  Lemma~\ref{lem:regRest}.
 
  To prove the bounds on the norms it suffices, by
  Lemma~\ref{lem:XdotG} and the triangle inequality, to prove them for
  \begin{align*}
    \big\|\Psi_{k,i,\text{\rm
    cut}}\big\|_{C^{\alpha}(B_{3N}({\mathbf
    x_0}, \rho))}
    \quad\text{and}\quad
   \big\|\Phi_{(j,k),i,\text{\rm
    cut}})\big\|_{C^{\alpha}(B_{3N}({\mathbf
    x_0}, \rho))}.
  \end{align*}
  For \( \Psi_{k,i,\text{\rm cut}}\), the estimate follows from
  \eqref{eq:alpha-bound_PsiCut}.

To bound \(\Phi_{(j,k),i,\text{\rm cut}}\), 
denote by
  \(t_{j,k}:\R^{3N}\to\R^{3N}\) the linear transformation (see also
  \eqref{def:PhiJK}), 
  \begin{align*}
    &t_{j,k}(\mathbf x)=\\&(x_{1},\ldots,x_{j-1},\tfrac{1}{2}(x_{j}+x_{k}),
  x_{j+1},\ldots,x_{k-1},\tfrac{1}{2}(x_{j}+x_{k}),x_{k+1},
  \ldots,x_{N}),
  \end{align*}
  so that 
  \begin{align*}
    \Phi_{(j,k),i,\text{\rm cut}}(\mathbf x)=\Psi_{j,i,\text{\rm
    cut}}(t_{j,k}(\mathbf x))-\Psi_{k,i,\text{\rm
    cut}}(t_{j,k}(\mathbf x)).
  \end{align*}
  Then, since \(|t_{j,k}(\mathbf z)|\leq|\mathbf z|\), 
  \begin{align}
   \label{eq:cut-alpha-norms}
    &\frac{\big|\Phi_{(j,k),i,\text{\rm cut}}(\mathbf x)
    -\Phi_{(j,k),i,\text{\rm 
    cut}}(\mathbf y)\big|}{|\mathbf x-\mathbf y|^{\alpha}}
   \leq \frac{\big|\Psi_{j,i,\text{\rm cut}}(t_{j,k}(\mathbf
    x))-\Psi_{j,i,\text{\rm cut}}(t_{j,k}(\mathbf
    y))\big|}{|t_{j,k}(\mathbf 
    x)-t_{j,k}(\mathbf y)|^{\alpha}}
    \nonumber
    \\
    &\qquad\qquad\qquad\qquad+\frac{\big|\Psi_{k,i,\text{\rm
          cut}}(t_{j,k}(\mathbf  
    x))-\Psi_{k,i,\text{\rm cut}}(t_{j,k}(\mathbf
    y))\big|}{|t_{j,k}(\mathbf x)-t_{j,k}(\mathbf y)|^{\alpha}} 
  \end{align}
  Due to the localisation \(\theta\) in the definition of
  \(\Psi_{k,i,\text{\rm cut}}\) (see \eqref{eq:3.11bis}),
  both of the terms on the RHS of \eqref{eq:cut-alpha-norms} are
  bounded by
\begin{align*}
  C(\rho)\|\zeta_{3,\text{\rm
    cut}}\|_{C^{1,\alpha}(B_{3N}({\mathbf x_0}, \sqrt{2}R))}.
\end{align*}
The bound \eqref{bound:Lemma3.5bis} for \(\Phi_{(j,k),i,\text{\rm
    cut}}\) now follows using \eqref{eq:whatever2}. 
This finishes the proof of  the bound \eqref{bound:Lemma3.5bis} for
  the functions \((\Psi_{j,i,\text{\rm cut}}-\Psi_{k,i,\text{\rm
  cut}})-\Phi_{(j,k),i,\text{\rm cut}}\).

 The proof for the functions
  \(\Psi_{j,i,\text{\rm cut}}-\hat\Psi_{j,i,\text{\rm cut}}\) is
  similar (see also the proof of Lemma~3.4' above),
  so we omit the details.
\end{pf}
To finish the proof of Theorem~\ref{thm:main:apriori},
define $U_{{\rm cut}}$ analogously to \eqref{def:U}, using the functions
$u_{j,i,\text{\rm cut}}$, $v_{(j,k),i,\text{\rm cut}}$ from
Lemma~3.4'. 
 Then, by Lemma~3.4', for any \(\rho<\sqrt{2}R<\rho'\), 
\begin{align}
  \label{eq:Ucut}
  \Delta &U_{{\rm cut}}=
  \sum_{j=1}^{N}\frac{Z}{2}\frac{x_{j}}{|x_{j}|}\cdot
  \hat\Psi_{j,{{\rm cut}}}\ 
  -\sum_{1\leq j<k\leq N}\frac{1}{4}\frac{x_{j}-x_{k}}{|x_{j}-x_{k}|}
  \cdot\Phi_{(j,k),{{\rm cut}}},\\
  \label{eq:apriori_U_cut}
  \| &U_{\rm cut}\|_{C^{1,1}(B_{3N}({\mathbf x_0}, \rho))}  \leq
  C \|\zeta_{3,{\rm cut}}\|_{L^{\infty}(B_{3N}({\mathbf
  x_0}, \rho'))}.
\end{align}
Define (\(\theta\equiv\theta_{R,\sqrt{2}}\))
\begin{align}
  \label{def:W_cut}
  W_{{\rm cut}} = \theta \zeta_{3,{\rm cut}}-
  U_{{\rm cut}},
\end{align}
then, using
\eqref{eq:3rd_eq_for_zeta}, \eqref{eq:3.11bis}, \eqref{eq:Ucut},  and the form of
\(\nabla F_{2}\) (see \eqref{eq:grad_F}), we get the following
equation for 
\(W_{{\rm cut}}\):
\begin{align}
  \label{eq:Lambda}
  \Delta W_{{\rm cut}}
  &={}-\frac{Z}{2}\sum_{j=1}^{N}\frac{x_{j}}{|x_{j}|}\cdot
  \Big\{\Psi_{j,\text{\rm cut}}-\hat\Psi_{j,\text{\rm
      cut}}\Big\}\nonumber\\ 
  &-\frac{1}{4}\sum_{1\leq j<k\leq
  N}\frac{x_{j}-x_{k}}{|x_{j}-x_{k}|}\cdot
  \Big\{\big(\Psi_{j,\text{\rm cut}}-\Psi_{k,\text{\rm
  cut}}\big)-\Phi_{(j,k),\text{\rm cut}}\Big\}\nonumber\\
  &+\sum_{j=1}^{N}\nabla_{j}\big(F_{2}-F_{2,\text{\rm cut}}\big)\cdot
   \Psi_{j,\text{\rm cut}}
   \nonumber\\
  &-\Big\{r_{1,\text{\rm cut}}\cdot\nabla(\theta\zeta_{3,\text{\rm
  cut}})+r_{2,\text{\rm cut}}(\theta\zeta_{3,\text{\rm
   cut}})\Big\}\equiv \Lambda. 
\end{align}
Here, \(\Lambda\) belongs to \(C^{\alpha}\) for all
\(\alpha\in(0,1)\), and, for all \(\rho<\sqrt{2}R<\rho'\),
\begin{align}
  \label{eq:est_R}
  \|\Lambda\|_{C^{\alpha}(B_{3N}({\mathbf x_0}, \rho))}\leq C
  \|\zeta_{3,\text{\rm 
    cut}}\|_{L^{\infty}(B_{3N}({\mathbf x_0}, \rho'))}
\end{align}
with \(C=C(\rho,\rho',R)\)
independent of \(\mathbf x_{0}\in\R^{3N}\). 
For the first two terms in \eqref{eq:Lambda} this follows from
Lemma~3.5'. For the third term, it follows using  the form of
\(F_{2}-F_{2,\text{\rm cut}}\) (see \eqref{F2}, \eqref{eq:def_cutoff},
and \eqref{eq:def_F2cut}) and \eqref{eq:alpha-bound_PsiCut}. For the
last term we use 
\eqref{eq:est_r_j} and \eqref{eq:whatever2}.

By Proposition~\ref{prop:GT} this means that \(W_{{\rm cut}}\) belongs
to \(C^{2,\alpha}\), and 
we have the estimate
\begin{align}
 \label{eq:a_priori_W}
  \| W_{{\rm cut}}&\|_{C^{1,1}(B_{3N}({\mathbf x_0}, R))}
  \leq \| W_{{\rm cut}}\|_{C^{2,\alpha}(B_{3N}({\mathbf x_0}, R))}
  \\
  &\leq
  C(R)\big( \| W_{{\rm cut}}\|_{L^{\infty}(B_{3N}({\mathbf x_0},
    \sqrt[3]{2}R))} 
  + \|\Lambda\|_{C^{\alpha}(B_{3N}({\mathbf x_0}, \sqrt[3]{2}R))}\big).
  \nonumber
\end{align}
Using \eqref{def:W_cut}, the triangle inequality,  and then
\eqref{eq:apriori_U_cut} (with 
\(\rho=\sqrt[3]{2}R\) and \(\rho'=2R\)), we have
\begin{align*}
  \| W_{{\rm cut}}\|_{L^{\infty}(B_{3N}({\mathbf x_0}, \sqrt[3]{2}R))}
  \leq C(R)\|\zeta_{3,{\rm cut}}\|_{L^{\infty}(B_{3N}({\mathbf
   x_0}, 2R))}.
\end{align*}
This, \eqref{eq:a_priori_W}, and \eqref{eq:est_R} with \(\rho=\sqrt[3]{2}R\) and
\(\rho'=2R\), gives the estimate
\begin{align}
   \label{eq:a_priori_Wfinal}
   \| W_{{\rm cut}}\|_{C^{1,1}(B_{3N}({\mathbf x_0}, R))} \leq
   C(R)\|\zeta_{3,{\rm cut}}\|_{L^{\infty}(B_{3N}({\mathbf
   x_0}, 2R))}.
\end{align}
 Using \(\theta \zeta_{3,{\rm cut}}=W_{{\rm cut}}+U_{{\rm cut}}\), 
\eqref{eq:apriori_U_cut} (with \(\rho=R\) and \(\rho'=2R\))
and \eqref{eq:a_priori_Wfinal}, the estimate 
\eqref{eq:a_priori_chizeta} follows. 
\end{pf*}
This finishes the proof of 
Theorem~\ref{thm:main:apriori}. \hspace*{\fill}\qed
\appendix
\section{Construction of the function \(\kappa\)}
\label{chap:kappa}
 In this appendix we construct the function \(\kappa\) that gives rise
 to the terms of order \(r^{2}\ln(r)\) in the function \(K_3\) solving
 \(\Delta K_3=-|\nabla F_2|^{2}\) (see the previous section,
 Remark~\ref{rem:sing_types} in particular). Therefore, \(\kappa\) is
 responsable for the \(C^{1,\alpha}\)-singularities in the
 wavefunction \(\psi\). 

More precisely, we prove the following:
\begin{lemma}
  \label{lem:construct_kappa}
  Let the function \(\gamma_2:\R^6\to \R\) be given by
  \begin{align}
    \label{def2:gamma2}
    \gamma_2(x,y)=\Big(\frac{x}{|x|}-\frac{y}{|y|}\Big)
  \cdot\frac{x-y}{|x-y|}\quad,\quad x,y\in\R^3.
  \end{align}
  Then there exists a function \(\kappa:\R^6\to\R\) of the form
  \begin{align}
  \label{eq:formula_kappa}
    \kappa(x,y)&= 
    \frac{2-\pi}{3\pi}\ (x\cdot y)\log(x^{2}+y^{2})
    +(x^2+y^2)\, G_{\kappa_1}\Big(\frac{(x,y)}{|(x,y)|}\Big)\nonumber\\
    &\equiv k(x,y)+\kappa_1(x,y)\quad,\quad G_{\kappa_1}\in
    C^{1,1}(\sphere^{5})
  \end{align}
  satisfying \(\Delta\kappa=\gamma_2\).
\end{lemma}
\begin{remark}
  Note that by Lemma~\ref{lem:r2G}, \(\kappa_{1}\in C^{1,1}(\R^{6})\).
\end{remark}
\begin{proof}
Recall that \(\mathfrak{h}_{2}^{(6)}=
\Ran(\mathcal{P}_{2}^{(6)})\) is given by the linear span
of the harmonic, homogeneous polynomials of degree \(2\) in
\({\mathbb R}^6\) restricted to
\(\sphere^{5}\).

By Lemma~\ref{prop:proj_gamma2} we have that 
\begin{align*}
  (\mathcal{P}_{2}^{(6)} \gamma_2)(r\omega)= c_{1} 
  \frac{x\cdot y}{x^{2}+y^{2}}
    \quad,\quad
   c_{1}=\frac{16(2-\pi)}{3\pi},
\end{align*}
where \(r^{2}=x^{2}+y^{2},
\omega=(x,y)/r\in\sphere^{5}\).
Let \(k(x,y)=\frac{c_{1}}{16}(x\cdot y)\log(x^{2}+y^{2})\). Then
\begin{align*}
  \big(\Delta_{x}+\Delta_{y}\big)k(x,y)=c_{1}\frac{x\cdot
    y}{x^{2}+y^{2}}=(\mathcal{P}_{2}^{(6)} \gamma_2)(r\omega).
\end{align*}
Letting \(\kappa_{1}=\kappa-k\) this reduces the problem (of finding
\(\kappa\) such that \((\Delta_{x}+\Delta_{y})\kappa=\gamma_{2}\)) to
finding \(\kappa_{1}\) such that
\begin{align}
  \label{eq:tilde_f_3}
  (\Delta_{x}+\Delta_{y})\kappa_{1}=\hat \gamma_{2}
\end{align}
 with
\begin{align}
  \label{eq:def_gamma2_tilde}
  \hat\gamma_{2}=\gamma_{2}-c_{1}\frac{x\cdot y}{x^{2}+y^{2}}.
\end{align}
Due to the above, 
\begin{align*}
  (\mathcal{P}_{2}^{(6)} \hat\gamma_{2})(r\omega)= 0.
\end{align*}
Therefore, by Proposition~\ref{prop:homogen}, there exists a solution
\(\kappa_{1}\) to 
\eqref{eq:tilde_f_3} such that \(\kappa_{1}(r\omega)=r^{2}
G_{\kappa_{1}}\!(\omega)\), with \(G_{\kappa_{1}}\!\in
C^{1,\alpha}(\sphere^{5})\) for all \(\alpha\in(0,1)\).  

To verify~\eqref{eq:formula_kappa} we need to prove that in fact
\(G_{\kappa_{1}}\!\in C^{1,1}(\sphere^{5})\). 
We will do this by proving that \(\kappa_1\in
C^{1,1}(\R^{6}\setminus\{0\})\), since then
\(G_{\kappa_{1}}\!=\kappa_{1}/ r^{2}\!\in C^{1,1}(\sphere^{5})\). 

To prove  \(\kappa_1\in
C^{1,1}(\R^{6}\setminus\{0\})\), we analyze the equation
\eqref{eq:tilde_f_3} for \(\kappa_{1}\) in the vicinity of singular
points of the function \(\hat\gamma_{2}\) on the sphere \(\sphere^{5}\).
There are two types of singular points: (a) \((x_{0}, x_{0})\in\sphere^{5}\),
(b) \((0,y_{0})\in\sphere^{5}\) (resp.\ \((x_{0},0)\in\sphere^{5}\)). 
The function \(\kappa_{1}\) is \(C^{\infty}\) in a neighbourhood of
all other points on \(\sphere^{5}\) due to Proposition~\ref{prop:GT}
(since, for \(r>0\), \(\hat\gamma_2\) is \(C^{\infty}\) away from 
points of type (a) and (b), see \eqref{def2:gamma2} and
\eqref{eq:def_gamma2_tilde}).   

(a): Let \(U_{a}\subset\R^{6}\) be a neighbourhood of a point
\((x_{0},x_{0})\in\sphere^{5}\) (i.e. \(2|x_{0}|^{2}=1\))
such that for some \(c>0\), \(|x|\geq c, |y|\geq c\) for \((x,y)\in U_{a}\).
Choose new coordinates: Let
\begin{align*}
  (x_{1}, x_{2})=t(x,y)=(x-y, x+y).
\end{align*}
Then
\begin{align*}
  \big(\gamma_{2}\circ t^{-1}\big)(x_{1},x_{2})=\frac{x_{1}}{|x_{1}|}\cdot
  \Big(\frac{x_{1}-x_{2}}{|x_{1}-x_{2}|}
  +\frac{x_{1}+x_{2}}{|x_{1}+x_{2}|}\Big)   
  \equiv\frac{x_{1}}{|x_{1}|}\cdot G_{a}(x_{1},x_{2})   
\end{align*}
with \(G_{a}\in C^{\infty}\big(t(U_{a})\big)\). Since
\(G_{a}(0,x_{2})=0\) for \(x_{2}\neq0\) (that is,  
for \(x=y\neq0\) in the original coordinates), we have, by Lemma~\ref{lem:XdotG},
that \(\gamma_{2}\circ t^{-1}\in C^{0,1}\big(t(U_{a})\big)\), and
therefore \(\gamma_{2}\in C^{0,1}(U_{a})\subset C^{\alpha}(U_{a})\) for all
\(\alpha\in(0,1)\).  
Since  \((x\cdot y)/(x^{2}+y^{2})\in
C^{\infty}(U_{a})\), we have (see \eqref{eq:def_gamma2_tilde})
\(\hat\gamma_{2}\in C^{\alpha}(U_{a})\) for all
\(\alpha\in(0,1)\). By Proposition~\ref{prop:GT} we get from
\eqref{eq:tilde_f_3} that
\(\kappa_{1}\in C^{2,\alpha}(U_{a})\). 

(b): Let \(U_{b}\subset\R^{6}\) be a neighbourhood of a point
\((0,y_{0})\in\sphere^{5}\) (i.e.\ \(|y_{0}|=1\)) such that for some
\(c>0\), \(|y|\geq c, |x-y|\geq c\) for \((x,y)\in U_{b}\). Then
\begin{align}
  \gamma_{2}(x,y)
  &=\Big(\frac{x}{|x|}-\frac{y}{|y|}\Big)
  \cdot\frac{x-y}{|x-y|}
  \nonumber\\&
  ={}-\frac{x}{|x|}\cdot\frac{y}{|y|}+\frac{x}{|x|}
  \cdot\Big(\frac{y}{|y|}-\frac{y-x}{|y-x|}\Big)
  -\frac{y}{|y|}\cdot\frac{x-y}{|x-y|}.
  \nonumber
\end{align}
Note that 
\begin{align*}
  {}-\frac{y}{|y|}\cdot\frac{x-y}{|x-y|}
  \in C^{\infty}(U_{b})
\end{align*}
and that
\begin{align*}
  \frac{x}{|x|}
  \cdot\Big(\frac{y}{|y|}- \frac{y-x}{|y-x|}\Big)
  \equiv\frac{x}{|x|}\cdot G_{b}(x,y),
\end{align*}
with \(G_{b}\in C^{\infty}(U_{b})\), \(G_{b}(0,y)=0\) for \(y\neq0\).
Therefore, by Lemma~\ref{lem:XdotG} and \eqref{eq:def_gamma2_tilde}, 
\begin{align*}
  \hat\gamma_{2}(x,y)-\Big({}-\frac{x}{|x|}\cdot\frac{y}{|y|}\Big) 
  \in C^{0,1}(U_{b})\subset C^{\alpha}(U_{b}) \text{ for all }\alpha\in(0,1).
\end{align*}

Let \(\kappa_{2}\) be such that 
\begin{align}
  \big(\Delta_{x}+\Delta_{y}\big)\kappa_{2}
  ={}-\frac{x}{|x|}\cdot\frac{y}{|y|}\quad,\quad
  \kappa_{2}\in C^{1,1}(U_{b}).
  \nonumber
\end{align}
The existence of such a function is ensured by Theorem~\ref{thm:abstract},
since \(y\neq0\) for \((x,y)\in U_{b}\), and 
\(\mathcal{P}_{2}^{(3)}\big(\frac{x}{|x|}\big)=0\) due to the anti-symmetry 
of \(\frac{x}{|x|}\).

Then (see \eqref{eq:tilde_f_3}) \(\kappa_{3}=\kappa_{1}-\kappa_{2}\) solves
\begin{align}
  \big(\Delta_{x}+\Delta_{y}\big)\kappa_{3}=
  \hat\gamma_{2}(x,y)-\Big({}-\frac{x}{|x|}\cdot\frac{y}{|y|}\Big)  
  \in C^{\alpha}(U_{b}) \text{ for all }\alpha\in(0,1),
  \nonumber
\end{align}
so  by elliptic regularity \(\kappa_{3}\in C^{2,\alpha}(U_{b})\subset
C^{1,1}(U_{b})\). Since \(\kappa_{2}\in C^{1,1}(U_{b})\), 
this proves \(\kappa_{1}=\kappa_{2}+\kappa_{3}\in
C^{1,1}(U_{b})\). Together with \(\kappa_{1}\in C^{2,\alpha}(U_{a})\) from
above, this implies \(G_{\kappa_{1}}\!=\kappa_{1}/r^{2}\in
C^{1,1}(\sphere^{5})\), 
and so \(\kappa_{1}=r^{2}\,G_{\kappa_{1}}\in C^{1,1}(\R^{6})\). 

This finishes the proof of the existence of \(\kappa\) solving 
\eqref{eq:eq_for_kappa}, and having the form \eqref{eq:formula_kappa},
with \(G=G_{\kappa_{1}}\).
\end{proof}
\section{Construction of the function \(\nu\)}
\label{chap:nu}

In this appendix we construct a function \(\nu\) solving 
\eqref{eq:eq_for_nu}.
\begin{lem}
  \label{lem:construct_nu}
There exists a solution $\nu = \nu(x,y,z)$ to the equation \eqref{eq:eq_for_nu}
satisfying
\begin{enumerate}[\rm (i)]
\item \label{invariance}
$\nu$ is invariant under cyclic permutation, i.e.,
$\nu(x,y,z) = (\nu \circ \sigma)(x,y,z)$ for all $x,y,z \in {\mathbb R}^3$,
where $\sigma (x,y,z) = (z,x,y)$.
\item $\nu \in C^{1,1}({\mathbb R}^9)$.
\end{enumerate}
\end{lem}

The idea is to change coordinates, to the centre-of-mass frame for
\((x,y,z)\). In these new coordinates, the problem of solving
\eqref{eq:eq_for_nu} turns 
out to reduce to a problem in \(6\) variables only. By an extra symmetry of
the function \(\gamma_{3}\) (see \eqref{def:gamma_2&gamma_3}), namely
permutation of the three electron-coordinates \(x, y\), and \(z\), the
logarithmic term that occured in the function \(\kappa\) 
(see \eqref{eq:formula_kappa}) does not occur here. This is because 
the projection on \(\mathfrak{h}_{2}^{(6)}\)
of \(\tilde\gamma_{3}\) (the function
that \(\gamma_{3}\) transforms into in the new coordinates, see
\eqref{eq:inv_gamma3} below) vanishes,
due to this extra symmetry.

\begin{pf}
Make the following change of coordinates
(each entry below is a diagonal \(3\times3\)-matrix with the listed 
number in the diagonal; we will use this notation repeatedly; here,
\(x,y,z\in\R^{3}\)) 
\begin{align}
  \label{eq:change_coord}
  \left(\begin{array}{c}
     x \\ y \\ z 
  \end{array}\right)
  = \mathcal{T}\,
  \left(\begin{array}{c}
     x_{1} \\ x_{2} \\ x_{3} 
  \end{array}\right)
  =\left(
    \begin{array}{ccc}
      \frac{1}{\sqrt{3}} & 0                   &  \frac{2}{\sqrt{6}} \\
      \frac{1}{\sqrt{3}} &  \frac{1}{\sqrt{2}} & -\frac{1}{\sqrt{6}} \\
      \frac{1}{\sqrt{3}} & -\frac{1}{\sqrt{2}} & -\frac{1}{\sqrt{6}} \\
    \end{array}\right)
  \left(\begin{array}{c}
     x_{1} \\ x_{2} \\ x_{3} 
  \end{array}\right).
\end{align}
Then
\begin{align}
  \label{eq:inv_gamma3}
  &\big(\gamma_{3}\circ \mathcal{T}\big)(x_{1},x_{2},x_{3})
  =
  \\
  \nonumber
  &\frac{x_{2}}{|x_{2}|}\cdot\frac{x_{2}+\sqrt{3}x_{3}}{|x_{2}+\sqrt{3}x_{3}|}
  +\frac{x_{2}}{|x_{2}|}\cdot\frac{x_{2}-\sqrt{3}x_{3}}{|x_{2}-\sqrt{3}x_{3}|}
  -\frac{x_{2}+\sqrt{3}x_{3}}{|x_{2}+\sqrt{3}x_{3}|}\cdot
  \frac{x_{2}-\sqrt{3}x_{3}}{|x_{2}-\sqrt{3}x_{3}|}
  \\
  &\equiv\tilde\gamma_{3}(x_{1},x_{2},x_{3}).
  \nonumber
\end{align}
That \(\tilde\gamma_{3}\) is independent of \(x_{1}\) is the fact that
\(\gamma_{3}\) only depends on the inter-electron coordinates 
(\(x-y\), \(y-z\), \(z-x\) respectively), and not on the centre-of-mass
coordinate (\(x_{CM}=\frac{1}{\sqrt{3}}(x+y+z)=x_{1}\)).

The function \(\gamma_{3}\) is invariant under
cyclic permutation of the elec\-tron-coordinates \(x, y\) and \(z\), that
is, \(\big(\gamma_{3}\circ\sigma\big)(x,y,z)=\gamma_{3}(x,y,z)\)
for all \(x, y, z\in \R^{3}\) with \(\sigma(x,y,z)=(z,x,y)\). This
gives that 
\begin{align}
  \label{eq:inv_R}
  \big(\tilde\gamma_{3}\circ\mathcal{R}\big)(x_{1},x_{2},x_{3})
  =\tilde\gamma_{3}(x_{1},x_{2},x_{3}) 
  \text{ for all } x_{1}, x_{2}, x_{3}\in\R^{3},
\end{align}
with \(\mathcal{R}\) the orthogonal transformation given by
\(\mathcal{R}=\mathcal{T}^{-1}\circ\sigma\circ \mathcal{T}\), that is by the 
\(9\times9\)-matrix (again, each entry is a diagonal \(3\times3\)-matrix)
\begin{align}
  \mathcal{R}=
  \left(\begin{array}{ccc}
     1 & 0                    & 0                     \\
     0 & \cos(\frac{2\pi}{3}) & \sin(\frac{2\pi}{3}) \\
     0 & -\sin(\frac{2\pi}{3}) & \cos(\frac{2\pi}{3})  \\
  \end{array}\right).
  \nonumber
\end{align}
Note that \(\mathcal{R}\) is a rotation of \((x_{2},x_{3})\) by
\(\frac{2\pi}{3}\) around \(x_{1}\) (all in \(\R^{9}\)), that is,
\(\mathcal{R}^{3}= I_{9}\), where \(I_{9}\) is 
the identity on \(\R^{9}\).

Define the function \(\bar\gamma_{3}\) by
\begin{align}
  \label{eq:def_gamma3Bar}
  \bar\gamma_{3}(x_{2},x_{3})
  =\tilde\gamma_{3}(x_{1},x_{2},x_{3})\quad,\quad
  (x_{2},x_{3})\in\R^{6}
\end{align}
(since \(\tilde\gamma_{3}\) is independent of \(x_{1}\), this is  
well defined). Then, due to \eqref{eq:inv_R}, 
\begin{align}
  \label{eq:R_invBis}
  \big(\bar\gamma_{3}\circ\bar{\mathcal{R}}\big)(x_{2},x_{3})
  =\bar\gamma_{3}(x_{2},x_{3}) 
  \text{ for all } x_{2}, x_{3}\in\R^{3},
\end{align}
with (each entry still being a diagonal \(3\times3\)-matrix)
\begin{align}
 \label{def:barR}
 \bar{\mathcal{R}}=
 \left(\begin{array}{cc}
     -\frac{1}{2}        & \frac{\sqrt{3}}{2} \\
     -\frac{\sqrt{3}}{2} & -\frac{1}{2}        \\ 
  \end{array}\right)
  =
  \left(\begin{array}{cc}
     \cos(\frac{2\pi}{3}) & \sin(\frac{2\pi}{3}) \\
     -\sin(\frac{2\pi}{3}) & \cos(\frac{2\pi}{3})  \\
  \end{array}\right).
\end{align}

Observe that if \(\bar\nu=\bar\nu(x_{2},x_{3})\) solves
(for \(\bar\gamma_{3}\), see \eqref{eq:inv_gamma3} and
\eqref{eq:def_gamma3Bar}) 
\begin{align}
  \label{eq:gamma4}
  \big(\Delta_{x_{2}}+\Delta_{x_{3}}\big)\bar\nu&=\bar\gamma_{3},
\end{align}
then trivially the function \(\tilde\nu\) defined by
\(\tilde\nu(x_{1},x_{2},x_{3})=\bar\nu(x_{2},x_{3})\) solves
\begin{align*}
  \big(\Delta_{x_{1}}+\Delta_{x_{2}}+\Delta_{x_{3}}\big)\tilde\nu
  =\tilde\gamma_{3}.
\end{align*}
Since \(\mathcal{T}\) is orthogonal, the function 
\(\nu=\tilde\nu\circ \mathcal{T}^{-1}\)
will then solve (recall that \(\tilde\gamma_{3}=\gamma_{3}\circ \mathcal{T}\))
  \(\big(\Delta_{x}+\Delta_{y}+\Delta_{z}\big)\nu=\gamma_{3}\),
that is, \eqref{eq:eq_for_nu}. The problem of solving
\eqref{eq:eq_for_nu} therefore reduces to solving \eqref{eq:gamma4}.

Observe next that (see \eqref{eq:inv_gamma3} and
\eqref{eq:def_gamma3Bar})
\begin{align}
   \bar\gamma_{3}(\mathcal{O}x_{2},\mathcal{O}x_{3})
   =\bar\gamma_{3}(x_{2},x_{3})
   \text{ for all } \mathcal{O}\in SO(3), x_{2}, x_{3}\in\R^{3}.
   \nonumber
\end{align}
This and \eqref{eq:R_invBis} gives, by \eqref{lem:NO_l=2} of
Lemma~\ref{lemma:first_invariance}, that
\(\mathcal{P}^{(6)}_{2}\bar\gamma_{3}=0\) .
Therefore, by
Proposition~\ref{prop:homogen}, there exists a solution \(\bar\nu\)
to \eqref{eq:gamma4} with
\begin{align}
  \bar\nu(x_{2},x_{3})=(x_{2}^{2}+x_{3}^{2})\,
  G_{\bar\nu}\left(\frac{(x_{2},x_{3})}{|(x_{2},x_{3})|}\right)&, \nonumber\\
  G_{\bar\nu}\in &C^{1,\alpha}(\sphere^{5}) \text{ for
  all } \alpha\in(0,1).
  \nonumber
\end{align}

We proceed to prove that in fact \(G_{\bar\nu}\in
C^{2,\alpha}(\sphere^{5})\) for all \(\alpha\in(0,1)\). We do this by
showing that \(\bar\nu\in 
C^{2,\alpha}(\R^{6}\setminus\{0\})\), using \eqref{eq:gamma4} and elliptic
regularity (Proposition \ref{prop:GT}). 

Note that there are two kinds of singular points of
\(\bar\gamma_{3}\) on \(\sphere^{5}\): (a) \(x_{2}=0\) (and so
\(x_{3}\neq0\)), (b) \(x_{2}=\sqrt{3}x_{3}\) (and so
\(x_{2}\neq0\neq x_{3}\)) (resp.  \(x_{2}=-\sqrt{3}x_{3}\)). The
function \(\bar\nu\) (and therefore, 
\(G_{\bar\nu}\)) is \(C^{\infty}\) in a neighbourhood of 
all other points on \(\sphere^{5}\) due to elliptic regularity
(Proposition \ref{prop:GT}).

(a): Let \(U_{a}\subset\R^{6}\) be a neighbourhood of a point
\((0,x_{3}^{0})\in\sphere^{5}\) (i.e.,
\(x_{3}^{0}\neq0\)), such that for some \(c>0\),
\(|x_{2}+\sqrt{3}x_{3}|\geq c\), 
\(|x_{2}-\sqrt{3}x_{3}|\geq c\) for \((x_{2},x_{3})\in U_{a}\). Note
that
\begin{align}
  \label{eq:gamma_3_bar}
  \bar\gamma_{3}(x_{2},x_{3})&=
  \frac{x_{2}}{|x_{2}|}\cdot\left(\frac{x_{2}
  +\sqrt{3}x_{3}}{|x_{2}+\sqrt{3}x_{3}|}
  +\frac{x_{2}-\sqrt{3}x_{3}}{|x_{2}-\sqrt{3}x_{3}|}\right)
  \nonumber\\ 
  &\quad-\frac{x_{2}+\sqrt{3}x_{3}}{|x_{2}+\sqrt{3}x_{3}|}\cdot
  \frac{x_{2}-\sqrt{3}x_{3}}{|x_{2}-\sqrt{3}x_{3}|}.
\end{align}
Write
\begin{align*}
  \frac{x_{2}}{|x_{2}|}\cdot\left(\frac{x_{2}
  +\sqrt{3}x_{3}}{|x_{2}+\sqrt{3}x_{3}|}
  +\frac{x_{2}-\sqrt{3}x_{3}}{|x_{2}-\sqrt{3}x_{3}|}\right)
  \equiv\frac{x_{2}}{|x_{2}|}\cdot G_{a}(x_{2},x_{3})
\end{align*}
where \(G_{a}\in C^{\infty}(U_{a})\), \(G_{a}(0,x_{3})=0\).
Furthermore,
\begin{align*}
\frac{x_{2}+\sqrt{3}x_{3}}{|x_{2}+\sqrt{3}x_{3}|}\cdot
  \frac{x_{2}-\sqrt{3}x_{3}}{|x_{2}-\sqrt{3}x_{3}|} \in C^{\infty}(U_{a}).
\end{align*}
Therefore, due to Lemma~\ref{lem:XdotG}, \(\bar\gamma_{3}\in
C^{0,1}(U_{a})\subset  
C^{\alpha}(U_{a})\) for all \(\alpha\in (0,1)\), and
so, by \eqref{eq:gamma4} and elliptic
regularity (Proposition \ref{prop:GT}), \(\bar\nu\in C^{2,\alpha}(U_{a})\). 

(b): Let \(U_{b}\) be a neighbourhood of a point
\((x_{2}^{0},x_{3}^{0})\in\sphere^{5}\) with
\(x_{2}^{0}=\sqrt{3}x_{3}^{0}\) (i.e., \(x_{2}^{0}\neq0\neq
x_{3}^{0}\)), such that 
for some \(c>0\), \(|x_{2}|\geq c\), \(|x_{2}+\sqrt{3}x_{3}|\geq c\)
for \((x_{2},x_{3})\in U_{b}\). Choose new coordinates: Let 
\begin{align*}
  (u,v)=\tau(x_{2},x_{3})=(x_{2}-\sqrt{3}x_{3},x_{2}+\sqrt{3}x_{3}).
\end{align*}
Then
\begin{align*}
 \big(\bar\gamma_{3}\circ \tau^{-1}\big)(u,v)=
  \frac{u}{|u|}\cdot\left(\frac{u+v}{|u+v|}-\frac{v}{|v|}\right)
  +\frac{u+v}{|u+v|}\cdot\frac{v}{|v|}.
\end{align*}
We proceed as above. Write
\begin{align*}
  \frac{u}{|u|}\cdot\left(\frac{u+v}{|u+v|}-\frac{v}{|v|}\right)
  \equiv\frac{u}{|u|}\cdot G_{b}(u,v)
\end{align*}
where \(G_{b}\in C^{\infty}\big(\tau(U_{b})\big)\) (since \(v\neq0, u+v\neq0\) in
\(\tau(U_{b})\)), \(G_{b}(0,v)=0\) for \(v\neq0\). Furthermore,
\begin{align*}
  \frac{u+v}{|u+v|}\cdot\frac{v}{|v|}\in C^{\infty}(U_{b}).
\end{align*}
Lemma~\ref{lem:XdotG} implies that \(\bar\gamma_{3}\circ \tau^{-1}\in
C^{0,1}\big(\tau(U_{b})\big)\), and so 
\(\bar\gamma_{3}\in C^{0,1}(U_{b})\) \(\subset
C^{\alpha}(U_{b})\) for all \(\alpha\in(0,1)\). By
\eqref{eq:gamma4} and elliptic 
regularity (Proposition \ref{prop:GT}) follows that 
\(\bar\nu\in C^{2,\alpha}(U_{b})\).

Singular points of the form \(x_{2}^{0}=-\sqrt{3}x_{3}^{0}\) 
are treated analogously.

From the above follows that \(\bar\nu\in
C^{2,\alpha}(\R^{6}\setminus\{0\})\), and therefore \(G_{\bar\nu}\in
C^{2,\alpha}(\sphere^{5})\), for all \(\alpha\in(0,1)\).

This finishes the construction of a function \(\bar\nu\in
C^{1,1}(\R^{6})\) that solves \eqref{eq:gamma4}, and has the form
\begin{align}
  \label{eq:form_nu_bar}
  \bar\nu(x_{2},x_{3})=(x_{2}^{2}+x_{3}^{2})\,
  G_{\bar\nu}\left(\frac{(x_{2},x_{3})}{|(x_{2},x_{3})|}\right)&, \\
  G_{\bar\nu}\in &C^{2,\alpha}(\sphere^{5}) \text{ for
  all } \alpha\in(0,1).\nonumber
\end{align}

As discussed above $\overline{\nu}$ defines a function $\nu$ solving the
equation \eqref{eq:eq_for_nu}. Clearly, since $\overline{\nu} \in C^{1,1}({\mathbb
R}^6)$, we get $\nu \in C^{1,1}({\mathbb R}^9)$.
The solution $\nu$ constructed in this manner does not necessarily satisfy
the invariance property (\ref{invariance}). In order to force this invariance,
we consider
$$
\nu_{\text{sym}} = \frac{1}{3} \sum_{j=1}^3 (\nu \circ \sigma^j)(x,y,z).
$$
Since the Laplace operator commutes with $\sigma$, and $\gamma_3$ is
invariant under $\sigma$, $\nu_{\text{sym}}$ satisfies the conclusion of
Lemma~\ref{lem:construct_nu}. 
\end{pf}
With the notation from the proof of Lemma~\ref{lem:construct_nu}, we define
$$
  \overline{\nu}_{{\rm cut}}(x_2,x_3) = \chi(x_2^2 + x_3^2)\,
  \overline{\nu}(x_2,x_3),
$$
with $\chi$ as in \eqref{eq:def_cutoff}, and
\(\tilde\nu_{{\rm cut}}(x_1,x_2,x_3)\equiv\overline{\nu}_{{\rm
    cut}}(x_2,x_3)\) (as already defined). 
As discussed above (for $\nu$) the function
$\tilde\nu_{{\rm cut}}$ defines a function $\nu_{{\rm
cut}}=\tilde\nu_{{\rm cut}}\circ\mathcal T^{-1} : {\mathbb R}^9 \rightarrow
{\mathbb R}$ (by the linear 
transformation \(\mathcal T\) in \eqref{eq:change_coord}).
We then get:
\begin{lem}
\label{lem:construct_nu_cut}
The function $\nu_{{\rm cut}}$ satisfies
$$
  \Delta\nu_{{\rm cut}} = \gamma_3+h,
$$
with  $\gamma_3$ as in \eqref{def:gamma_2&gamma_3}
and $h\in
C^{\alpha}({\mathbb R}^9)$ for all $\alpha \in (0,1)$.
Furthermore, we have the estimate
\begin{align}
  \label{eq:construct_nu_cut}
  \| \nu_{{\rm cut}} \|_{C^{1,1}(B_{9}((x_0,y_0,z_0),R))}+
  \| h \|_{C^{\alpha}(B_{9}((x_0,y_0,z_0),R))} \leq C,
\end{align}
with \(C\) independent of \((x_0,y_0,z_0)\in\R^{9}\) and \(R>0\).
\end{lem}
\begin{pf}
We calculate, using \eqref{eq:gamma4},
\begin{align*}
   \big(\Delta_{x_1}+\Delta_{x_2}+\Delta_{x_3}\big)\tilde\nu_{{\rm
    cut}}&=
   \big(\Delta_{x_2}+\Delta_{x_3}\big) \overline{\nu}_{{\rm cut}}
   \equiv\Delta \overline{\nu}_{{\rm cut}}\\
   &=\overline{\gamma}_3+
   \big\{(\Delta \chi) \overline{\nu} + 2 \nabla \chi \cdot \nabla
   \overline{\nu}\big\}
    - (1-\chi) \overline\gamma_3\\
   &\equiv \tilde\gamma_3+\tilde h.
\end{align*}
Using \eqref{eq:gamma_3_bar} and \eqref{eq:form_nu_bar}
we see that the term in $\{\cdot \}$ is $C^{\alpha}$ and has compact
support. The function $(1-\chi) \overline{\gamma}_3$ is
$C^{\alpha}$ (this was proved in the proof of Lemma~\ref{lem:construct_nu})
and homogeneous of degree zero outside \(B_{6}(0,2)\). Therefore,
\begin{align*}
  \| \tilde h \|_{C^{\alpha}(B_{9}((x_1^0,x_2^0, x_3^0),R))} \leq C,
\end{align*}
with \(C\)  independent of \((x_1^0,x_2^0,x_3^0)\in\R^{9}\) and
\(R>0\). Since \(\chi\) has compact support, and \(\overline{\nu}\in
C^{1,1}(\R^{6})\), we have
\begin{align*}
  \| \tilde \nu_{{\rm cut}} \|_{C^{1,1}(B_{9}((x_1^0,x_2^0, x_3^0),R))} \leq C,
\end{align*}
with \(C\)  independent of \((x_1^0,x_2^0,x_3^0)\in\R^{9}\) and
\(R>0\).

Since \(\mathcal T\) is an orthogonal transformation,
\eqref{eq:construct_nu_cut} follows. This finishes the proof of the lemma.
\end{pf}
\section{Computation of  \(\mathcal{P}_{2}^{(6)}\gamma_2\)}
\label{chap:P_2_two_elec}
In this appendix we compute \(\mathcal{P}_{2}^{(6)}\gamma_{2}\),
the singular part of the two-particle terms in
\(|\nabla F_{2}|^{2}\), see \eqref{eq:grad_F2_squared} and
\eqref{def:gamma_2&gamma_3}. This is Lemma~\ref{prop:proj_gamma2}
below. It follows from general results on 
\(\mathcal{P}_{2}^{(6)}\eta\) when \(\eta\) has certain 
symmetry-properties (Lemma~\ref{lemma:first_invariance}). The 
latter is also responsable for the non-occurence of terms
of order \(r^{2}\ln(r)\) (of regularity \(C^{1,\alpha}\) only)
in the function \(\nu\) constructed in the previous appendix;
see Lemma~\ref{lem:construct_nu}.
\begin{lem}
  \label{prop:proj_gamma2}
  Let 
\begin{align}
  \label{eq:gamma3_again}
  \gamma_{2}(x,y)=\Big(\frac{x}{|x|}-\frac{y}{|y|}\Big)
  \cdot\frac{x-y}{|x-y|},\quad(x,y)\in\R^{3}\times\R^{3}.
\end{align}
Then
\begin{align*}
  \big(\mathcal{P}_{2}^{(6)}\gamma_{2}\big)(x,y)
  =\frac{16(2-\pi)}{3\pi}\frac{x\cdot y}{x^{2}+y^{2}},
  \quad(x,y)\in\R^{3}\times\R^{3}. 
\end{align*}
\end{lem}
\begin{pf}This will follow from Lemma~\ref{lemma:first_invariance}
and Lemma~\ref{lem:value_c0} below. Na\-me\-ly,
by (\ref{lem:invarianceI}) and (\ref{cor:x*y})
in Lemma~\ref{lemma:first_invariance}
we get that, due to symmetry,   
\begin{align*}
 \big(\mathcal{P}_{2}^{(6)}\gamma_{2}\big)(x,y)
  =c_{1}\,\frac{x\cdot y}{x^{2}+y^{2}}\ \text{ for some }\ c_{1}\in\R,
\end{align*}
that is, only the function \(x\cdot y\) (restricted to
\(\sphere^{5}\)) contributes to the projection onto \(\mathfrak
h_{2}^{(6)}\) of the function
\(\gamma_{2}\) in \eqref{eq:gamma3_again}.
That \(c_{1}=\frac{16(2-\pi)}{3\pi}\) is the
result of Lemma~\ref{lem:value_c0} (which is merely two computations).
\end{pf}
\begin{lem}
  \label{lemma:first_invariance}
  Assume \(\eta\in L^{2}({\mathbb S}^5)\) satisfies
  \begin{align} 
    \label{eq:inv_SO3}
    \eta({\mathcal O}x, {\mathcal O}y)&=\eta(x,y)
  \end{align}
  for all \(\mathcal{O} \in SO(3)\) and almost all
  \((x, y)\in\sphere^{5}\subset\R^{3}\times\R^{3}\).
  Let \(\mathcal{Q}_{1}\) be the orthogonal projection (in
  \(L^{2}(\sphere^{5})\)) onto
  \begin{align*}  
    \Span\left\{\left.P_{1}\right|_{\sphere^{5}},
    \left.P_{2}\right|_{\sphere^{5}}\right\},
  \end{align*}
  and \(\mathcal{Q}_{2}\) the orthogonal projection  onto
  \begin{align*}
    \Span\left\{\left.P_{1}\right|_{\sphere{^5}}\right\},
  \end{align*}
  where \(P_{1}(x,y)=x\cdot y\), \(P_{2}(x,y)=x^{2}-y^{2}\),
  \((x,y)\in\R^{3}\times\R^{3}\). 

  Then
\begin{enumerate}[\rm (i)]
   \item\label{lem:invarianceI}
    \(\mathcal{P}_{2}^{(6)}\eta = \mathcal{Q}_{1}\eta\).
  \item
  \label{cor:x*y}
  Let \(\eta\) satisfy
  \begin{align}
    \label{eq:symmetry1}
    \eta(x, y) = \eta(y, x) 
    \text{ for almost all } (x,  
    y)\in\sphere^{5}\subset\R^{3}\times\R^{3}. 
  \end{align}
  Then \(\mathcal{P}_{2}^{(6)}\eta = \mathcal{Q}_{2}\eta\).
\item
\label{lem:NO_l=2}
 Let \(\bar{\mathcal{R}}\) be as in \eqref{def:barR}.
 Assume
\(\eta\) satisfies 
  \begin{align}
    \label{eq:symmetry2}
    \!\!\!\!\!\!\!\!
    \eta(\bar{\mathcal{R}}(x, y)) = \eta(x, y) 
    \text{ for almost all } (x,  
    y)\in\sphere^{5}\subset\R^{3}\times\R^{3}.
  \end{align}
Then \(\mathcal{P}_{2}^{(6)}\eta = 0\).
\end{enumerate}
\end{lem}
\begin{pf*}{Proof of Lemma~\ref{lemma:first_invariance}}
Suppose  (\ref{lem:invarianceI}) is proven then the 
proofs of (\ref{cor:x*y}) and (\ref{lem:NO_l=2})
are simple:
\begin{pf*}{Proof of {\rm\eqref{cor:x*y}}}
Due to (\ref{lem:invarianceI}) we only need to prove that
  \begin{align*}
    \int_{{\mathbb S}^5} \eta(x,y) (x^2-y^2)\,d\omega = 0.
  \end{align*}
  This follows using the symmetry \eqref{eq:symmetry1} of \(\eta\)
  (which preserves the measure \(d\omega\) of \(\sphere^{5}\)): 
  \begin{align*}
    \int_{{\mathbb S}^5} \eta(x,y) P(x,y)\,d\omega =
    \frac{1}{2} \int_{{\mathbb S}^5} \eta(x,y)
    \big(P(y,x)+P(x,y)\big)\,d\omega, 
  \end{align*}
  and when \(P(x,y)=P_{2}(x,y)=x^2-y^2\), then 
  \(P(y,x)+P(x,y)=0.\) This proves (\ref{cor:x*y}).
\end{pf*}
\begin{pf*}{Proof of {\rm\eqref{lem:NO_l=2}}}
 Using  (\ref{lem:invarianceI}) and \eqref{eq:symmetry2}
it is enough to show that
\begin{align*}
P(x,y) + P(\bar{\mathcal R}(x,y)) + P(\bar{\mathcal R}^2(x,y)) = 0,
\end{align*}
when \(P(x,y) = x \cdot y\) or \( x^2 - y^2\) (since \(\bar{\mathcal R}\)
preserves the measure \(d\omega\) of \(\sphere^{5}\)).
This follows by direct calculation. 
\end{pf*}
It remains to prove  (\ref{lem:invarianceI}):
\begin{pf*}{Proof of {\rm\eqref{lem:invarianceI}}}
  Recall that \({\mathfrak h}_{2}^{(6)}=\Ran(\mathcal{P}_{2}^{(6)})\). 
  Define \({\mathfrak h}_{2,inv}\) by
  \begin{align*}
    {\mathfrak h}_{2,inv}=\Span\left\{ f\in{\mathfrak h}_{2}^{(6)} \,\big|\, 
    f({\mathcal O} x, {\mathcal O} y) = f(x,y) \, 
    \text{ for all } {\mathcal O} \in SO(3) \right\}.
  \end{align*}
  Note that \(\mathcal{P}_{2}^{(6)}\eta\in {\mathfrak h}_{2,inv}\)
because of \eqref{eq:inv_SO3}.
We need to prove that
  \begin{align*}
    {\mathfrak h}_{2,inv}=\Span\left\{\left.P_{1}\right|_{\sphere^{5}},
    \left.P_{2}\right|_{\sphere^{5}}\right\}.
  \end{align*}
  Since every function in \(\mathfrak{h}_{2,inv}\) can be written as a
finite sum of spherical harmonics of degree \(2\) it suffices to consider
  a real, harmonic polynomial \(P\) which is homogeneous
  of degree \(2\), and which is invariant under the action of
  \(SO(3)\):
  \begin{align}
    \label{eq:inv_P}
    P({\mathcal O} x, {\mathcal O} y)=P(x,y)
    \text{ for all } {\mathcal O} \in SO(3).
  \end{align}
  Identifying \(P\) with a quadratic form on \(\R^{6}\), there exist
  real symmetric matrices \(A, B\), and \(C\), such that 
  \begin{align}
    \label{eq:A_B_C}
    P(x,y)= x\cdot Ax + y\cdot B y + x\cdot C y.
  \end{align}
  The condition of
  harmonicity of \(P\) becomes \(\Tr[ A + B ] = 0\). We prove that
  \(A,B\), and \(C\)
  have to be multiples of the identity matrix \(I_{3}\) on \(\R^{3}\). 
  To do so, let us first restrict to \(x=0\). Using 
  \eqref{eq:inv_P} and \eqref{eq:A_B_C} we get  
  \begin{align*}
    y\cdot B y = P(0,y) = P({\mathcal O}0,{\mathcal O}y) =
    \mathcal{O}y\cdot B{\mathcal O} y,
  \end{align*}
  for all \({\mathcal O} \in SO(3)\). Let \(\lambda\) be a (real) eigenvalue of
  \(B\), with corresponding eigenvector \(v\):
    \(Bv=\lambda v\).
  Let \(y\) be any vector in \({\mathbb R}^3\). Then there exists an
  \({\mathcal{O}_{y}} \in SO(3)\) such that \(\mathcal{O}_{y} y = \mu_{y} v\) for
  some \(\mu_{y}\in\R\), and
  therefore 
    \(y\cdot B y=\mathcal{O}_{y}y\cdot B\mathcal{O}_{y}y=\lambda\|y\|^2\).
  Since this is true for all \(y \in {\mathbb R}^3\), we get \(B=\lambda
  I_{3}\). A similar argument (with \(y=0\), and letting \(x\) vary) shows that
  also \(A\) is a multiple of the identity. Finally, the condition of
  harmonicity, \(\Tr[ A + B] = 0\), implies that \(A = -B = -\lambda
  I_{3}\). 

  Finally the term \(x\cdot C y\). This will be treated
  similarly. Due to the above (see \eqref{eq:A_B_C}), 
  \(x\cdot C y=P(x,y)-\lambda(y^{2}-x^{2})\). Therefore,
  \eqref{eq:inv_P} implies 
  \begin{align*}
    x\cdot C y = \mathcal{O}x\cdot C {\mathcal O} y \quad
    \text{ for all } {\mathcal O} \in SO(3). 
  \end{align*}
  By arguments similar to
  the above, we find that \(C\) is also a
  multiple of the identity \(I_{3}\). 
  Since \(P(x,y)=\lambda(x^{2}-y^{2})+x\cdot C y\),
  this finishes the proof of (\ref{lem:invarianceI}).
\end{pf*}
This finishes the proof of Lemma~\ref{lemma:first_invariance}.
\end{pf*}
\begin{lem}
  \label{lem:value_c0}
 Let \(\mathcal{Q}_{2}\) be the orthogonal projection (in
  \(L^{2}(\sphere^{5})\)) onto
  \begin{align*}
    \Span\left\{\left.P_{1}\right|_{\sphere{^5}}\right\}\quad,\quad
   P_{1}(x,y)=x\cdot y\quad,\quad (x,y)\in\R^{3}\times\R^{3},
  \end{align*}
  and let 
  \begin{align*}
    \gamma_{2}(x,y)=\left(\frac{x}{|x|}
  -\frac{y}{|y|}\right)\cdot\frac{x-y}{|x-y|}
  \quad,\quad (x,y)\in\R^{3}\times\R^{3}.
  \end{align*}
 Then
 \begin{align}
   \label{eq:value_c0}
   \mathcal{Q}_{2}\gamma_{2}
   = c_{1}\frac{x\cdot y}{x^{2}+y^{2}}\quad,\quad c_{1}=\frac{16(2-\pi)}{3\pi}.
 \end{align}
\end{lem}
\begin{pf}
Note that, with 
\begin{align*}
  Y(\omega)=\frac{\left.
                  P_{1}
              \right|_{\sphere^{5}}\!(\omega)
                }{ \big\| \left.
                  P_{1}
              \right|_{\sphere^{5}}
           \big\|_{L^{2}(\sphere^{5})}
          }
  \quad,\quad \omega=\frac{(x,y)}{\sqrt{x^{2}+y^{2}}},
\end{align*}
we have \(\|Y\|_{L^{2}(\sphere^{5})}=1\), and so
\begin{align}
  \label{eq:proj_Y}
  \mathcal{Q}_{2}\gamma_{2}(\omega)
  &=Y(\omega)\int_{\sphere^5}Y(\omega)\gamma_{2}(\omega)\,d\omega
  \\&
  =\left\{\frac{1}
         { \big\| \left.
                P_{1}
               \right|_{\sphere^{5}}
           \big\|_{L^{2}(\sphere^{5})}^{2}
    }
    \cdot 
    \int_{\sphere^5}
    \left.
                 P_{1}
               \right|_{\sphere^{5}}\!(\omega)\,
    \gamma_{2}(\omega)\,d\omega
    \right\}\cdot
    \frac{x\cdot y}{x^{2}+y^{2}}.
    \nonumber
\end{align}
We need to compute the two integrals in the brackets.

Since \(P_{1}\) is homogeneous of order \(2\) and \(\gamma_{2}\) of order \(0\)
(as functions on \(\R^{6}\)), we have
\begin{align*}
  \int_{B_{6}(0,R)} \!\!\!\!\!\!\!\!&P_{1}(x,y)\gamma_{2}(x,y)\,dx\,dy
  =\frac{R^{8}}{8}\,
   \int_{\sphere^5}
    \left.
        P_{1}
    \right|_{\sphere^{5}}\!(\omega)\,
    \gamma_{2}(\omega)\,d\omega.
\end{align*}
Therefore,
\begin{align}
 \label{eq:scalingP2} 
 \int_{\sphere^5}
    \left.
        P_{1}
    \right|_{\sphere^{5}}\!(\omega)\,
    \gamma_{2}(\omega)\,d\omega
  =8\int_{B_{6}(0,1)} 
  \!\!\!\!\!\!\!\!\!
  P_{1}(x,y)\gamma_{2}(x,y)\,dx\,dy.
\end{align}
Choose coordinates \((|x|, |y|, |x-y|, \Omega)\) for \(\R^{6}\) (with
\(\Omega\) three necessary angles). Note that
\begin{align*}
  P_{1}(x,y)&=x\cdot y = \frac{1}{2}\big(|x|^{2}+|y|^{2}-|x-y|^{2}\big)
  \quad,\quad (x,y)\in\R^{3}\times\R^{3},
  \intertext{and
  }
  \gamma_{2}(x,y)
  &=\frac{|x|+|y|}{|x-y|}
  \Big(1-\frac{|x|^{2}+|y|^{2}-|x-y|^{2}}{2|x||y|}\Big)
  \ ,\ (x,y)\in\R^{3}\times\R^{3}.
\end{align*}
 Then 
(see Hylleraas \cite[(45d)]{Hylleraas}; let \(s=|x|, t=|y|, r=|x-y|\))
\begin{align}
  \label{eq1:projection}
  &\int_{B_{6}(0,1)} P_{1}(x,y)\gamma_{2}(x,y)\,dx\,dy
  \nonumber
  =\frac{1}{4}\Big(\int\,d\Omega\Big)
  \times
  \nonumber
  \\&\times
  \int_{0}^{1}\!\int_{0}^{\sqrt{1-s^{2}}}\!\!\!\int_{|s-t|}^{s+t}
  (s^{2}+t^{2}-r^{2})(s+t)
  \big(2st-(s^{2}+t^{2}-r^{2})\big)\,dr\,dt\,ds\nonumber\\
  &= \frac{1}{4}\frac{(2-\pi)}{48}\int\,d\Omega. 
\end{align}
Using \eqref{eq:scalingP2}   and \eqref{eq1:projection} this means that
\begin{align}
  \label{eq:firstIntegral}
   \int_{\sphere^5}
    \left.
       P_{1}
    \right|_{\sphere^{5}}\!(\omega)\,
    \gamma_{2}(\omega)\,d\omega
   =\frac{2-\pi}{24}\int\,d\Omega. 
\end{align}

Next, observe that, again due to homogeneity, we have
\begin{align*}
   \int_{B_{6}(0,R)}(x\cdot y)^{2}\,dx\,dy
    =\frac{R^{10}}{10}\left.\big\|P_{1}\right|_{\sphere^{5}}
    \big\|_{L^{2}(\sphere^{5})}^{2}  
\end{align*}
and so
\begin{align}
  \label{eq:homgen2}
  \left.\big\|P_{1}\right|_{\sphere^{5}}\big\|_{L^{2}(\sphere^{5})}^{2}
  =10\int_{B_{6}(0,1)}(x\cdot y)^{2}\,dx\,dy.
\end{align}
Since \(x\cdot y=\frac12\big(|x|^{2}+|y|^{2}-|x-y|^{2}\big)\) we get
(using coordinates as above)
\begin{align*}
  &\int_{B_{6}(0,1)}(x\cdot y)^{2}\,dx\,dy
  \nonumber
  \\&
  =\frac14\Big(\int
  d\Omega\Big)\int_{0}^{1}\int_{0}^{\sqrt{1-s^{2}}}\int_{|s-t|}^{s+t}
  \big(s^2+t^2-r^2\big)^{2}srt\,dr\,dt\,ds
  \nonumber
  \\&
  =\frac{\pi}{1280} \int\,d\Omega.  
\end{align*}
This means (see \eqref{eq:homgen2}) that
\begin{align}
  \label{eq:normL2}
  \left.\big\|P_{1}\right|_{\sphere^{5}}\big\|_{L^{2}(\sphere^{5})}^{2}
  =\frac{\pi}{128}\int\,d\Omega. 
\end{align}
Now \eqref{eq:value_c0}
follows from \eqref{eq:proj_Y}, \eqref{eq:firstIntegral},
and \eqref{eq:normL2}.
This finishes the proof of Lemma~\ref{lem:value_c0}.
\end{pf}
\begin{acknowledgement}
All four authors thank the organizers of the program {\it Partial
  Differential Equations and Spectral Theory} for invitations 
to the Mittag-Leffler Institute in 2003 where part of the work was done.
Furthermore, parts of this work have been carried out at various
institutions, whose hospitality is gratefully acknowledged: Aalborg
University (SF, MHO, THO), The Erwin Schr\"{o}dinger Institute (T\O
S), Universit\'{e} Paris-Sud (T\O S), and the IH\'ES (T\O S).
Financial support from the Danish Natural Science Research Council, 
European Science Foundation Programme {\it Spectral Theory and Partial
  Differential Equations} (SPECT), and EU IHP network 
{\it Postdoctoral Training Program in Mathematical Analysis of
Large Quantum Systems},
contract no.\
HPRN-CT-2002-00277 is
gratefully acknowledged.\\
SF has been supported by a Marie Curie Fellowship of the European
Community Programme `Improving the Human Research Potential and the
Socio-Economic Knowledge Base' under contract number HPMF-CT-2002-01822,
and by a grant from the Carlsberg
Foundation.
\\
Finally, SF and T\O S wish to thank I. Herbst for useful discussions
at the Mittag-Leffler Institute.
\end{acknowledgement}
\bibliographystyle{hamsplain}

\providecommand{\bysame}{\leavevmode\hbox to3em{\hrulefill}\thinspace}

\end{document}